\documentclass[twocolumn,aps,prd,floatfix,preprintnumbers,a4paper,nofootinbib,superscriptaddress]{revtex4-1}
\usepackage{graphicx}				
\usepackage{placeins}
\usepackage{amssymb}
\usepackage{amsmath}
\usepackage[dvipsnames]{xcolor}
\usepackage{subcaption}
\usepackage{xcolor,soul}

\usepackage{verbatim}
\usepackage{rotating}
\usepackage{comment}
\usepackage{ragged2e}

\usepackage[colorlinks]{hyperref}

\begin{document}

\newcommand\red[1]{{\color[rgb]{0.75,0.0,0.0} #1}}
\newcommand\green[1]{{\color[rgb]{0.0,0.60,0.08} #1}}
\newcommand\blue[1]{{\color[rgb]{0,0.20,0.65} #1}}
\newcommand\cyan[1]{{\color[HTML]{00c3ff} #1}}
\newcommand\bluey[1]{{\color[rgb]{0.11,0.20,0.4} #1}}
\newcommand\gray[1]{{\color[rgb]{0.7,0.70,0.7} #1}}
\newcommand\grey[1]{{\color[rgb]{0.7,0.70,0.7} #1}}
\newcommand\white[1]{{\color[rgb]{1,1,1} #1}}
\newcommand\darkgray[1]{{\color[rgb]{0.3,0.30,0.3} #1}}
\newcommand\orange[1]{{\color[rgb]{.86,0.24,0.08} #1}}
\newcommand\purple[1]{{\color[rgb]{0.45,0.10,0.45} #1}}
\newcommand\note[1]{\colorbox[rgb]{0.85,0.94,1}{\textcolor{black}{\textsc{\textsf{#1}}}}}
\def\gw#1{gravitational wave#1}
\def\oed#1{QA frame#1}
\def\nr#1{numerical relativity
 (NR)#1\gdef\nr{NR}}
\def\bh#1{black-hole
 (BH)#1\gdef\bh{BH}}
 \def\bbh#1{binary black hole#1
  (BBH#1)\gdef\bbh{BBH}}
\def\qa#1{quadrupole-aligned
(QA)#1\gdef\qa{QA}}
\def\pn#1{post-Newtonian
 (PN)#1\gdef\pn{PN}}
 \def\qnm#1{Quasinormal Mode
    (QNM)#1\gdef\qnm{QNM}}
   \def\eob#1{effective-one-body
      (EOB)#1\gdef\eob{EOB}}
\def\imr#1{inspiral-merger-ringdown
 (IMR)#1\gdef\imr{IMR}}
 \def\fig#1{Fig.~(\ref{#1})}
 \def\cfig#1{Fig.~\ref{#1}}
 \newcommand{\figs}[2]{Figures~(\ref{#1}-\ref{#2})}
 \def\eqn#1{Eq.~(\ref{#1})}
 \def\ceqn#1{Eq.~\ref{#1}}
 \newcommand{\Eqns}[2]{Equations~(\ref{#1})-(\ref{#2})}
 \newcommand{\eqns}[2]{Eqs.~(\ref{#1})-(\ref{#2})}
 \newcommand{\ceqns}[2]{Eqs.~\ref{#1}-\ref{#2}}
 \def\lm{{\ell m}}
 \def\tpsi{\tilde{\psi}}
\def\dcp{{\sc{PhenomDCP}}}
\def\d{{\sc{PhenomD}}}
\def\fdamp{f_1}
\def\fring{f_0}

\definecolor{brown-ish}{RGB}{181,180,20}

\newcommand{\mdh}[1]{{\color{blue}{ Mark: #1}}}
\newcommand{\JT}[1]{{\color{orange}{ JT: #1}}}
\newcommand{\PK}[1]{{\color{RubineRed}{ Penny: #1}}}

\newcommand{\Caltech}{Theoretical Astrophysics Group, California Institute of Technology, Pasadena, CA 91125, U.S.A.}
\newcommand{\Cardiff}{School of Physics and Astronomy, Cardiff University, Cardiff, CF24 3AA, United Kingdom}
\newcommand{\BHAM}{School of Physics and Astronomy and Institute for Gravitational Wave Astronomy, University of Birmingham, Edgbaston, Birmingham, B15 2TT, United Kingdom
}

\title{Impact of anti-symmetric contributions to signal multipoles in the measurement of black-hole spins}

\author{Panagiota Kolitsidou} \affiliation{\BHAM}\affiliation{\Cardiff}
\author{Jonathan E. Thompson} \affiliation{\Caltech}\affiliation{\Cardiff}
\author{Mark Hannam} \affiliation{\Cardiff}

\begin{abstract}
Many current models for the gravitational-wave signal from precessing black-hole binaries neglect an asymmetry in the 
$\pm m$ multipoles. The asymmetry is weak, but is responsible for out-of-plane recoil, which for the final black hole
can be several thousand km/s. In this work we show that the multipole asymmetry is also necessary to accurately measure
the black-hole spins. We consider synthetic signals calculated from the numerical relativity surrogate model \texttt{NRSur7dq4}, which includes the
multipole asymmetry, and measure the signal parameters using two versions of the same model, one with and one without
the multipole asymmetry included. We find that in high signal-to-noise-ratio observations where the spin
magnitude and direction can in principle be measured accurately, neglecting the multipole asymmetry can result in
biased measurements of these quantities. Measurements of the black-hole masses and the standard aligned-spin combination 
$\chi_{\rm eff}$ are not in general strongly affected. As an illustration of the impact of the multipole asymmetry on a real signal we consider 
the LVK observation GW200129\_065458, and find that the inclusion of the multipole asymmetry is necessary to 
 identify the binary as unequal-mass and a high in-plane spin in the primary. 
\end{abstract}

\maketitle

\section{Introduction}
\label{sec:intro}

Gravitational-wave (GW) observations of binary black holes (BBHs) have begun to uncover the astrophysical population of
stellar-mass black holes (BHs) in the universe~\cite{LIGOScientific:2020kqk,KAGRA:2021duu}. The distribution of BH masses 
and their angular momenta (spins) also provides hints
of the dominant binary formation mechanisms (Refs.~\cite{LIGOScientific:2016vpg,LIGOScientific:2018jsj,LIGOScientific:2020kqk,KAGRA:2021duu} 
and references therein). In the $\sim$80 BBH observations in the first three LIGO-Virgo-KAGRA
(LVK)~\cite{2015DEFrange,acernese:2014advanced,2019kagra}  
observing runs from 2015 to 2020, most signals have been too weak to allow measurements of the full spin information for both 
BHs in a binary, and astrophysical inference has relied primarily on the distribution of black-hole masses, and the most accurately
measured combination of the two spins, a mass-weighted sum of the spin components aligned with the binary's orbital angular
momentum, $\chi_{\rm eff}$~\cite{Ajith:2009bn}. As detector sensitivities improve, and we accrue more observations, more signals 
will be loud enough for us to also measure the in-plane spin components, and to distinguish both spins. 

Accurate spin
measurements will also require sufficiently accurate theoretical signal models. All current models rely on a combination of 
approximate semi-analytic calculations and/or numerical solutions of Einstein's equations, and their physical fidelity is limited by the
accuracy of each of these inputs, and also physical approximations used to simplify the model construction. As we will discuss,
one 
simplification effectively neglects an asymmetry in the $\pm m$ spherical-harmonic multipoles. The purpose of this work is to study the 
impact of that asymmetry on the measurement of BH properties, in particular the spins.

The BHs in a binary are characterised by their masses, $m_1$ and $m_2$ (the total mass is $M = m_1 + m_2$), 
and their spin angular momenta, $\mathbf{S}_1$
and $\mathbf{S}_2$, which are usually expressed as the dimensionless vectors $\boldsymbol{\chi}_i = \boldsymbol{a}_i / m_i = \mathbf{S}_i / m_i^2$. 
When $\boldsymbol{\chi}_i$ are aligned with the binary's orbital angular momentum $\hat{\mathbf{L}}$ the orbital plane and spin 
directions remain constant. For these ``aligned-spin'' or ``non-precessing'' binaries, if the gravitational-wave signal is decomposed
in the spin-weighted spherical harmonics with weight $s=-2$, \begin{equation}
h(t, \theta, \phi) = \sum_{\ell, m} h_{\ell m}(t) {}^{-2}Y_{\ell m}( \theta, \phi),
\end{equation} the multipoles obey the reflection symmetry, \begin{equation}
h_{\ell m} = (-1)^{\ell} h^*_{\ell -m}.  \label{eq:symmetry}
\end{equation} 
This symmetry simplifies the construction of aligned-spin waveform models: we need only model the $+m$ 
multipoles and can then directly calculate the $-m$ multipoles from symmetry. 

When the spins are mis-aligned with the orbital angular momentum, the binary's orbital plane and spins 
precess~\cite{Apostolatos:1994sio,Kidder:1995cbsv}.
This complicates the modelling process, and many models make use of a convenient approximation: during the inspiral we
can consider the signal from a co-precessing frame that tracks the precession of the orbital angular momentum. All current 
precessing-binary models employ the idea of a co-precessing frame. In this
frame the signal equals, to a good approximation, that from a non-precessing system with the same aligned-spin 
components $\boldsymbol{\chi}_i \cdot \hat{\mathbf{L}}$~\cite{Schmidt:2012tmg}. One way to produce an approximate precessing-binary
model, then, is to take a non-precessing-binary model and apply a time-dependent rotation to introduce any 
precession dynamics. Some variant of that approximation is used in all of the {\tt Phenom} and {\tt SEOBNR} models
used in LVK analyses to 
date~\cite{Hannam:2013oca,Khan:2018fmp,Khan:2019kot,Pratten:2020ceb,Estelles:2021gvs,Hamilton:2021pkf,Pan:2014yaq,Taracchini:2014hgd,Ossokine:2020thd,Ramos-Buades:2023ehm}. 

One consequence of using aligned-spin multipoles to approximate the co-precessing-frame signal is that these by
construction obey the symmetry in Eq.~(\ref{eq:symmetry}), which no longer holds for precessing binaries. (As noted
in Ref.~\cite{Boyle:2014ioa}, rotations cannot restore this symmetry.) The anti-symmetric contribution in the
co-precessing frame is weak, and in many cases, for example for signals with typical signal-to-noise ratios (SNRs) in
the LVK observations to date, we may expect that this approximation
is valid. 
(Ref.~\cite{Islam:2021PhI} has verified this for GW190412, which has a SNR of $\sim$19 and no evidence for precession.)
However, since we generally require loud signals to measure mis-aligned spins and precession~\cite{Green:2020ptm},
these are precisely the kinds of signals where the anti-symmetric contribution may be important. This point was
previously made in Ref.~\cite{Kalaghatgi:2020gsq}, which showed that neglecting the anti-symmetric contribution might lead
to parameter biases in observations with SNRs as low as 15, depending on the binary's orientation and 
polarisation. 

The most striking physical consequence of the multipole asymmetry is out-of-plane recoil of the final black 
hole~\cite{Kidder:1995cbsv,Gonzalez:2007srv,Campanelli:2007ew}.
Ref.~\cite{Bruegmann:2007bri} shows that in the ``superkick'' configuration (an equal-mass binary with equal spin on 
each back hole, but with the spins lying in the orbital plane and in opposite directions), the magnitude of the out-of-plane
recoil depends sinusoidally on the direction of the spins relative to the separation vector of the two black holes at some
reference frequency. The dependence of the phasing of the anti-symmetric contribution on the in-plane-spin direction is
also discussed in detail in Ref.~\cite{Ghosh:2023mhc}. Given the dependence of the anti-symmetric contribution on the spin
direction, and the results in Ref.~\cite{Kalaghatgi:2020gsq}, which looked directly at the distinguishability of waveforms from
systems with different in-plane-spin directions, we might expect that the anti-symmetric contribution will be important for
measuring in-plane spins. 

To study this effect, we make use of the surrogate model \texttt{NRSur7dq4}~\cite{Varma:2019csw}. This model \emph{does} include the 
multipole asymmetry, but we also consider a version with the anti-symmetric contribution set to zero. 
In Sec.~\ref{section:Waveform model} we discuss the model in more detail. In Sec.~\ref{section:Binary configurations} we 
outline our procedure to measure the properties from
a series of synthetic signals at high SNR (100), using both the full and symmetric-only versions of the \texttt{NRSur7dq4} model. 
The results are presented in Sec.~\ref{section:Results}, where we also consider the LVK observation 
GW200129\_065458~\cite{LIGOScientific:2021djp} (hereafter referred to as GW200129), which is the first 
observation for which claims have been made of strong evidence for precession~\cite{Hannam:2021pit} and large 
recoil~\cite{Varma:2022pld}. In the next section, however, we will first summarise the features of the multipole asymmetry 
and the questions we will address in this paper.

\section{Multipole asymmetry and questions for study}
\label{sec:background}

The multipole asymmetry is discussed in more detail in Ref.~\cite{Ghosh:2023mhc}, but we summarise the main features here, and 
our expectations for how the asymmetry might impact parameter measurements, to be tested in this work. 

The GW multipoles $h_{\ell m}(t)$ may be split into symmetric and anti-symmetric contributions. As an example, we write the $(\ell=2, |m|=2)$
multipoles as  \begin{eqnarray}
h_{2,2}(t) & = & A(t) e^{-i \phi_s(t)} + a(t) e^{-i \phi_a(t)}, \label{eq:h22} \\
h_{2,-2}(t) & = & A(t) e^{i \phi_s(t)} - a(t) e^{i \phi_a(t)} \label{eq:h2m2},
\end{eqnarray} where $A(t)$ and $\phi_s(t)$ are the symmetric amplitude and phase, and $a(t)$ and $\phi_a(t)$ are the anti-symmetric
amplitude and phase, and $a(t)/A(t) \ll 1$; see Ref.~\cite{Ghosh:2023mhc} for examples. The amplitude of the anti-symmetric contribution $a(t)$
is approximately proportional to the magnitude of the in-plane spin, and $a(t) = 0$ for aligned-spin systems. If we consider single-spin
systems in a co-precessing frame, then during the inspiral $\phi_a(t) = \Phi(t) + \alpha(t) + \phi_0$, where $\Phi(t)$ is the binary's orbital phase, $\alpha(t)$ is the 
precession angle of the black hole's spin and $\phi_0$ is an overall constant. These details will be different for the anti-symmetric contribution to 
other multipoles, but a general feature of the anti-symmetric contribution to all multipoles
is that an overall in-plane spin rotation of $\Delta \alpha$ will introduce a shift of $\Delta \alpha$ into each of the anti-symmetric phases. 
This phase shift manifests itself in out-of-plane recoil. This is discussed in Ref.~\cite{Bruegmann:2007bri} for the superkick configurations, 
where the magnitude of the recoil varies sinusoidally with $\Delta \alpha$. 

Given this basic phenomenology of the anti-symmetric contribution, we may consider how we expect it to influence 
measurements of the black-hole masses and spins. We make four points, each of which we will return to in the results of our 
parameter-estimation study. 

1. Since the anti-symmetric part depends on the in-plane spins, we do not expect it (or its absence from a waveform model) to influence parameter measurements unless the signal is strong enough for in-plane spin information to be measurable. This 
motivates our choice of high SNR signals in our injection study, since in those cases we can be confident that the magnitude
and tilt or misalignment angle of each spin with the direction of the orbital angular momentum, $\theta_{LS_{1,2}}$, 
should be measurable. Conversely, we expect that the absence of the anti-symmetric contribution
in a model will lead to biases in the spin measurements, but are less likely to bias parameters that are independent of
the in-plane spin components, like the total mass, mass ratio, and aligned-spin combination 
$\chi_{\rm eff} = (m_1 \boldsymbol{\chi}_1 \cdot \hat{\mathbf{L}} + m_2 \boldsymbol{\chi}_2 \cdot \hat{\mathbf{L}})/M$. 
($\chi_{\rm eff}$ affects the inspiral rate~\cite{Cutler:1994ys,Poisson:1995ef,Ajith:2011ec}, and therefore the overall binary phasing, 
and so is likely to be measured well regardless of the multipole asymmetry, which has minimal, if any, effect on the rate of inspiral.)
This 
implies, for example, that the individual aligned-spin components, $\boldsymbol{\chi}_1 \cdot \hat{\mathbf{L}}$ and 
$\boldsymbol{\chi}_2 \cdot \hat{\mathbf{L}}$, may exhibit significant biases, but the $\chi_{\rm eff}$ combination will be
fairly well constrained. 

2. As noted above, changes in the initial in-plane-spin direction will introduce an overall phase offset into the anti-symmetric contribution, and
this affects the out-of-plane recoil. We might expect that the power in the anti-symmetric contribution also varies with
the in-plane-spin direction, and perhaps there is a correlation: the importance of the anti-symmetric contribution in parameter measurements
(and the extent of the bias when the anti-symmetric contribution is neglected) may be large for cases with large recoil, and small for cases
with small recoil. However, the signal's SNR depends on $|h|^2$ as observed at the detector (i.e., from one direction), 
while the recoil depends on $|\dot{h}|^2$ integrated over the entire sphere. (See Ref.~\cite{Ruiz:2008GRe} for useful expressions for
radiated linear and angular momenta.) There is therefore no reason to expect, for example,
that a large recoil in general corresponds to a larger importance of the anti-symmetric contribution on the parameter measurements. 

3. The power in the signal is dominated by the symmetric part of the $(\ell=2, |m| = 2)$ multipole in all of the cases we consider. Even when
the signal is nominally edge-on, the majority of the signal power is in the plus polarisation, where the total power in the $(\ell=2, |m| = 2)$
multipoles is comparable in face-on or face-off configurations. Since the overall amplitude of the anti-symmetric (2,2) contribution is a ratio 
$a(t)/A(t)$ of the symmetric contribution that depends only on the intrinsic parameters of the binary, the fraction of the total power in the 
anti-symmetric contribution will be roughly the same regardless of the orientation. We therefore expect that any biases due to neglecting
the anti-symmetric contribution will be of similar magnitude regardless of the binary's orientation, for fixed total SNR. 
This may initially seem counter-intuitive from the definition of the asymmetry (we might expect the asymmetry contributions to 
cancel out for edge-on systems); we explain why this is not the case in Sec.~\ref{sec:dependence}. 

4. Since the magnitude of the anti-symmetric contribution $a(t)$ depends on the in-plane-spin magnitude, we \emph{do} expect
the bias due to neglecting the anti-symmetric contribution to be larger for configurations with larger in-plane spins. 

These considerations provide us with a series of predictions to test in our injection study: we expect that neglecting the 
anti-symmetric contribution will lead to a bias in the spin measurements, within the broad constraint on the measurement of 
$\chi_{\rm eff}$, and minimal or no bias in the masses, and that the extent
of the bias will be broadly independent of the binary's orientation and recoil, but \emph{will} be roughly proportional to the 
magnitude of the in-plane spins. We will consider each of these predictions in our results, and find that they hold for 
most (but not all) of our high-SNR injections. For the GW signal GW200129, however, we find that neglecting the anti-symmetric contribution
\emph{does} affect the measurement of the mass ratio, although this is a signal where imprints of in-plane spins (i.e., precession)
on the signal are only just beyond the threshold of measurability. We also note that our study is limited to single-spin systems;
the phenomenology is likely to be more complex in two-spin configurations.

\section{Waveform model}
\label{section:Waveform model}

Two families of models have been used for most of the LVK measurements of binary properties,
{\tt Phenom} and {\tt SEOBNR}~\cite{Hannam:2014ghd,Khan:2020hdg,Pratten:2020ceb,Pan:2014yaq,Taracchini:2014hgd,Purrer:2015tud,Bohe:2017hgs,Ossokine:2020thd}. As noted above, the versions of these models available through the first three LVK 
observing runs did not include the multipole asymmetry.  A third class of precessing-binary models, NR surrogates, \emph{do} include
the multipole asymmetry~\cite{Blackman:2017sdr,Blackman:2017nrw,Varma:2019csw}. 
We will use two variants of the \texttt{NRSur7dq4} model~\cite{Varma:2019csw} to determine the impact of neglecting the multipole asymmetry 
in BBH measurements. 

The \texttt{NRSur7dq4} model has been built from numerical relativity simulations with mass ratios $1\leq q = m_1/m_2 \leq 4$, 
generic spin directions and spin magnitudes up to $0.8$ and includes all $l \leq 4$ spin-weighted spherical-harmonic multipoles.
In addition, these NR simulations start at $\sim$20 orbits or $\sim$4300$M$ prior to merger. 
Therefore, the surrogate models are restricted to waveforms of this length and are inadequate whenever longer waveforms are required.
Assuming for example a waveform with a starting frequency of 20\,Hz, the surrogate will only be valid for binaries with total masses $M\gtrsim 65\,M_{\odot}$ depending on the mass ratio and the spins of the system~\cite{Varma:2019csw}.    
However, within its range of validity \texttt{NRSur7dq4} is currently the most accurate waveform model available.
   
   To perform a systematics study we isolated the effect of the multipole asymmetry on the parameter estimation results by using two versions of the \texttt{NRSur7dq4} model, the ``full'' \texttt{NRSur7dq4} and the ``symmetric'' \texttt{NRSur7dq4}. The full \texttt{NRSur7dq4} is the original \texttt{NRSur7dq4} waveform model without any alterations. 
The symmetric \texttt{NRSur7dq4} is a modified version of this model with the anti-symmetric contribution removed, as follows. 
In the surrogate model, the following contributions are modelled in the co-orbital frame, 
   \begin{equation}\label{eq:hlmplus}
    h_{lm}^{\pm}=\frac{h_{\ell m}^{coorb} \pm (-1)^\ell h_{\ell-m}^{coorb*}}{2}.
    \end{equation}
 For even $\ell$ the symmetric contribution is $h_{\ell m}^+$ and the anti-symmetric contribution is $h_{\ell m}^-$, and for odd $\ell$ it is
 the reverse. We expect that only the $\ell=2$ anti-symmetric contribution is significant for our results, since the anti-symmetric 
 contribution to higher multipoles is in general weaker than the symmetric $\ell=4$ contributions. Nonetheless, in the symmetric version of
 the model we set $h_{\ell m}^-$ to zero for $\ell = 2, 4$ and set $h_{\ell m}^+$ to zero for $\ell = 3$. The symmetric model
was constructed from the implementation of  \texttt{NRSur7dq4}  in the \texttt{LALSuite} software library~\cite{lalsuite}. We refer to the symmetric \texttt{NRSur7dq4} model as \texttt{NRSur7dq4\_sym} to simplify notation.


\section{Parameter estimation analysis}
\label{section:Binary configurations}

We perform two investigations. In the first we consider synthetic (full) \texttt{NRSur7dq4} signals with SNR 100, and compare measurements of their parameters using both the full and symmetric versions of \texttt{NRSur7dq4}. In the second, 
to explore the impact that neglecting multipole asymmetry has on current and near-future observations, we also use the \texttt{NRSur7dq4} and \texttt{NRSur7dq4\_sym} models to analyse the public detector data of the precessing signal GW200129 that have undergone glitch removal, 
 which we refer to as ``de-glitched" data~\cite{LIGOScientific:2021djp,Hannam:2021pit}. One aspect of the GW200129 observation not considered 
 in Refs.~\cite{Hannam:2021pit,Varma:2022pld} was the impact of the method used to ``de-glitch'' the data.  Ref.~\cite{Payne:2022spz} argues that 
 incomplete glitch removal may lead to a spurious precession measurement. However, their analysis is limited by modelling the data as a 
 \emph{non-precessing} 
 signal plus a glitch; earlier tests on precessing injections~\cite{Hourihane:2022doe} may not be sufficient to show that the method can reliably distinguish
 between precession and glitches, because in each of the test injections the precession contributed insufficient power to 
 be measurable. A more recent analysis, which does not rely on these assumptions~\cite{Macas:2023wiw}, suggests that with a more accurate 
 glitch-subtraction procedure, the evidence for precession \emph{increases}. However, for the purposes of the analysis in this paper, where we are 
 concerned with how results vary with respect to different models used to analyse the same set of data, the details of how those data were produced 
 are less relevant.
	
\begin{figure}[ht!]
\includegraphics[width=1.\linewidth]{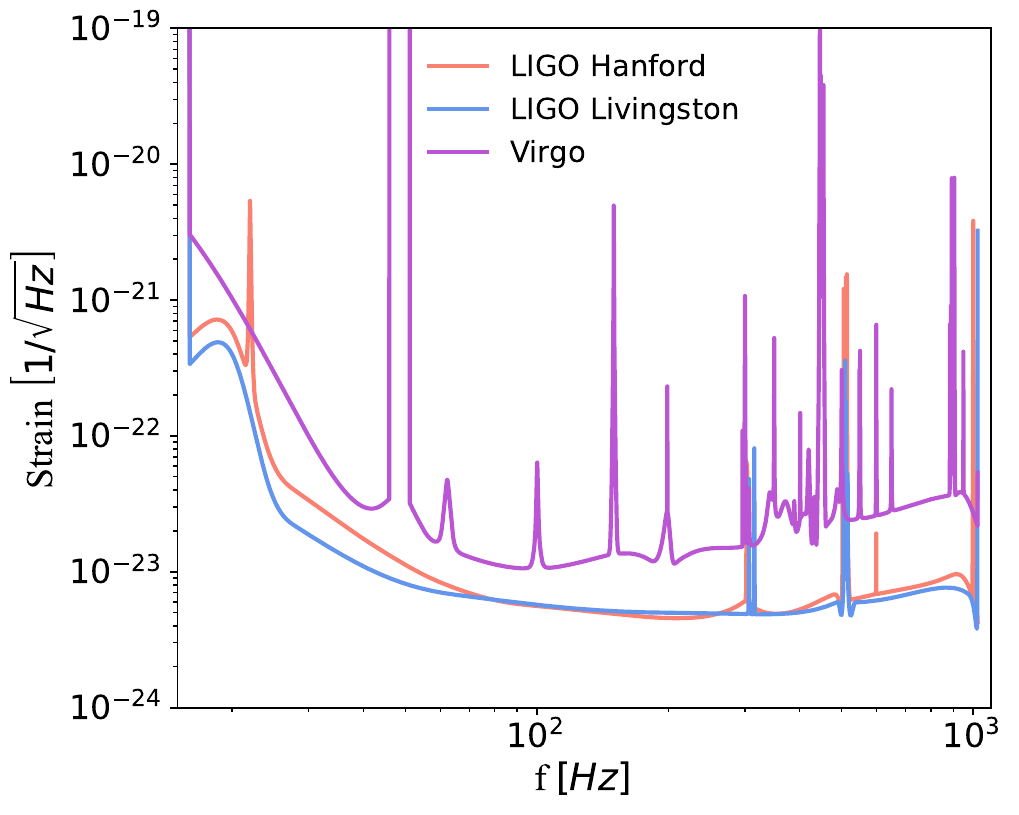}
\caption{Amplitude spectral density of the
three interferometers' strain sensitivity: LIGO Livingston, LIGO Hanford, Virgo. The square of the amplitude spectral density gives the power spectral density of the detectors.
}
\label{fig:asd}
\end{figure}

The analysis is performed using the Markov Chain Monte Carlo (MCMC) stochastic sampling technique from the \texttt{LALInference} software library presented in Ref.~\cite{veitch2015parameter} that was used for the first observing runs, O1-O2~\cite{LIGOScientific:2016aoc,LIGOScientific:2018mvr,LIGOScientific:2021djp, LIGOScientific:2020ibl,Abbott:2021GWTC21}. 
For our analysis, we use all three detectors and publicly available power spectral densities that were taken during the O3b observing run. These are the same power spectral densities that were used in the analysis of the GW200129 signal in Refs.~\cite{LIGOScientific:2021djp,Hannam:2021pit}. The corresponding sensitivity curves of the LIGO Hanford, LIGO Livingston and Virgo detectors are shown in Fig.~\ref{fig:asd}. 

In our parameter estimation analysis, we have chosen to use a flat prior over spin magnitude, the cosine of the tilt angle and the component masses.
The parameter estimation results can be significantly affected by the selected priors of the spin magnitudes and the tilt angles. Since there is no evident justification for employing a prior from the observed population or one motivated by other astrophysical factors, we have selected these particular priors that do not introduce strong assumptions about the underlying astrophysical population.  
These are the default priors that were also used in Refs.~\cite{LIGOScientific:2021djp,Hannam:2021pit}. 
Furthermore, the prior parameter space has been adjusted to not exceed significantly the validity range of the surrogate model, setting the total mass to be 
$M \geq 68 M_ \odot$, the chirp mass to be within $14.5 M_ \odot$ and $49 M_ \odot$ and the mass ratio to be less than 1:4 or 1:6 depending on 
the configuration. We chose the minimum frequency where the analysis starts to be 20\,Hz.
The \texttt{NRSur7dq4} waveforms were generated with starting time that corresponds to 11\,Hz for the $(\ell=2, |m|=2)$ multipole, to ensure that the 
highest-frequency multipoles, $(\ell=3,|m|=3)$, also start below 20\,Hz. 

In the case of the \texttt{NRSur7dq4} injections, the data were all injected with an SNR of $100$ and start at $20$ Hz using the same basic setup as the O3 catalog~\cite{LIGOScientific:2021djp}.
For their sky location, the declination is $\delta=1.4323$ rads and right ascension $\alpha=0.2896$ rads, while the polarisation is set to $\psi=1.4$ rads. 
Each production run produced approximately $\sim  10^5$ samples. Considering that for standard applications of the \texttt{LALInference} sampler $10^4$ is 
a typical amount of samples, we are confident that $10^5$ samples is a sufficient number. 
However, to further ensure the convergence of each run we took into account the behaviour 
of the autocorrelation function and the value of the Gelman-Rubin diagnostic~\cite{gelman1992inference}.

The \texttt{NRSur7dq4} data are injected in zero-noise meaning that the detector noise is set to zero while the power spectral densities of the detectors (see Fig.~\ref{fig:asd}) are used to compute the likelihood. In the zero-noise injection, the noise is removed, but the parameter estimation analysis is performed with the relative frequency-dependent sensitivity (noise curve) that corresponds to each detector and for sky location, orientation and polarisation values appropriately also adjusted to the detectors allowing the computation of an SNR.
We can interpret the results obtained from this type of injection as an average over many Gaussian noise realisations. 

The Gaussian likelihood~\cite{Cutler:1994gmc} is given by the noise-weighted inner product~\cite{Finn:1992dmg}
\begin{equation}
\log \mathcal{L} \propto - \langle d(t)-h_M(\mathbf{\theta})|d(t)-h_M(\mathbf{\theta}) \rangle,
\end{equation}
where $h_M(\mathbf{\theta})$ is the waveform model evaluated at parameters $\mathbf{\theta}$ and 
$d(t)$ is the data given as the sum of the signal $s(t)$ and $n(t)$ the noise.
For a zero-noise injection, since $n(t)=0$, the data becomes $d(t)=s(t)$ and $\log \mathcal{L} \propto - \langle s(t)-h_M(\mathbf{\theta})|s(t)-h_M(\mathbf{\theta}) \rangle$. 
From the definition of the inner product between two waveforms $h_1$ and $h_2$,
\begin{equation}
\langle h_1|h_2 \rangle = 4 \Re \int_0^{\infty} \frac{h_1(f)h_2^*(f)}{S_n(f)} df,
\end{equation}
where $S_n(f)$ is the power spectral density,
it becomes clear that in the case of the zero-noise injections, the frequency-dependent sensitivity of the detectors is used in the calculation of the likelihood.
From the definition of the $\log$ likelihood, we note that if the model produces a waveform $h_M(\mathbf{\theta})$ that matches well the signal $s(t)$, the 
$\log$ likelihood $|\log \mathcal{L}|$ has a lower value. 

In the case of the GW200129 de-glitched data, the parameter estimation analysis is performed using the same settings as those employed in LVK 
GWTC-3 analysis~\cite{LIGOScientific:2021djp}, while also applying the additional settings described in Ref.~\cite{Hannam:2021pit} such as reducing the prior parameter space to fit within the validity range of the \texttt{NRSur7dq4}. For our analysis 
the waveform is generated at 20\,Hz and we have included all the $l \leq 3$ spin-weighted spherical-harmonic multipoles.

\subsection{NRSur7dq4 theoretical waveforms}
\label{section:NRSur7dq4 injections}

In the first part of this work, we use the \texttt{NRSur7dq4} waveform model to investigate how the absence of the multipole asymmetry from the model affects parameter measurement for a number of theoretical signals of strongly precessing binaries with high SNRs. Furthermore, we consider specific configurations that allow us to explore how the biases depend on the recoil velocity of the final black hole, the inclination of the system, the magnitude of the primary black hole's spin and the mass ratio of the binary, to compare against our phenomenological expectations from Sec.~\ref{sec:background}. 
In each of these cases the signal is
generated from the full \texttt{NRSur7dq4} waveform model, and the parameter recovery uses the \texttt{NRSur7dq4} and \texttt{NRSur7dq4\_sym} models. 

Our fiducial example was a binary with total mass $M = 100\,M_\odot$, mass-ratio $q = 2$, and a dimensionless primary spin magnitude of $a_1 / m_1 = 0.7$, 
with the spin directed entirely in the orbital plane, to maximise precession effects and the anti-symmetric contribution. Starting from this
basic configuration, we identified initial orientations of the in-plane spin to produce the maximum and minimum possible recoils. 
   	
\begin{figure}[ht!]
    \centering\includegraphics[width=1.\linewidth]{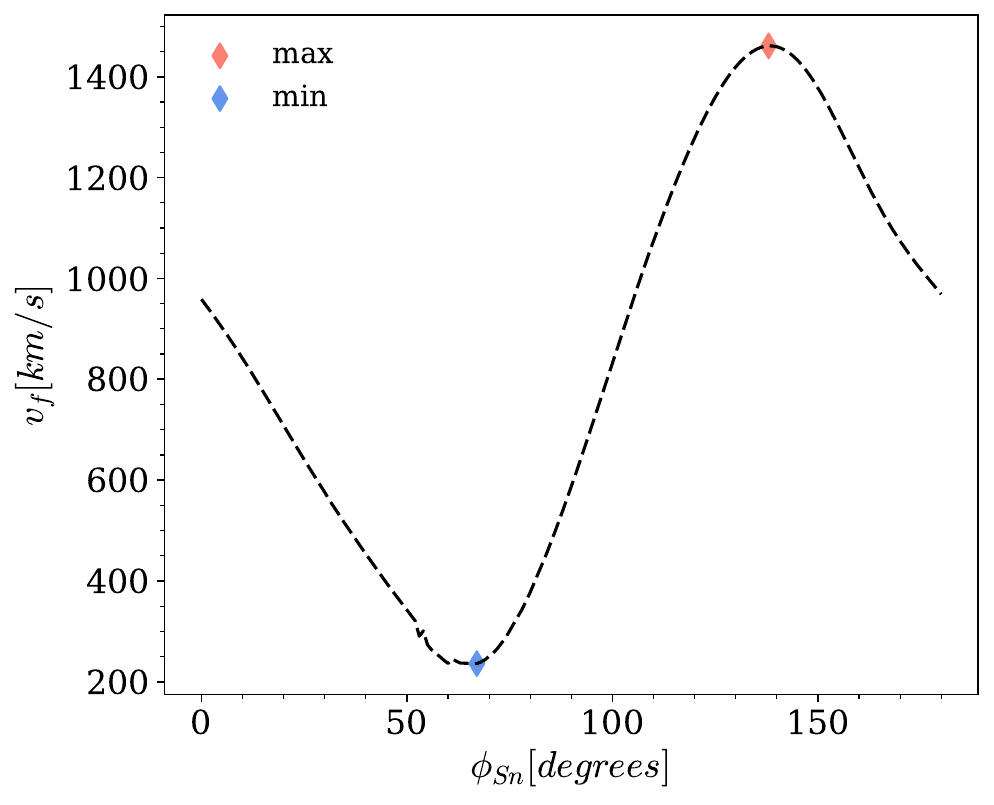}    
    \caption{The minimum (blue) and maximum (red) recoil velocity values and the corresponding in-plane spin direction angles that were selected for this study. }
\label{fig:recoil_values}	   
\end{figure}

We identified the maximum and minimum recoil by computing the recoil velocities for \texttt{NRSur7dq4} theoretical waveforms with varying 
in-plane spin directions of the binaries between $0^\circ$ and $180^\circ$. The in-plane spin  direction is denoted by the misalignment 
angle $\phi_{Sn}$ between the black holes' separation vector, $\hat{n}$, and the projection of the spin vector $\hat{S}$ on the orbital plane,
at the starting frequency. The waveforms were generated in the inertial $\mathbf{L_0}$-frame 
where $\hat{\mathbf{L}}=\hat{\mathbf{z}}$ at a reference time, satisfying \texttt{LAL} conventions 
using the \texttt{LALSimulation} function \texttt{SimInspiralChooseTDModes}~\cite{lalsuite,schmidt2017numerical}.
The recoil velocity was computed from the waveform multipoles~\cite{Ruiz:2007yx} in the final $\mathbf{J}$-frame where the 
$z$-axis is parallel to the total angular momentum, $\mathbf{J}$, of the remnant black hole. 
Fig.~\ref{fig:recoil_values} shows the measured recoil velocities for different $\phi_{Sn}$ angles. 
Based on these results, the lowest recoil velocity is 
$v_{f_\text{min}} = 236$\,km/s and the highest is $v_{f_\text{max}} = 1461$\,km/s. For these two cases the initial in-plane-spin directions $\phi_{Sn}$
are, respectively, $67^\circ$ and $138^\circ$.

In our signal injections we choose to characterise the binary inclination relative to the direction of maximum asymmetry emission at merger. 
(This is motivated in Sec.~\ref{sec:dependence}.)
In general in precessing systems the binary inclination can be defined in multiple ways: we can consider the orientation of the observer relative to the
direction of the total angular momentum, $\mathbf{J}$, which is the closest we have to a fixed direction in precessing binaries. However,
if the binary is precessing then by definition $\mathbf{J}$ is \emph{never} the normal to the orbital plane. A common alternative definition of
the inclination (adopted in the \texttt{LAL} infrastructure) is the direction of the observer relative to the orbital angular momentum, $\mathbf{L}_0$,
 at the frequency when the signal
enters the detector's sensitivity band. Since $\mathbf{L}$ precesses during inspiral, this definition describes the orientation of the orbital
plane to the observer at only one moment; at other points during the inspiral the actual orientation can in principle take on any value. 
Given these ambiguities, we choose a definition of the inclination relevant to the direction of maximum power in the antisymmetric
contribution to the signal. 

To do this, we use as a proxy for the merger time $t_m$ the time when the magnitude of the $\ell=2$ multipoles (added in quadrature) is maximum.
We then identify the direction that maximises the $(\ell=2, |m|=2)$ power at $t_m$ (motivated by the definition of the 
quadrupole-aligned frame~\cite{Schmidt:2011tpc}); this will also be the direction that maximises the power in the anti-symmetric $(2,2)$
contribution. We define inclination relative to this direction, i.e., $\iota = 0^\circ$ corresponds to the observer being face-on to the direction
of maximum emission at merger. In practice, to impose this in our injections using the \texttt{LAL} infrastructure, we first rotated our
signal multipoles so that the maximum emission at $t_m$ was along the $z$-axis, and then prevented \texttt{LAL} from performing a 
frame rotation by artificially setting $\hat{\mathbf{L}}$ to be along the $z$-axis in the waveform metadata.

The two \texttt{NRSur7dq4} waveforms (corresponding to maximum and minimum recoil) were injected with different inclinations, 
varying from $0^\circ$ to $90^\circ$ in steps of $30^\circ$. This allowed us to investigate how the inclination of the detected system affects the 
biases that the asymmetry's absence may introduce in the parameter estimation results.

In addition, to test how the \texttt{NRSur7dq4\_sym} model behaves for different mass ratios and spin magnitudes, 
we performed two additional injections. The selected configurations for that purpose are a binary black hole 
configuration with mass ratio $q=2$ and a smaller in-plane spin of magnitude $a_1/m_1=0.4$, 
and a binary 
with a higher mass ratio $q=4$ and slightly higher in-plane spin of magnitude $a_1/m_1=0.8$. In these last 
two cases, the in-plane spin direction is $\phi_{Sn}=0^\circ$ and the total mass of these binary is $M=100 M_{\odot}$. 
The selected inclination is $\iota = 60^\circ$ and they are both injected at SNR $100$. For these additional injections
we use the standard \texttt{LAL} definition of inclination.

To summarise, we performed 20 parameter-estimation analyses of 10 configurations using the \texttt{NRSur7dq4} and \texttt{NRSur7dq4\_sym} models: the maximum- and minimum-recoil versions of the fiducial configuration, at orientations $\iota = 0^\circ, 30^\circ, 60^\circ, 90^\circ$ and two additional single-spin configurations $(q=2, a_1/m_1=0.4,\theta_{\rm LS} = 90^\circ)$ 
and $(q=4, a_1/m_1=0.8, \theta_{\rm LS} = 90^\circ)$ 
at orientation $\iota = 60^\circ$. 
We will show results from a representative subset of these analyses in Sec.~\ref{section:Results}.

\subsection{GW200129 gravitational wave signal}
\label{section:GW200129 injection}

In the second part of this work, we consider the GW200129 gravitational wave signal that was first reported in Ref.~\cite{LIGOScientific:2021djp}. 
Ref.~\cite{Hannam:2021pit} presented strong evidence that GW200129 was the first GW observation of a precessing binary, with masses 
$m_1=39 M_{\odot}$ and $m_2=22 M_{\odot}$, and the primary black hole rapidly spinning with $a_1/m_1=0.9$,
 and the spin lying almost 
entirely in the orbital plane. The measured parameters of the signal calculated with the \texttt{NRSur7dq4} are displayed in Table 1 of 
Ref.~\cite{Hannam:2021pit}. The total network SNR of GW200129 is $26.5$ and the SNRs in each detector were  measured to be $14.6$ in 
Hanford, $21.2$ in Livingston and $6.3$ in Virgo.
Ref.~\cite{Varma:2022pld} also showed that the GW200129 has a large recoil velocity of $v_f=1542$\,km/s, which suggests that the anti-symmetric 
contribution to the signal was measurable and could significantly influence the parameter estimates. 

We test the importance of the anti-symmetric contribution by also analysing GW200129 with \texttt{NRSur7dq4\_sym}. As noted
in Sec.~\ref{section:Binary configurations}, besides the change in the model used in the analysis, all other settings are the same as in 
the analysis reported in Ref.~\cite{Hannam:2021pit}.


\section{Results}
\label{section:Results}

We present our results as follows.
We first consider the importance of the multipole asymmetry on measurements of our fiducial high-SNR configuration, in Sec.~\ref{section:Generic example};
this allows us to examine expectation (1) from Sec.~\ref{sec:background}. We then consider expectations (2), (3) and (4) in Sec.~\ref{sec:dependence} 
by considering variations in recoil, orientation and mass-ratio and spin magnitude. We then look at the importance of the multipole asymmetry 
on GW200129 in Sec.~\ref{sec:gw200129}.

\subsection{The impact of the anti-symmetric contribution}
\label{section:Generic example}

\begin{figure*}[ht!]
\includegraphics[width=0.3\linewidth]{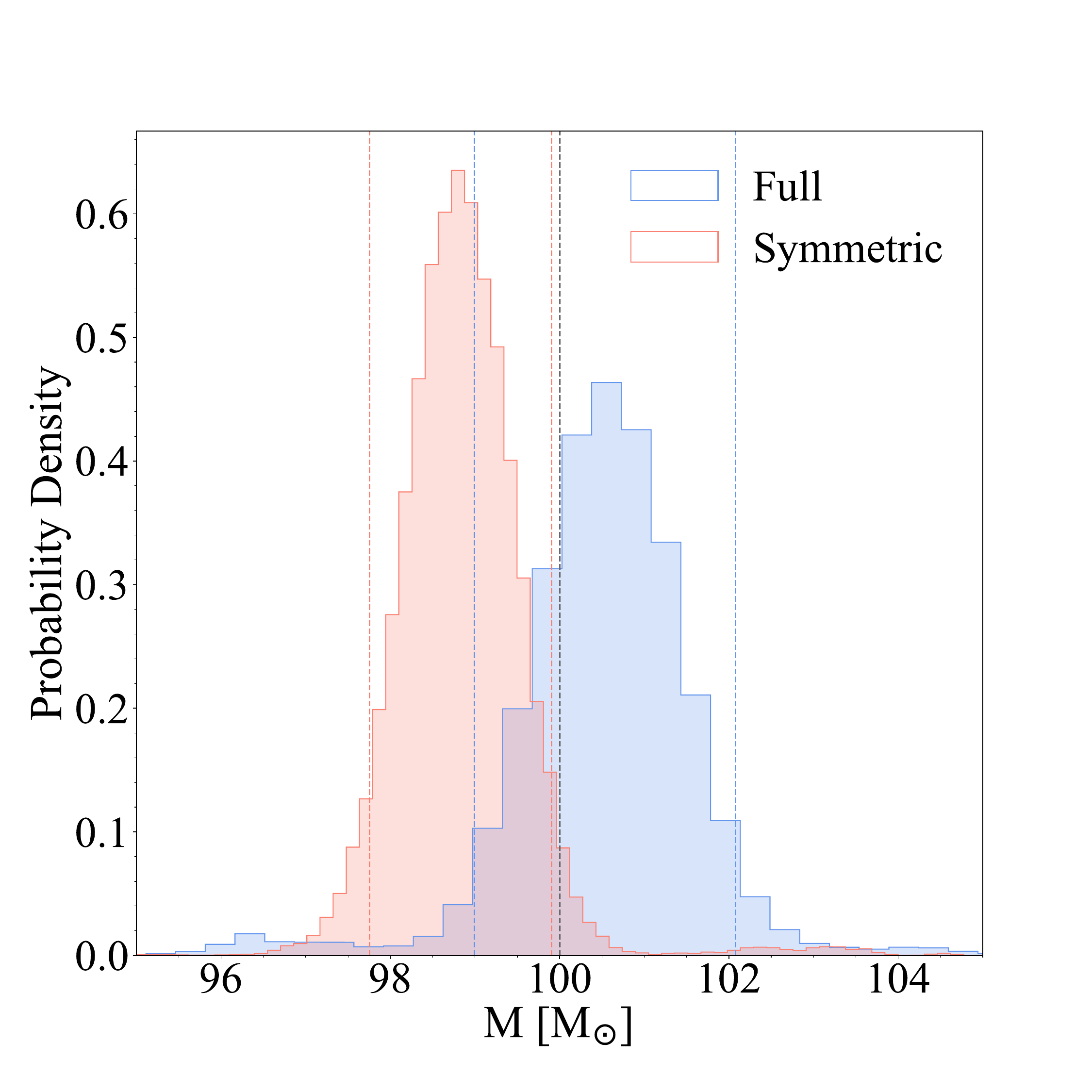}
\includegraphics[width=0.3\linewidth]{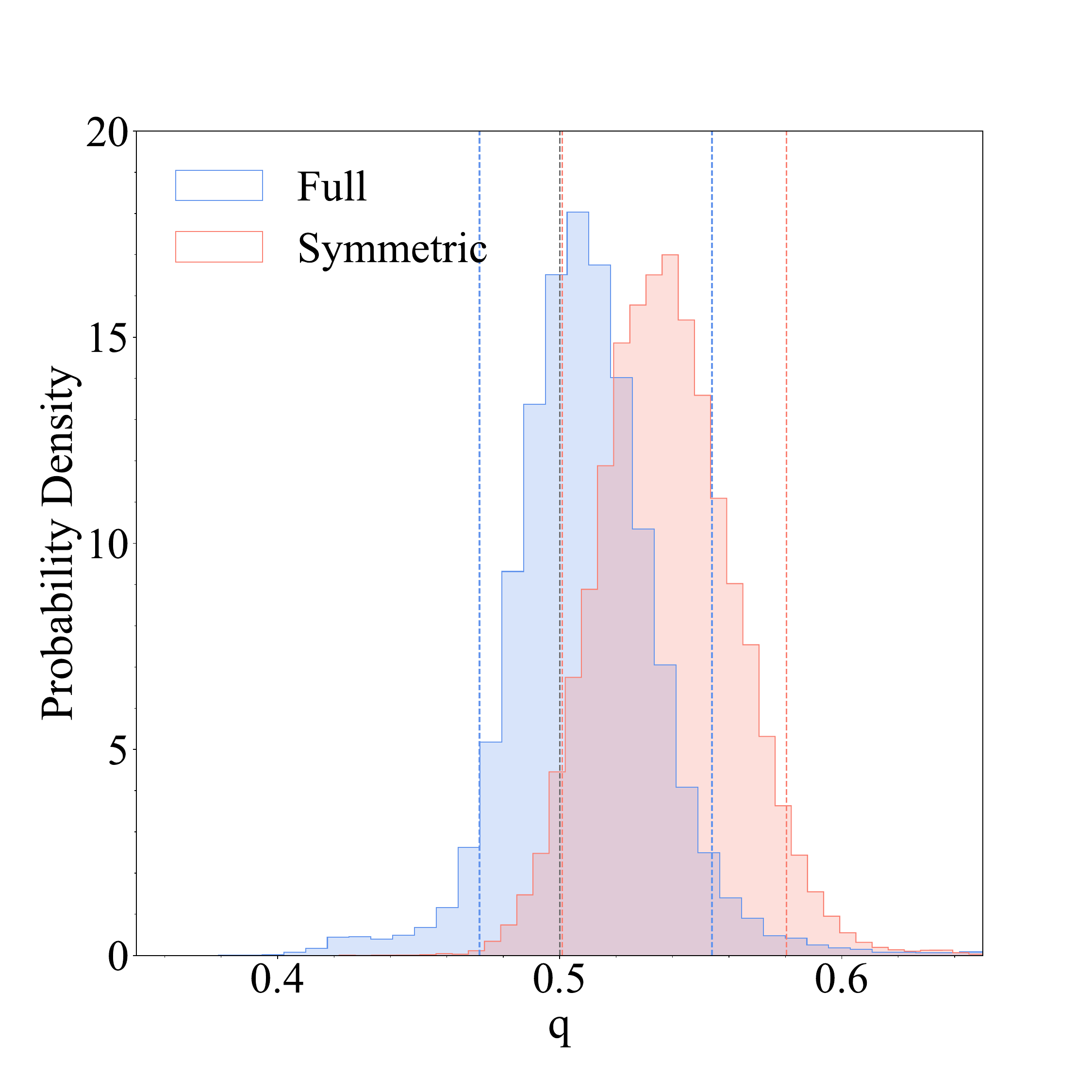}
\includegraphics[width=0.3\linewidth]{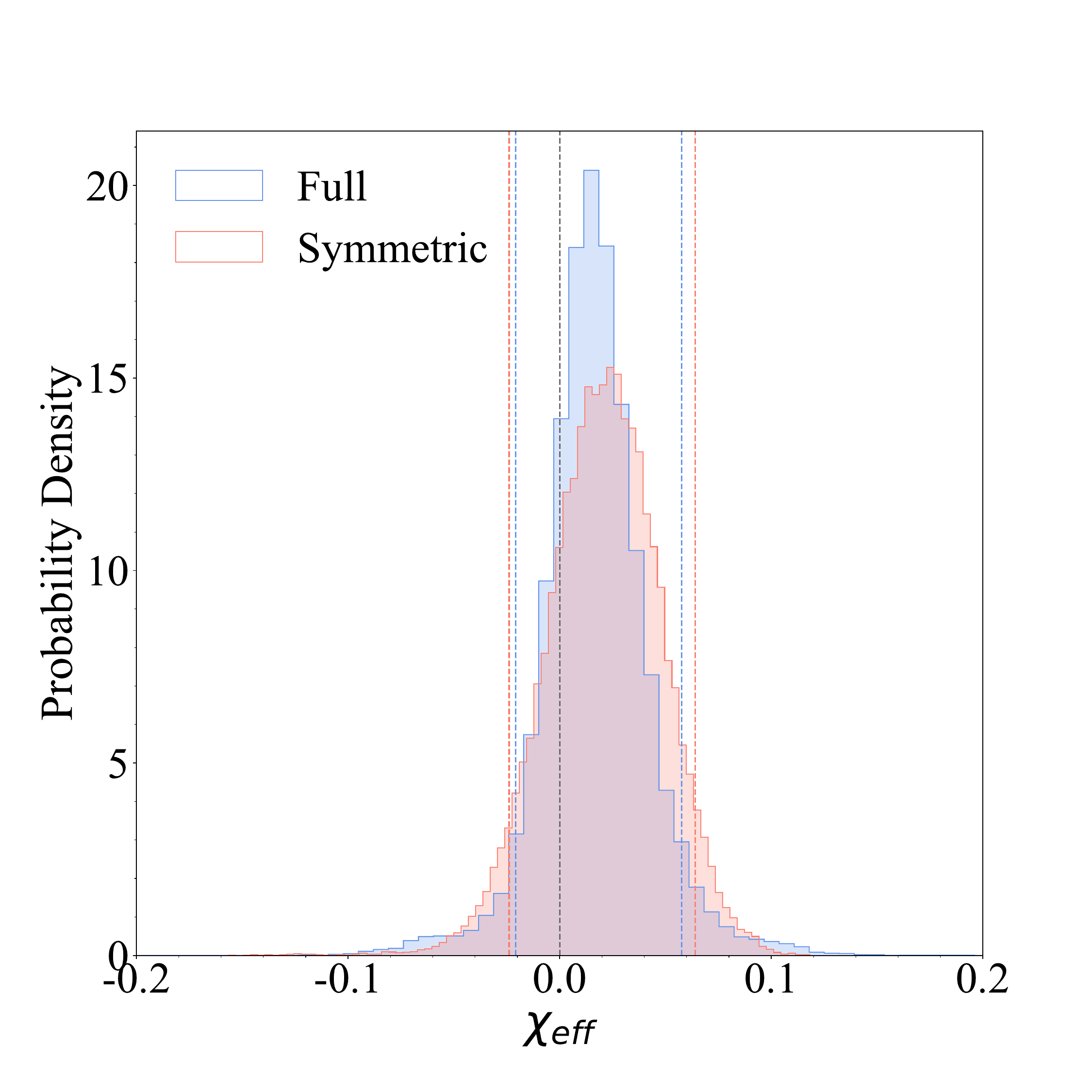}

\includegraphics[width=0.3\linewidth]{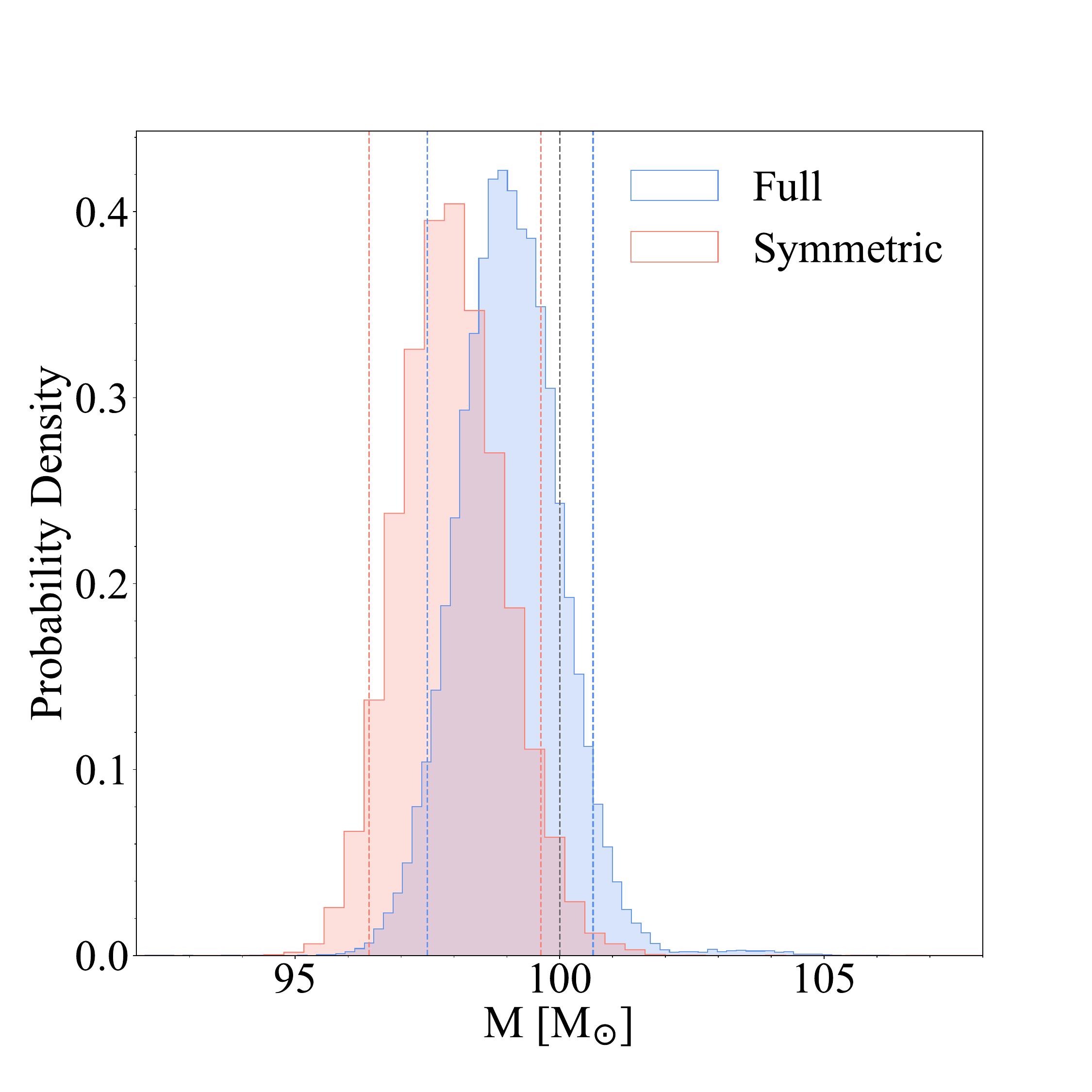}
\includegraphics[width=0.3\linewidth]{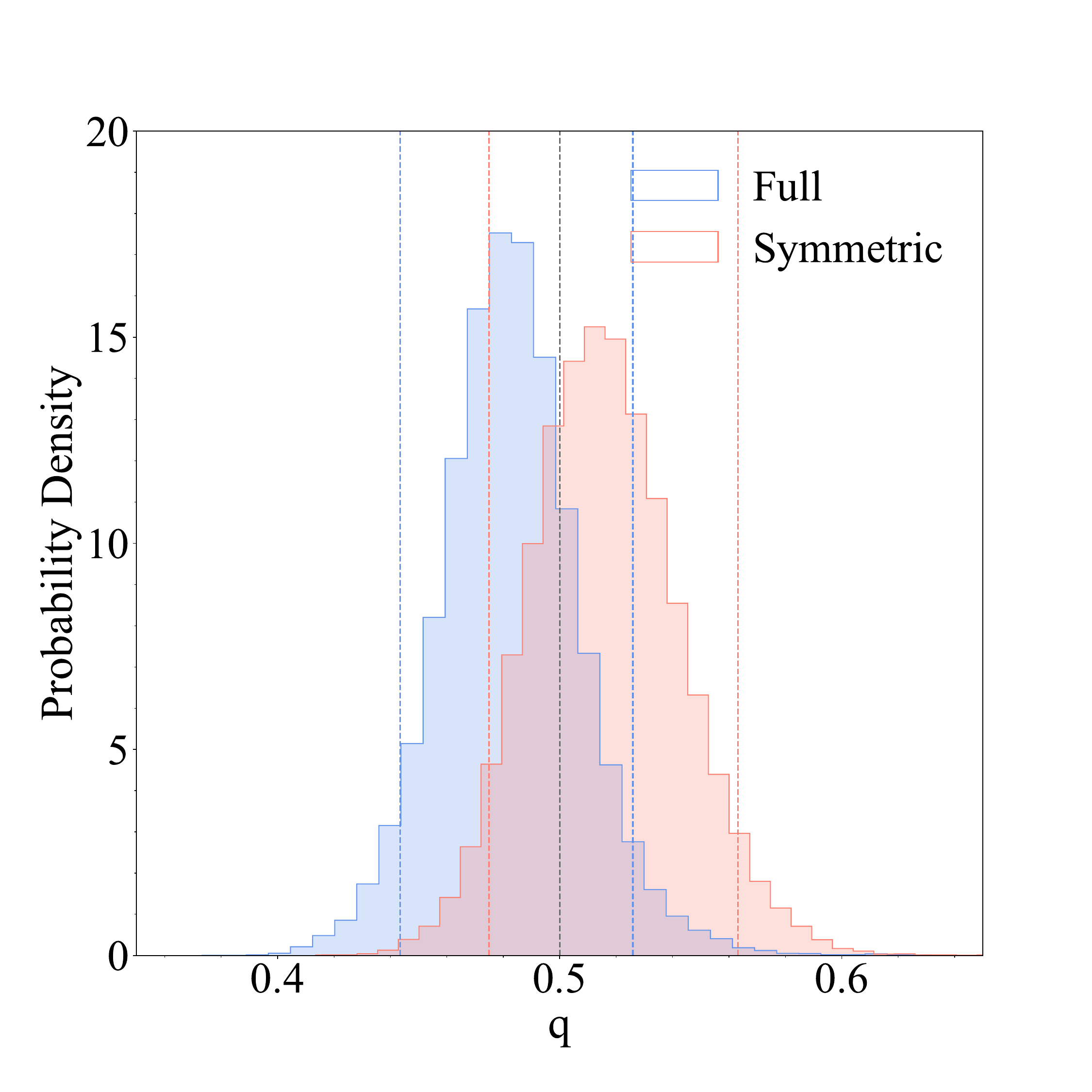}
\includegraphics[width=0.3\linewidth]{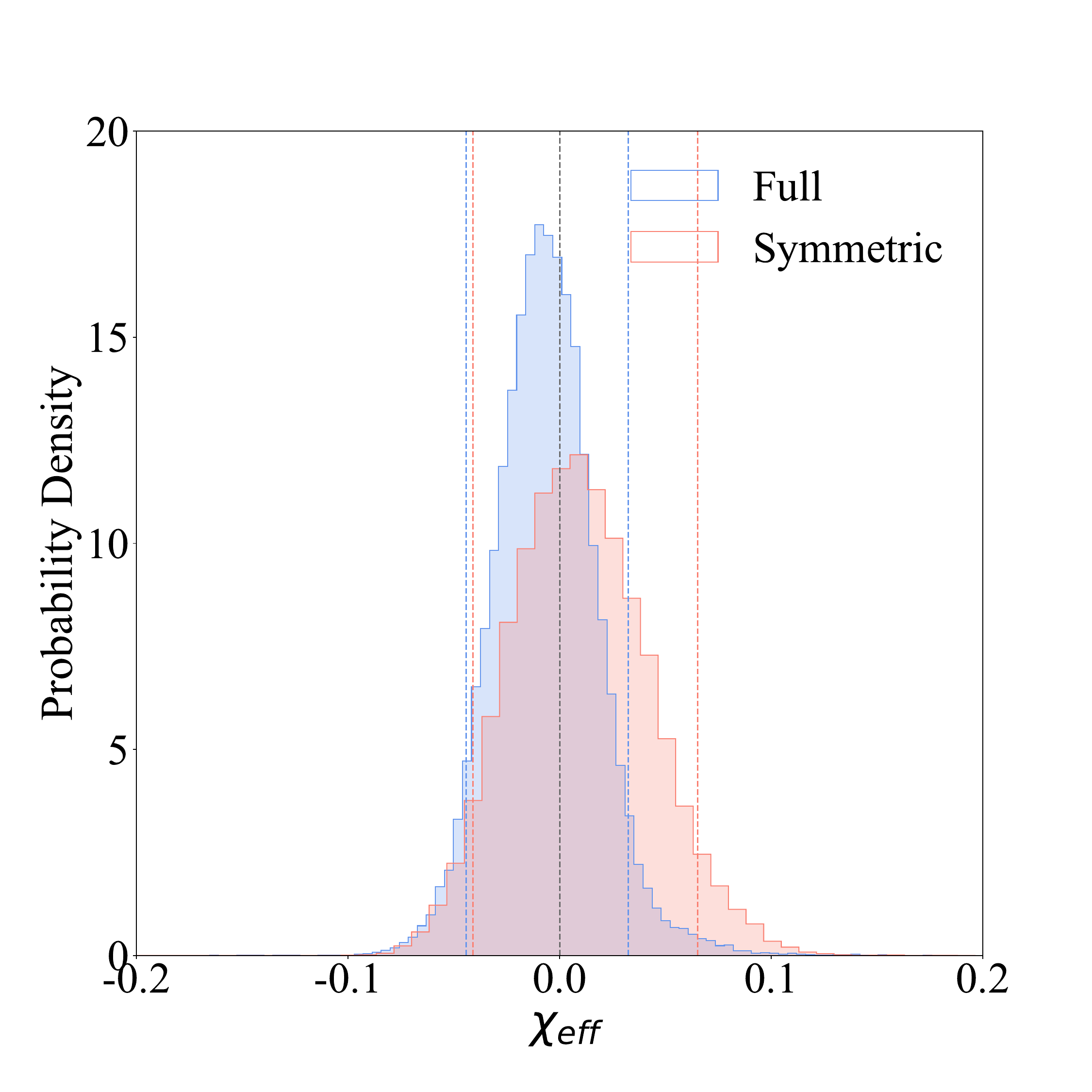}

\includegraphics[width=0.3\linewidth]{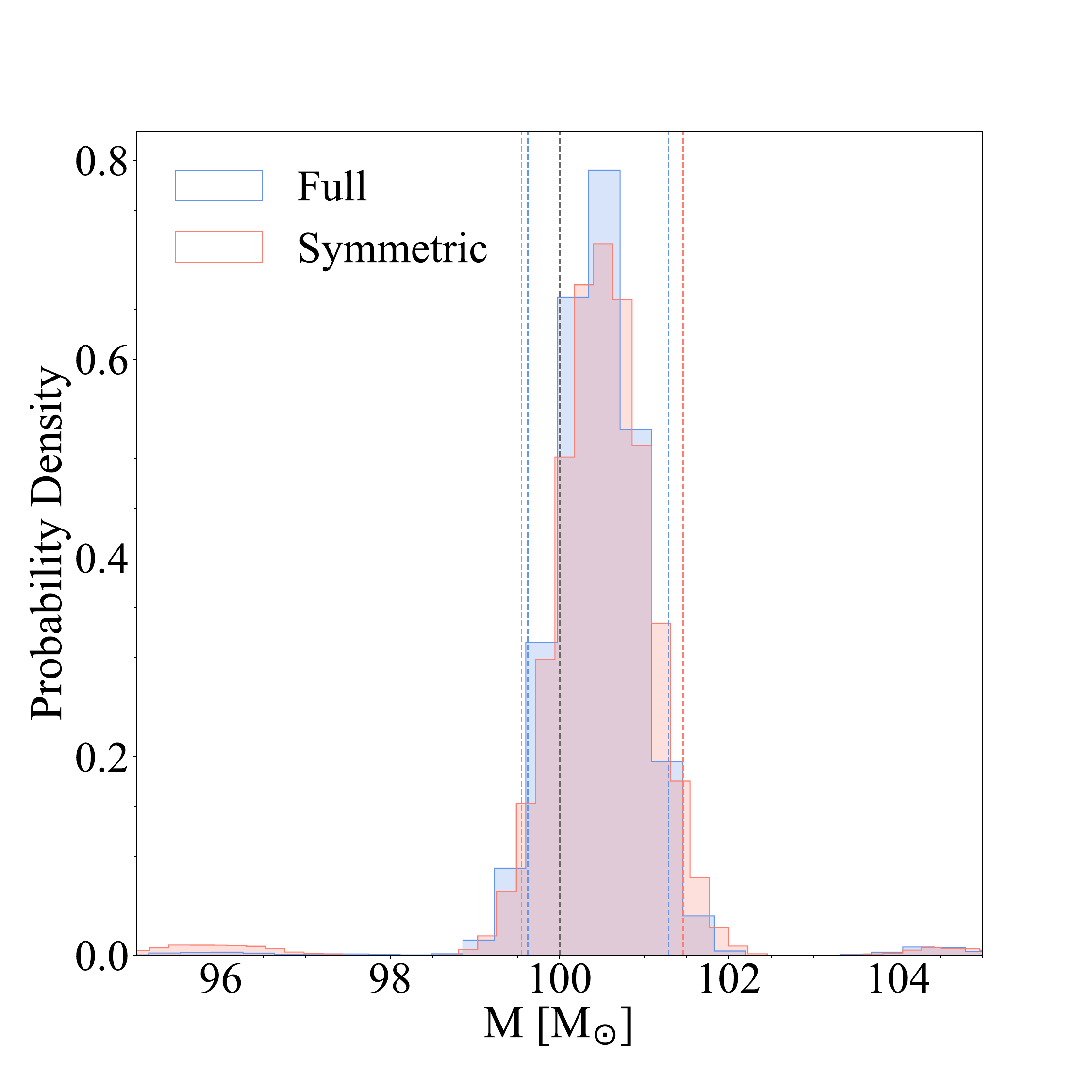}
\includegraphics[width=0.3\linewidth]{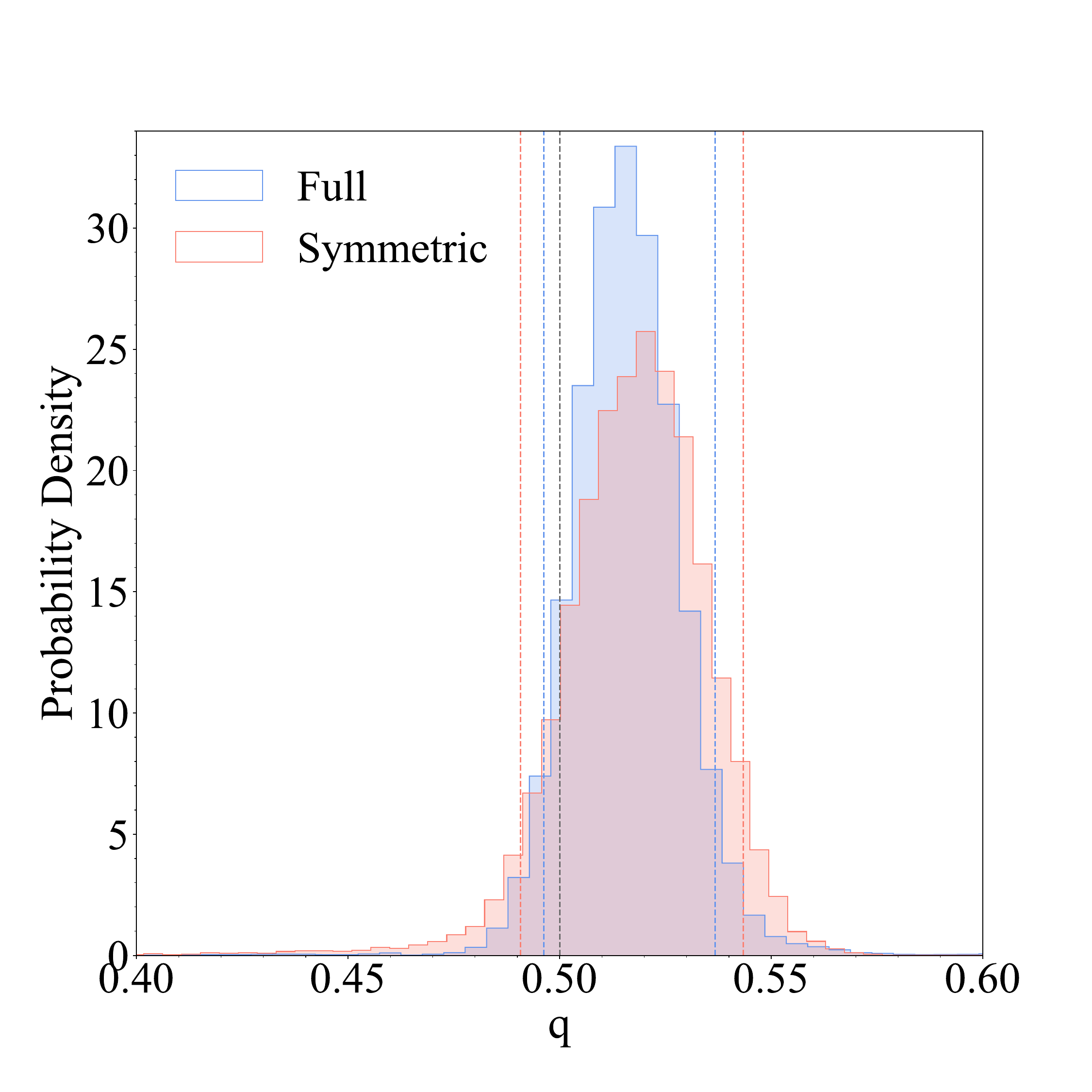}
\includegraphics[width=0.3\linewidth]{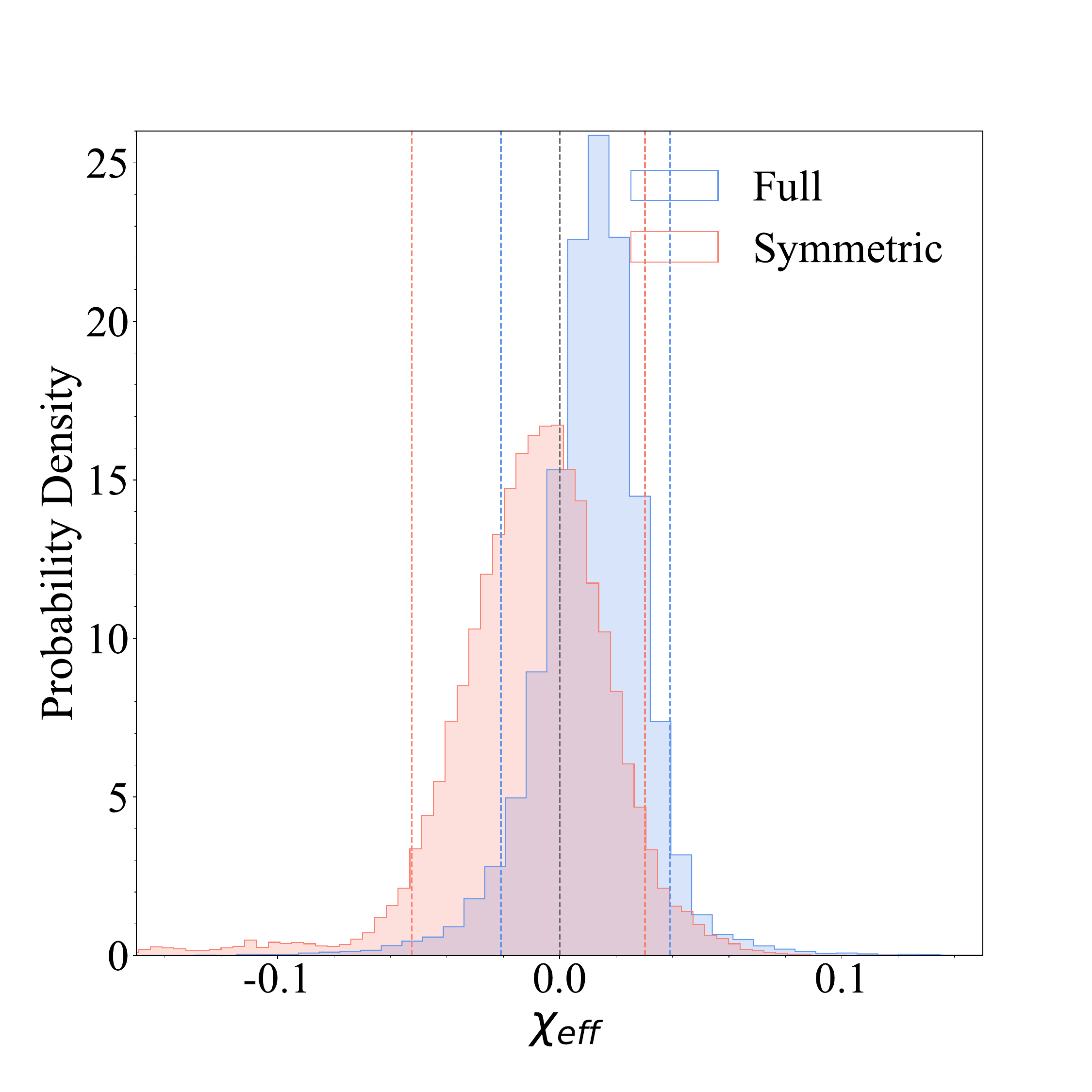}
\caption{Measurements of $M$, $q$ and $\chi_{\rm eff}$ for the $(q=2$, $a_1/m_1=0.7$, $\theta_{\rm LS} = 90^\circ)$ configurations with 
(top) $\iota = 30^\circ$, minimum recoil, (middle) $\iota = 30^\circ$, maximum recoil, and (bottom) $\iota = 90^\circ$, minimum recoil,  as they were measured by the \texttt{NRSur7dq4} (blue) and \texttt{NRSur7dq4\_sym} (red) models.
}
\label{fig:Mqchi}
\end{figure*}

In this section we will show a subset of results that illustrate the impact of the mode asymmetry that we observe from our parameter-estimation
analyses. Our fiducial configuration is $(q=2, a_1/m_1=0.7, \theta_{\rm LS} = 90^\circ)$, and in Fig.~\ref{fig:Mqchi} we show results for the 
initial in-plane spin orientation that leads to minimal recoil ($\phi_{Sn} = 67^\circ$, top row) and maximum recoil ($\phi_{Sn} = 138^\circ$, middle row), 
both at inclination $\iota = 30^\circ$ with respect to the direction of maximum emission at merger. The bottom row shows the
minimum recoil configuration viewed at $\iota = 90^\circ$.

Our first expectation from Sec.~\ref{sec:background} was that measurements of the masses and $\chi_{\rm eff}$ would not be biassed by 
neglecting the multipole asymmetry. Fig.~\ref{fig:Mqchi} shows the measurements for $M$, $q$ and $\chi_{\rm eff}$ for three configurations, 
and we see that to some extent our expectation holds, in that the measured values are only slightly affected by the symmetric approximation in
 \texttt{NRSur7dq4\_sym}. Nonetheless, we do see \emph{some} bias; in the top and middle panels the true value of the mass and/or mass-ratio lies 
 outside the 90\% confidence interval. In several measurements shown here (and similarly in the other configurations we studied) there is a less clear sign of bias.

\begin{figure*}[ht!] 
\includegraphics[width=0.2\linewidth]{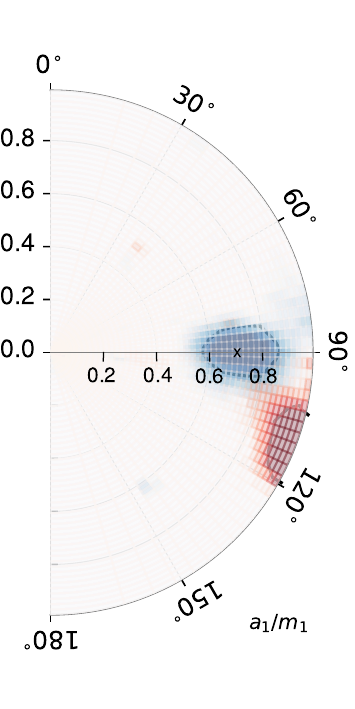}
\includegraphics[width=0.2\linewidth]{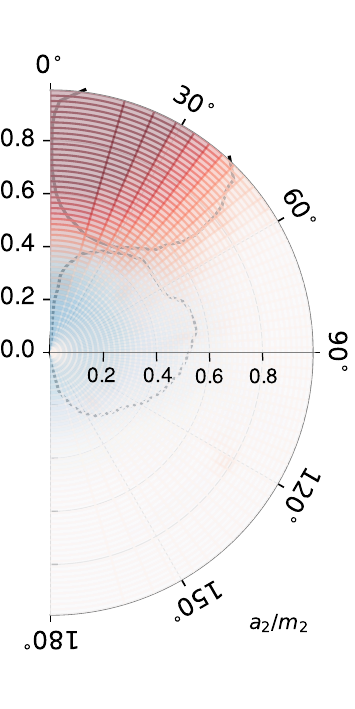}
\hspace{1cm}
\includegraphics[width=0.2\linewidth]{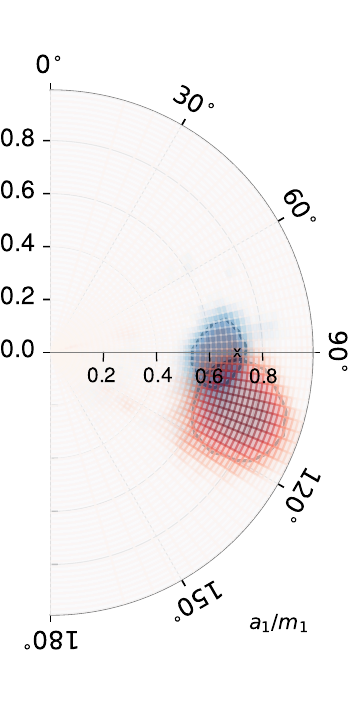}
\includegraphics[width=0.2\linewidth]{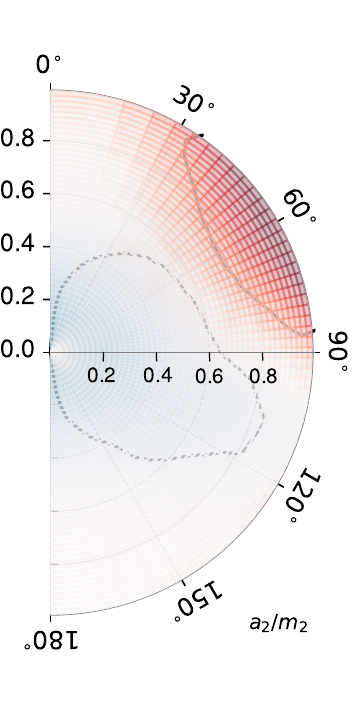}
  \caption{Magnitude and direction of each spin, $a_1/m_1$ and $a_2/m_2$, for $(q=2$, $a_1/m_1=0.7$, $\theta_{\rm LS} = 90^\circ)$ 
  configurations at inclination $\iota=30^\circ$ as they were measured by the \texttt{NRSur7dq4} (blue) and \texttt{NRSur7dq4\_sym} (red) models. Left: The configuration with initial in-plane-spin direction chosen to yield minimum recoil. Right: 
  The configuration with maximum recoil.}
\label{fig:disc_30}  
\end{figure*}

We now look at the individual spin magnitudes and tilt angles. These are shown for the minimum-recoil configuration in the left panel of
Fig.~\ref{fig:disc_30}.

\begin{figure}[ht!]
\includegraphics[width=0.4\linewidth]{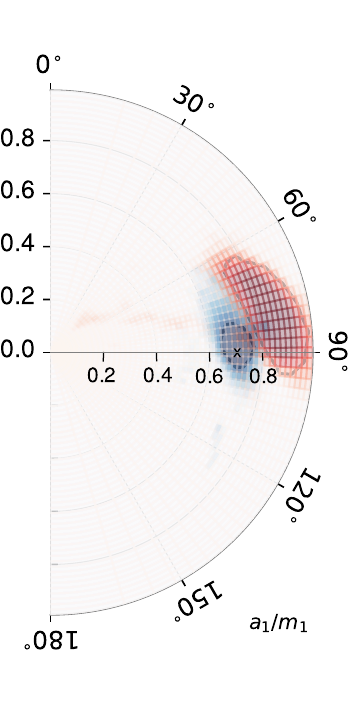}
\includegraphics[width=0.4\linewidth]{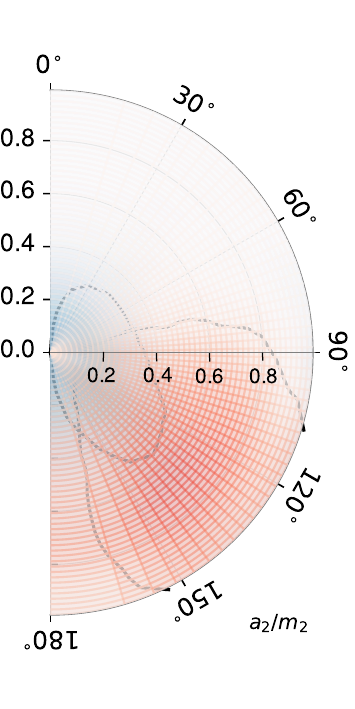}
\label{fig:disc90}

\caption{Spin magnitude and direction of (left) $a_1/m_1$ and (right) $a_2/m_2$ of the minimum recoil \\$(q=2$, $a_1/m_1=0.7$, $\theta_{\rm LS} = 90^\circ)$ configuration's black holes 
with inclination $90^\circ$  as they were measured by the \texttt{NRSur7dq4} (blue) and \texttt{NRSur7dq4\_sym} (red) models.}
\label{fig:disc_min_90}
\end{figure}

In each disc plot, the spin magnitude is between 0 and 1 while the tilt angle ranges between $0^\circ$ and $180^\circ$, where \(0^\circ\) corresponds to an 
aligned-spin system where the spins are in the same direction as the orbital angular momentum. 
The shading indicates the parameters' measured values and the different colours correspond to the results from the recovery with the two versions of the surrogate model.

We see that the recovered spin magnitude and the tilt angle of the primary black hole with \texttt{NRSur7dq4\_sym} have a higher value, 
indicating that the spin vector lies outside the plane of the binary. Furthermore, the recovered spin magnitude reaches the Kerr limit, $a_1/m_1=1$. 
In contrast, the measured parameters with the \texttt{NRSur7dq4} agree well with the true values. 
A similar behaviour can be observed for $a_2/m_2$. The true spin of the secondary black hole is zero, as recovered well with the \texttt{NRSur7dq4} model. 
However, the \texttt{NRSur7dq4\_sym} model measures a high spin value for the same black hole and a low tilt angle, i.e., the spin appears nearly aligned with the 
orbital angular momentum.

Despite the significant biases in the spin measurements with the \texttt{NRSur7dq4\_sym} model, we do see, as expected, that the combination $\chi_{\rm eff}$
is measured correctly; the biases counteract so that $\chi_{\rm eff}$ has the correct value. We saw similar results in all of the fiducial-configuration 
binaries: the \texttt{NRSur7dq4\_sym} recovery for $a_1/m_1$ and $a_2/m_2$ varied in magnitude and direction, but always such that 
$\chi_{\rm eff}$ was roughly correct. We might expect, however, that in larger-mass-ratio
binaries with sufficiently high spin on the primary, that the spin measurements will rail against the Kerr limit, and it will not be possible for the biases
to fully counteract each other to give a correct value of $\chi_{\rm eff}$. We will see examples of this in the next section.

\subsection{Dependence on recoil,  inclination, spin magnitude and mass ratio}
\label{sec:dependence}

We now consider how the impact of the multipole asymmetry varies with the recoil (or, equivalently, changes in the initial in-plane-spin
direction), the binary's inclination to the detector, the spin magnitude, and the mass ratio.

As noted in Sec.~\ref{sec:background}, although changes in the initial in-plane-spin direction will change the out-of-plane recoil of the final
black hole, we do not necessarily expect this to qualitatively change the bias due to neglecting the multipole asymmetry. This is borne out in the
right-hand panel of Fig.~\ref{fig:disc_30}, which shows the recovery of the spins for the same system, but now with $\phi_{Sn} = 138^\circ$
and maximum recoil. We see that the details of the spin measurements from the \texttt{NRSur7dq4\_sym} model differ --- the primary spin magnitude $a_1/m_1$ is 
closer to the correct value, but the secondary spin magnitude $a_2/m_2$ shows a stronger preference for extreme spins -- but qualitatively the
results are similar. 

We next consider how the impact of the asymmetry changes with inclination. We noted in Sec.~\ref{sec:background} that we do not expect
the effects to change significantly with inclination. Let us explain this further. Naively, the impact of the asymmetry \emph{does} have a
clear dependence on inclination. If we write the (2,2) multipoles as $h_{2,\pm2} = h_s \pm h_a$, where $h_s$ and $h_a$ are the 
symmetric and antisymmetric contributions (as in Eqs.~(\ref{eq:h22}) and (\ref{eq:h2m2})), then the strain as a function of the inclination with 
respect to the normal to the orbital plane $\theta$ and azimuthal angle $\varphi$ is given by 
\begin{equation}
h(\theta,\varphi) = h_{2,2} {} ^{-2}Y_{2,2}(\theta,\varphi) + h_{2,-2} {} ^{-2}Y_{2,-2}(\theta,\varphi).
\end{equation} The spherical harmonics depend on $\theta$ as $(1 \pm \cos \theta )^2$, and so the relative strength $R_{as}$ of the anti-symmetric
and symmetric contributions, compared to their relative strength at $\theta = 0$, is \begin{equation}
R_{as} = \frac{4 \cos\theta}{3 + \cos(2\theta)}.
\end{equation}

From this we see that edge-on to the binary, 
$\theta=\pi/2$, the anti-symmetric contributions will cancel out. However,
in a precessing system we can never be edge-on to the binary at all times. 

In our fiducial configuration, $(q=2, a_1/m_1=0.7, \theta_{\rm LS} = 90^\circ)$, the maximum opening angle between the orbital angular
momentum and the total angular momentum is $\beta_{\rm max} \approx 0.35$. (See Fig.~8 in Ref.~\cite{Hamilton:2021pkf}.) If we
were to define inclination with respect to $\mathbf{J}$, then a nominal inclination of $\pi/2$ would correspond to an inclination 
with respect to the orbital plane of $\pi/2 - 0.35$, and $R_{as} = 0.61$. If we were to define the inclination with respect to the 
orbital angular momentum when the signal enters the detector's sensitivity band (as is the standard \texttt{LAL} convention), 
then depending on where this point lies in the precession cycle, the inclination relative to the normal to the orbital plane at
merger could be as large as $2\beta_{max} \approx 0.7$, with $R_{as} = 0.91$. This illustrates the non-trivial importance of how
we define inclination. 

This motivated the inclination we have used for these analyses, where $\iota = 0$ corresponds to the direction of maximum emission
at merger. With this definition, we expect that $\iota = \pi/2$ will correspond to the binary being edge-on to the detector at
merger (i.e, the peak in the signal amplitude), and therefore zero contribution from the asymmetry at merger. At all other times
the signal is weaker and the opening angle $\beta$ is smaller, and so we may hope to minimise the impact of the asymmetry on 
parameter measurements. 

The lower panel of Fig.~\ref{fig:Mqchi} shows $M$, $q$, and $\chi_{\rm eff}$ for an inclination of $\iota = 90^\circ$. In this 
case we do not see any clear sign of bias, which suggests that we may have removed the impact of the asymmetric contribution.
(Similarly, we see slightly larger biases in cases in $\iota = 0$ cases.)
However, Fig.~\ref{fig:disc_min_90} shows the spin measurements for the $\iota = 90^\circ$ signal, and we see that some bias
remains. It appears to be smaller than in the $\iota = 30^\circ$ signals, but has not been significantly reduced. For any given 
configuration there will be some inclination that minimises the impact of the asymmetry, but given that the inclination oscillates
due to precession, and $R_{as}$ is approximately one up to $\theta \approx 1$\,rad, we conclude that the impact of the asymmetry
does not in general depend significantly on the binary orientation.  We did not attempt to identify a specific relationship between the 
details of the biases and the choices of inclination and total recoil, but this would be interesting to study further in the future.

\begin{figure*}[ht!]
\includegraphics[width=0.3\linewidth]{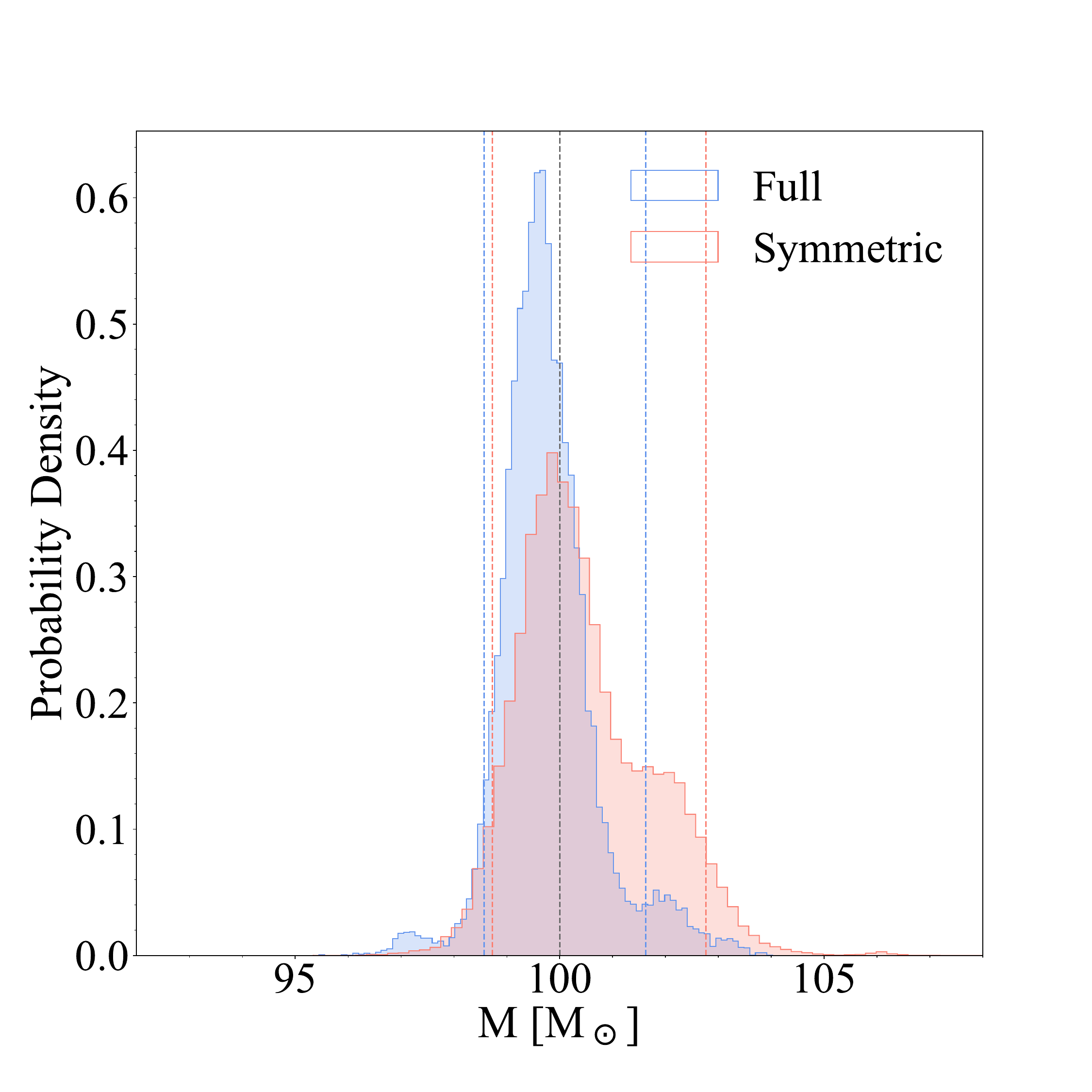}
\includegraphics[width=0.3\linewidth]{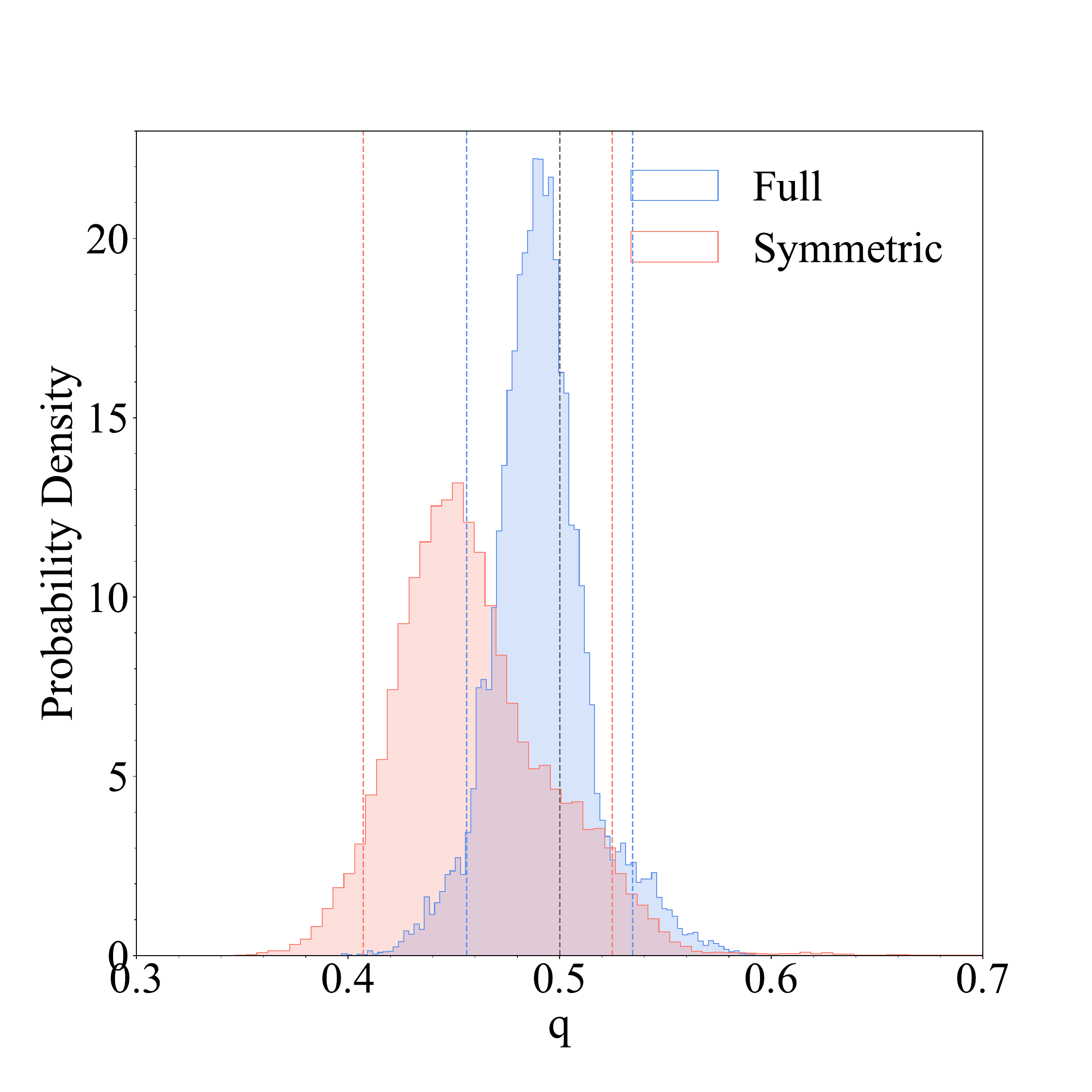}
\includegraphics[width=0.3\linewidth]{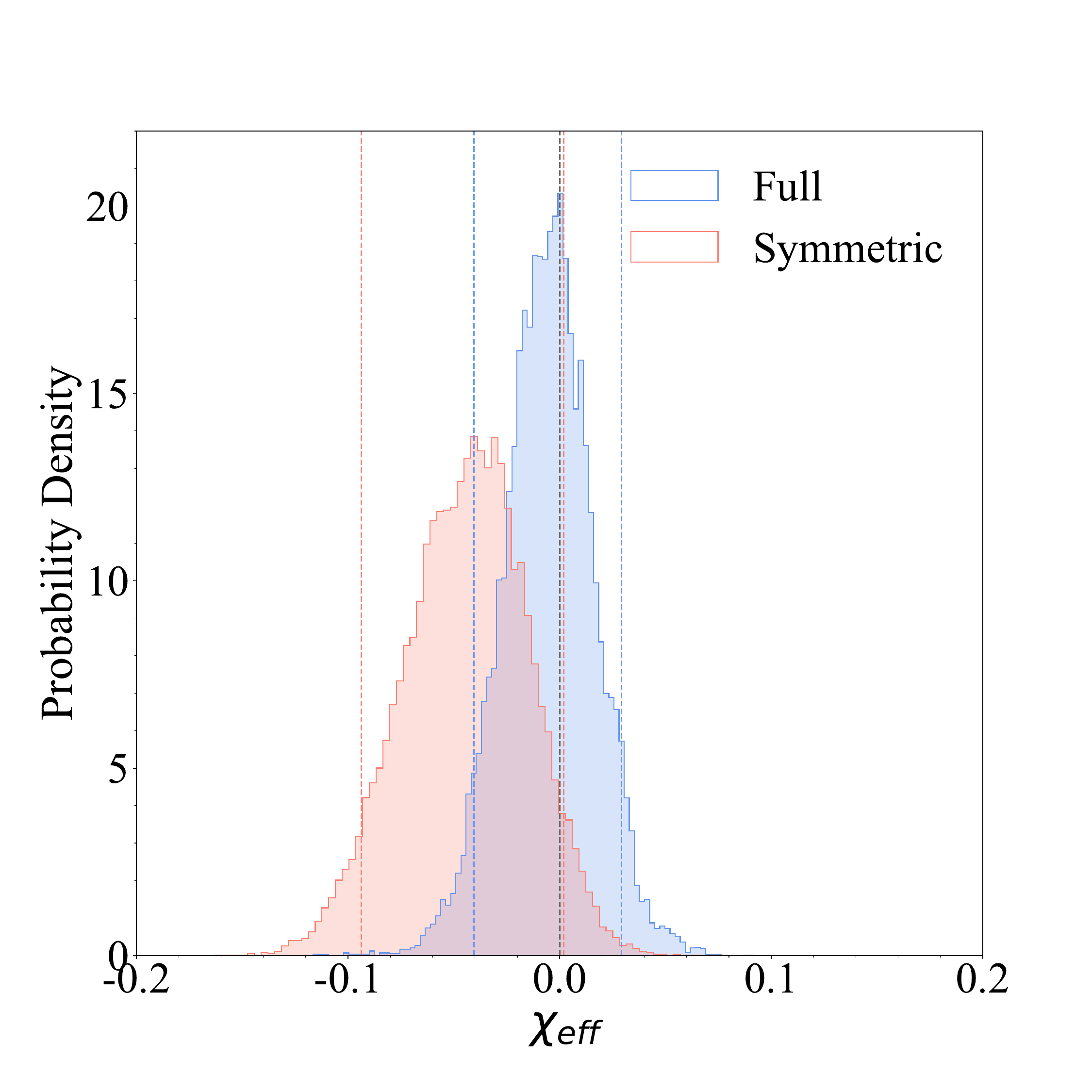}

\includegraphics[width=0.3\linewidth]{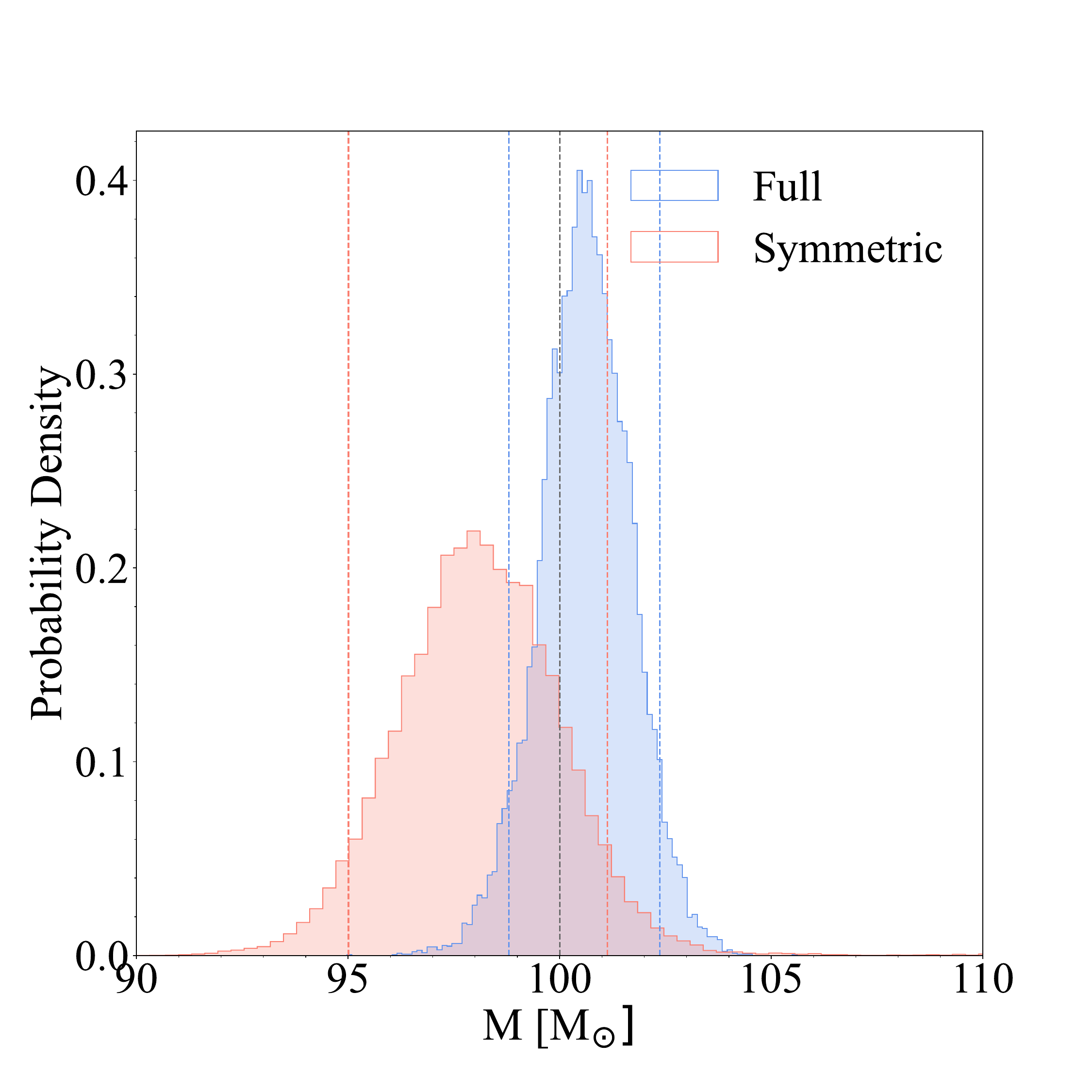}
\includegraphics[width=0.3\linewidth]{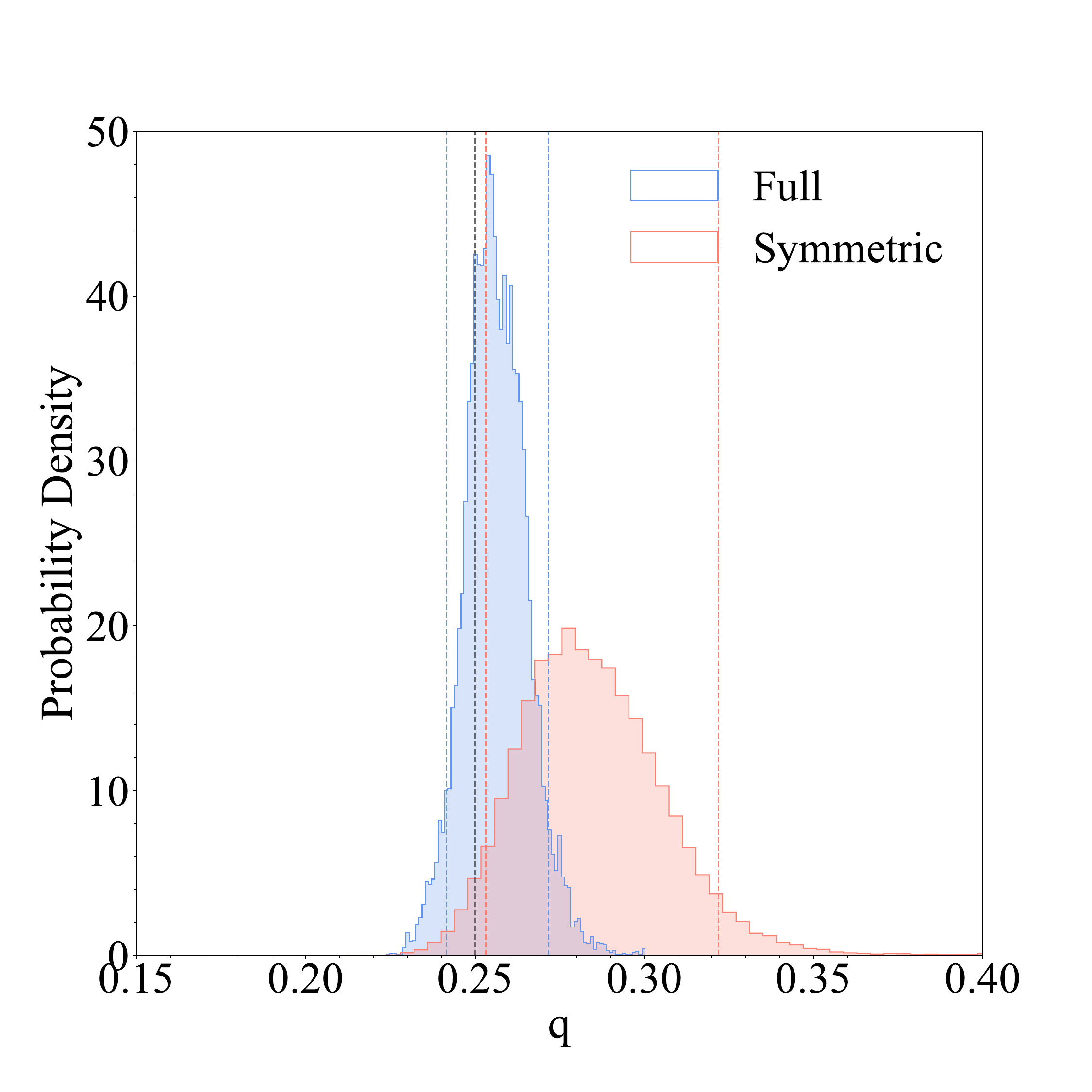}
\includegraphics[width=0.3\linewidth]{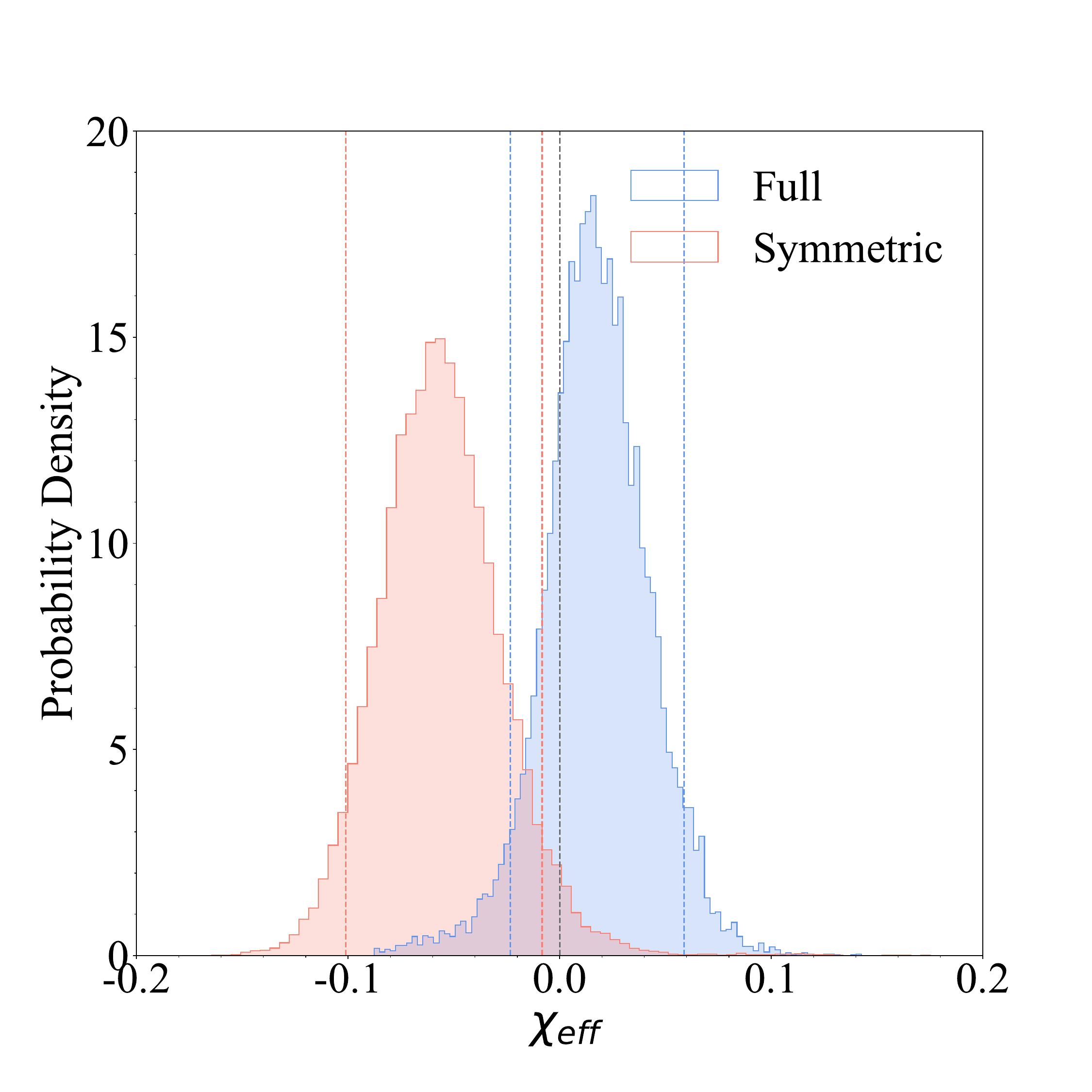}
\caption{Measurements of $M$, $q$ and $\chi_{\rm eff}$ for the (top) $(q=2$, $a_1/m_1=0.4$, $\theta_{\rm LS} = 90^\circ)$ and (bottoom)
$(q=4$, $a_1/m_1=0.8$, $\theta_{\rm LS} = 90^\circ)$ configurations as they were measured by the 
\texttt{NRSur7dq4} (blue) and \texttt{NRSur7dq4\_sym} (red) models.
}
\label{fig:variations}
\end{figure*}

Finally, we consider changes in the spin magnitude and mass ratio: a lower-spin system, $(q=2, a_1/m_1=0.4,\theta_{\rm LS} = 90^\circ)$ and
a system with larger mass ratio and larger spin, $(q=4, a_1/m_1=0.8, \theta_{\rm LS} = 90^\circ)$. 

\begin{figure*}[ht!]
\includegraphics[width=0.2\linewidth]{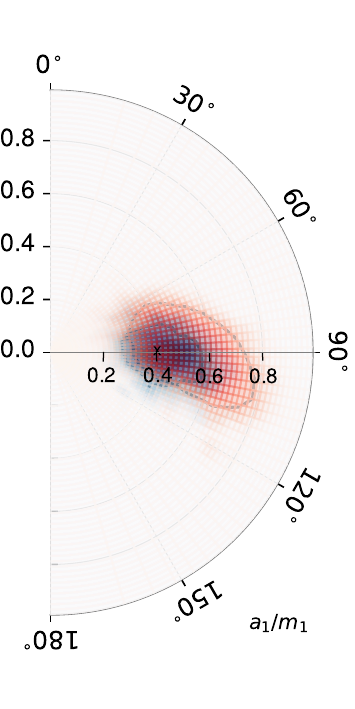}
\includegraphics[width=0.2\linewidth]{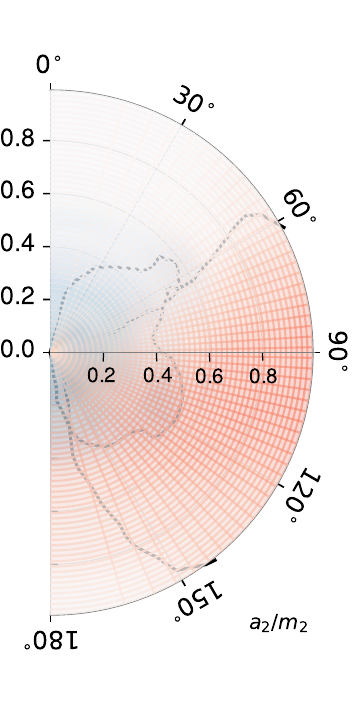}
\hspace{1cm}
\includegraphics[width=0.2\linewidth]{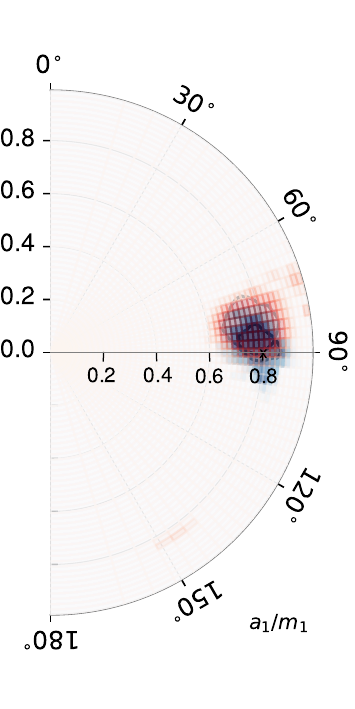}
\includegraphics[width=0.2\linewidth]{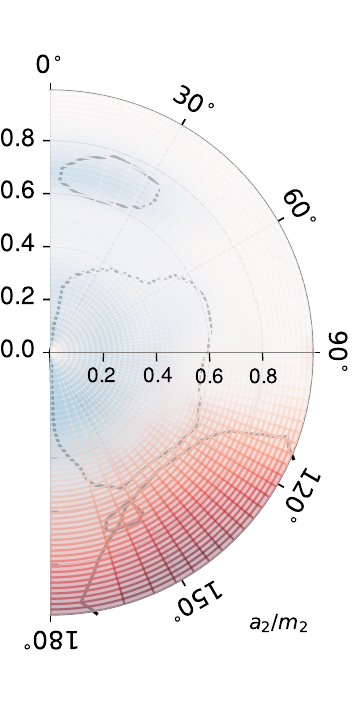}
\caption{Spin magnitude and direction of $a_1/m_1$ and $a_2/m_2$ of the (left) $(q=2$, $a_1/m_1=0.4$, $\theta_{\rm LS} = 90^\circ)$ and (right) 
$(q=4$, $a_1/m_1=0.8$, $\theta_{\rm LS} = 90^\circ)$ configurations, both with inclination $\iota = 60^\circ$ as they were measured by the \texttt{NRSur7dq4} (blue) and \texttt{NRSur7dq4\_sym} (red) models.}  
\label{fig:spin_disk_others}
\end{figure*}

We see in the top row of Fig.~\ref{fig:variations} that in the lower-spin case the posteriors for $M$, $q$ and $\chi_{\rm eff}$ are wider in the analysis with
the \texttt{NRSur7dq4\_sym} model, but we still do not see any significant bias, except for a shoulder in the $M$ posterior in one case. 
This is consistent with our expectation that a lower spin magnitude will 
also lower the impact of the multipole asymmetry. 
For the high-mass-ratio case (bottom row), there is more sign of biases. The posteriors from the
\texttt{NRSur7dq4\_sym} recovery are much broader that for the \texttt{NRSur7dq4}, especially for the total mass, where the width of the 90\% confidence region
has almost doubled. We also see that there is now a clear bias in $\chi_{\rm eff}$ when recovering with the \texttt{NRSur7dq4\_sym} model.

Fig.~\ref{fig:spin_disk_others} shows the spin magnitudes and tilt angles for the lower-spin and higher-mass-ratio cases. As we expect,
the bias is reduced when the spin magnitude is reduced, and in this case there is no clear bias in the measurement of the primary spin, 
and the secondary spin, although it appears biased in the disc plots, the real difference between the \texttt{NRSur7dq4} and \texttt{NRSur7dq4\_sym} models 
analyses is that with \texttt{NRSur7dq4} the second spin magnitude is constrained by less than 0.45, while with the \texttt{NRSur7dq4\_sym} the 
second spin is not constrained; the 90\% confidence interval covers 90\% of the parameter range. 

The high-mass-ratio case is more interesting. It now appears that the primary spin can be measured accurately with both
models, suggesting that the spin imprint on the symmetric contribution to the signal is strong enough to constrain the value. This is not
the case for the secondary spin, and without the anti-symmetric contribution to the model the secondary spin is biased. The bias in this
sector of the model also appears to be so strong that it is no longer counteracted by the inspiral phasing that plays the dominant role in
determining $\chi_{\rm eff}$, and so this is now also biassed.  We expect that this is a general
trend: at higher mass ratios ($q \gtrsim 4$) the measurement of the primary spin is more reliable than quantities that include both spins.
Since there is a partial degeneracy between the mass ratio and $\chi_{\rm eff}$~\cite{Cutler:1994ys,Poisson:1995ef,Baird:2013dbm}, 
the bias in $\chi_{\rm eff}$ also leads to a bias in the mass ratio.


\subsection{GW200129 signal}
\label{sec:gw200129}

\begin{figure}[ht!]
  \begin{subfigure}[t]{.7\linewidth}
    \includegraphics[width=0.85\linewidth]{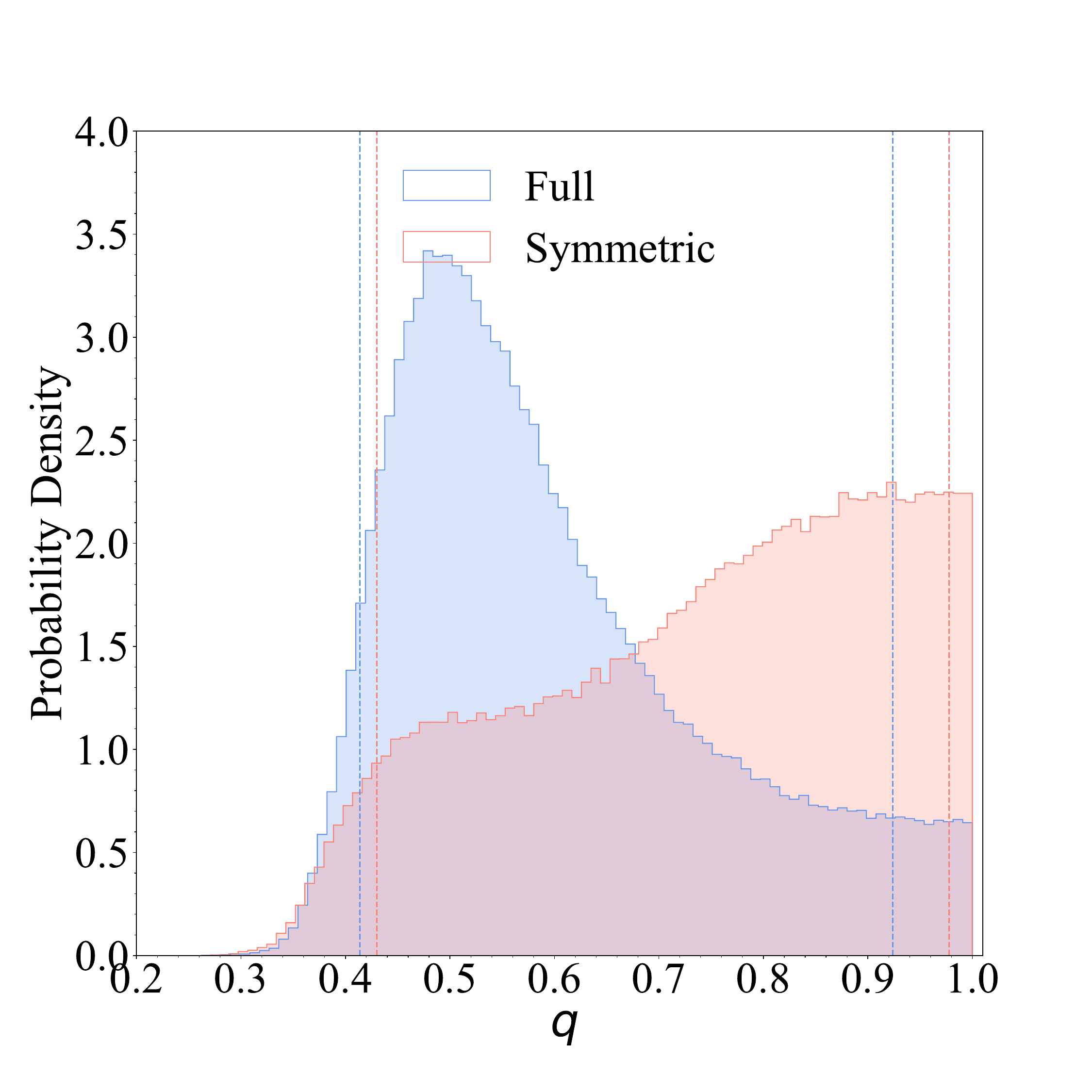}
    \label{}
  \end{subfigure}
  \hfill
  \begin{subfigure}[t]{.7\linewidth}
    \includegraphics[width=0.85\linewidth]{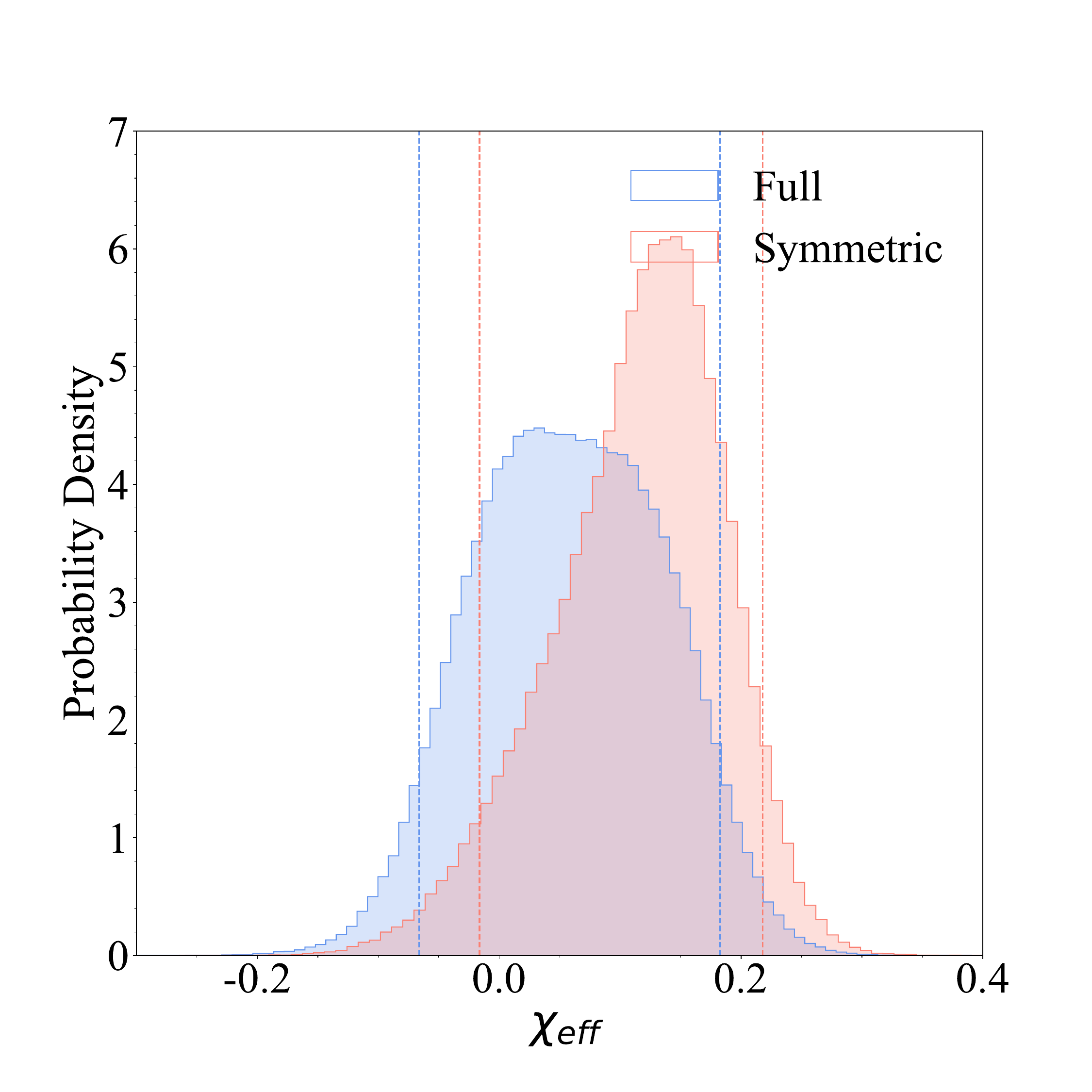}
    \label{}
  \end{subfigure}
  \hfill
  \begin{subfigure}[t]{.7\linewidth}
    \includegraphics[width=0.85\linewidth]{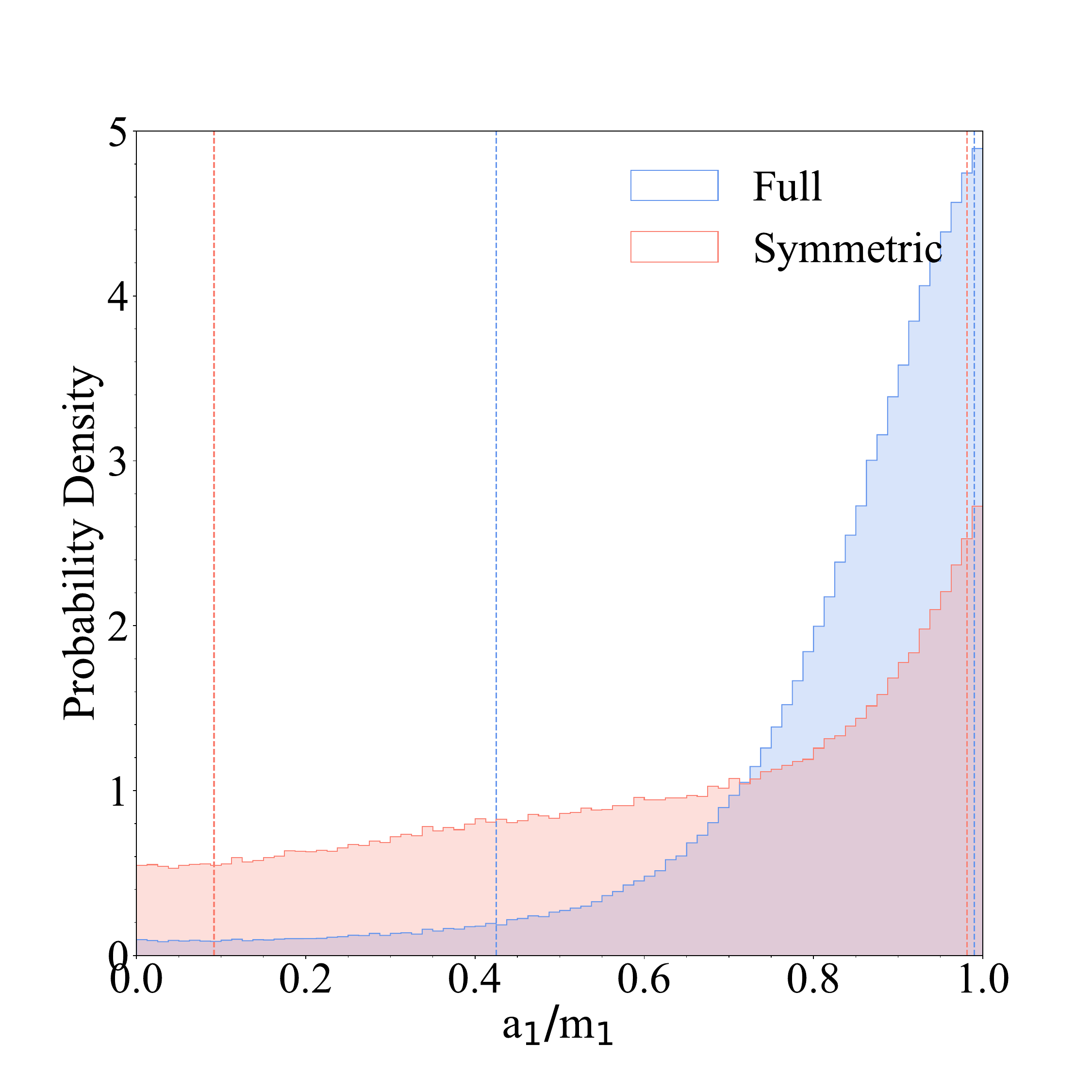}
    \label{}
  \end{subfigure}
  \hfill
  \begin{subfigure}[t]{.7\linewidth}
    \includegraphics[width=0.85\linewidth]{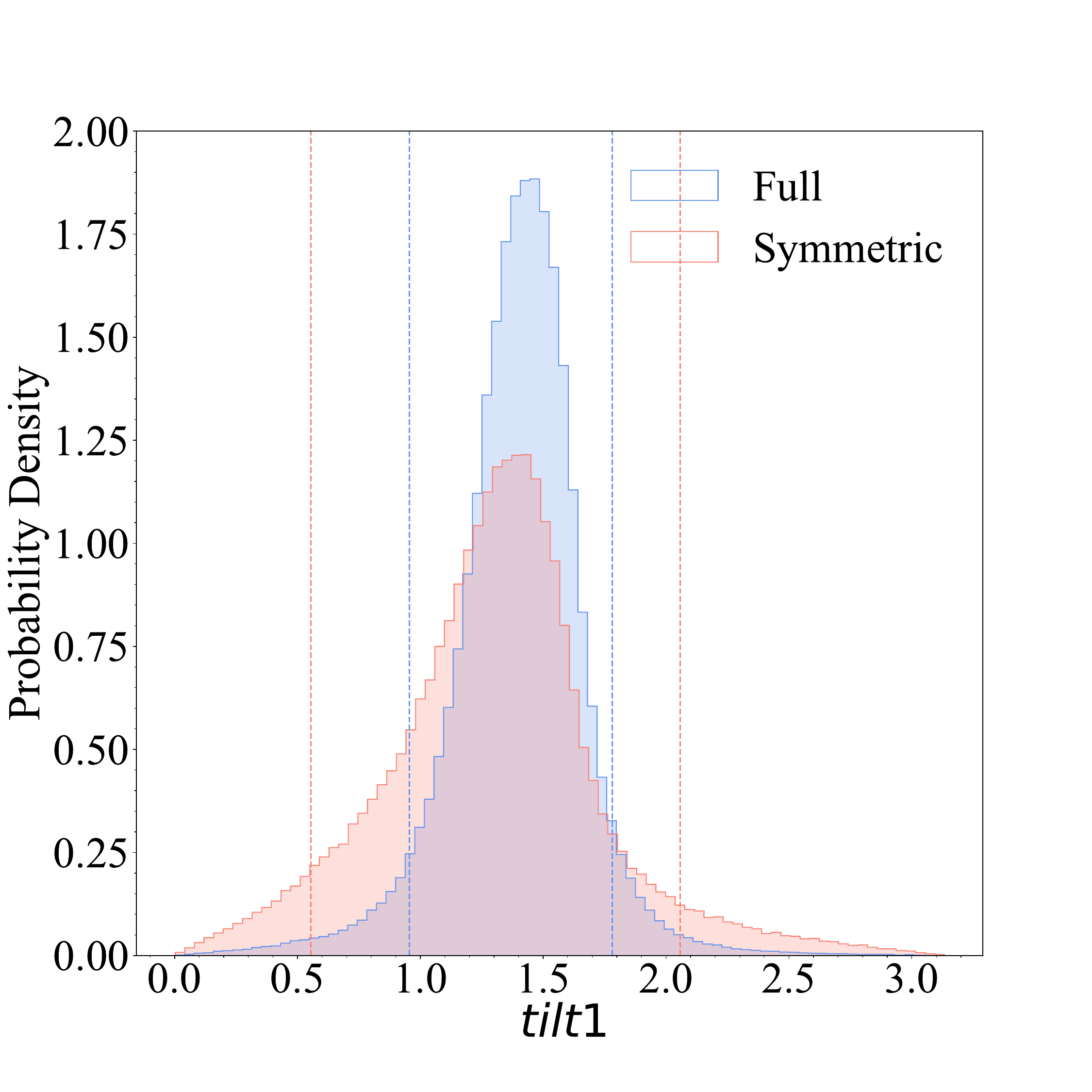}
    \label{}
  \end{subfigure}
  \caption{One-dimensional posterior distributions for the mass ratio, $\chi_{\rm eff}$, primary spin magnitude and tilt angle, for
  the \texttt{NRSur7dq4} (blue) and \texttt{NRSur7dq4\_sym} (red) recovery of GW200129.}
    \label{fig:posteriors_GW200129}
\end{figure}

We now consider the gravitational-wave signal GW200129. The measured parameters presented in Refs.~\cite{LIGOScientific:2021djp,Hannam:2021pit} indicate that this system is similar to some of the injected \texttt{NRSur7dq4} waveforms that were discussed in the previous section. However, interestingly in this case the SNR is only $26.5$ making this signal significantly weaker compared to the theoretical signals of the previous section.
As a result, we expect the effects of the absence of the asymmetry to be more subtle.

As previously, we analyse the signal with the \texttt{NRSur7dq4} and \texttt{NRSur7dq4\_sym} models. 
As shown in Table~\ref{tab:table1}, the total mass, $M$, is recovered consistently with the two version of the \texttt{NRSur7dq4}. 
However, the measurements of the mass ratio, $q$, and the individual masses, $m_1$ and $m_2$, differ between the two models.
The results presented in Fig.~\ref{fig:posteriors_GW200129} show that the full \texttt{NRSur7dq4} model measures that this is an unequal-mass system while the measurement of the mass ratio with the \texttt{NRSur7dq4\_sym} model is not well constrained.  
Furthermore, the primary spin measurements presented in Fig.~\ref{fig:posteriors_GW200129} show that the recovery with both versions of the surrogate lead to similar results for the tilt angle. However, in the case of the primary spin magnitude, this is poorly constrained with the \texttt{NRSur7dq4\_sym}, while it is clearly identified as a high spin by \texttt{NRSur7dq4}.

    \begin{table}[h!]
    \begin{tabular}{l|c c}
       & Full & Symmetric  \\  
      \hline \hline 
      Primary mass, $m_1 (M_{\odot})$ & $47.62 ^{+6.17} _{-8.88}$ & $ 42.48 ^{+11.0} _{-4.94}$ \\ 
      Secondary mass, $m_2 (M_{\odot})$ & $27.0 ^{+8.83} _{-4.96}$ & $ 32.54 ^{+4.64} _{-9.73}$ \\ 
      Mass ratio, $q=m_2/m_1$ & $0.57 ^{+0.36} _{-0.15}$ & $0.77 ^{+0.21} _{-0.34}$ \\ 
      Total mass, $M=m_1+m_2 (M_{\odot})$ & $74.83 ^{+3.06} _{-3.07}$ & $75.28 ^{+3.06} _{-3.27}$ \\ 
      Primary spin, $a_1/m_1$ & $0.88 ^{+0.11} _{-0.45}$ & $0.68 ^{+0.31} _{-0.58}$\\ 
      Primary spin tilt angle, $cos \theta_{LS_1}$ & $0.16 ^{+0.42} _{-0.36}$ & $0.25 ^{+0.6} _{-0.72}$ \\
      $\chi _{\rm eff}$ & $0.06 ^{+0.12} _{-0.12}$ & $ 0.12 ^{+0.09} _{-0.14}$ \\ 
      $\chi _p$ & $0.85 ^{+0.13} _{-0.37}$ & $ 0.66 ^{+0.31} _{-0.45}$ \\ 
      \hline \hline
    \end{tabular}
        \caption{
    The recovered parameters for the de-glitched GW200129 data with their $90 \% $ credible intervals. The results were recovered using the \texttt{NRSur7dq4} and \texttt{NRSur7dq4\_sym} models.
    }
        \label{tab:table1}
    \end{table}

From these results it becomes evident that even at relatively low SNR, including the asymmetry in the model was essential in identifying this system as 
an unequal-mass binary with large in-plane spin. We note that in the LVK analyses of this signal, which used the \texttt{IMRPhenomXPHM} and
\texttt{SEOBNRv4PHM} models, the \texttt{IMRPhenomXPHM} results showed some support for unequal masses and high spin. However, since
this model does not include the multipole asymmetry, it is possible that the apparent measurement of a high primary spin was due to uncertainties in the 
waveform model (as suggested in Refs.~\cite{Hannam:2021pit, Hoy:202203H}), and its partial agreement with the results from the more accurate and complete
\texttt{NRSur7dq4} model may have been coincidental. To fully clarify these questions would require a more detailed study of the uncertainties 
of all three models in this region of parameter space, but since the \texttt{Phenom} and \texttt{SEOBNR} models have now 
both been superseded by upgraded
versions~\cite{ramos2023seobnrv5phm,Thompson:2023},
these points may be moot. The broader and more important conclusion that we can draw from these results is that further improvement in 
symmetric models alone will not be sufficient to accurately measure the parameters of precessing 
systems, even at moderate SNRs; the inclusion of the multipole asymmetry is required in all waveform models.


\section{Conclusion}
\label{section:Conclusions}    

We have studied the impact of neglecting the multipole asymmetry in waveform modelling on the measurement of binary source parameters.
We focussed on loud signals (with SNR 100), to assess the impact of the multipole asymmetry in systems where the individual spins should be
measurable. We find that neglecting the multipole asymmetry introduces systematic errors into the measurement of the magnitude and 
direction of each spin. The parameters that are measured in the absence of precession ($M$, $q$, $\chi_{\rm eff}$) are only weakly affected
by neglecting the asymmetry, at least for systems with comparable masses or small spins.
	
Furthermore, we investigate how the biases depend on the inclination of the binary, 
the primary spin magnitude and the mass ratio of the system. We also test their dependence on the 
recoil velocity of the final black hole by injecting \texttt{NRSur7dq4} waveforms with 
different in-plane spin directions that correspond to the maximum and minimum recoil. Our results 
show no evidence of strong dependence between the biases and the recoil velocity  	
or the inclination of the system. We find that for the inclinations we consider, $\iota \in [ 0^\circ, 30^\circ, 60^\circ,90^\circ]$, there is 
no strong impact on the biases even if the system is oriented from face-on to edge-on. Similarly, in the case of the 
maximum and minimum recoil value, the magnitude of the biases remains largely unaffected by these extremes in the recoil values.
Across all of these cases, the bias in the spin magnitudes and directions will vary as these parameters are changed, but the biases do 
not become particularly larger or smaller. We leave a detailed understanding of the direction and magnitude of the biases as a function
of inclination and spin direction to future work. 
	
In contrast, the biases introduced by the \texttt{NRSur7dq4\_sym} model do depend on the primary spin magnitude and the mass ratio of the system. 
We investigate these effects for configurations with two different primary spin values $a_1/m_1=0.4,0.7$.
Since the effects of the multipole asymmetry are weaker for lower spins, the biases are more subtle in the analysis of the biniary with spin 
$a_1/m_1=0.4$. To test the dependency on the mass ratio, we considered binaries with mass ratios $q=2,4$. In addition, we consider a 
higher-mass-ratio, high-spin configuration, and here the primary spin is better constrained by the symmetric model, but the secondary spin rails 
against extremal values, and this in turn does lead to a bias in $\chi_{\rm eff}$. 
	
We have also considered the GW200129 signal, which is the only GW observation so far to show strong evidence for precession~\cite{Hannam:2021pit}. 
We find that without the multipole asymmetry it is not possible to reliably identify the high primary spin (the lower bound of the 90\% credible interval
drops from 0.43 to 0.1), and the mass ratio is less well constrained; see Fig.~\ref{fig:posteriors_GW200129} and Tab.~\ref{tab:table1}.
This illustrates the importance of the multipole asymmetry in measurements of precessing binaries, even at relatively low SNRs. 
This example also illustrates the confusing systematic errors that can be introduced by model uncertainty: in the LVK analysis the \texttt{IMRPhenomXPHM}
model may by spuriously identifying a high primary spin due to inaccuracies in the symmetric contribution (since we find that an accurate symmetric
model does not identify a high spin). 

These results have important consequences for future observations of binary black holes. As detector sensitivities improve, we will observe more systems
at SNRs where it is in principle possible to measure the full spin information (both ``aligned'' and ``in-plane'' components). Employing symmetric
waveforms for the analysis of these signals will lead to incorrect measurements, making it difficult to confidently identify precessing systems, and to measure
the spin magnitudes and orientations, and the recoil. This will likely also impact population studies and efforts to better understand binary formation mechanisms. 

The current study used the \texttt{NRSur7dq4} model, which does include multipole asymmetry. However, this model cannot be used for systems
with large mass ratios, or masses below 65\,$M_\odot$. Our results show that it is essential to include the multipole asymmetry in other waveform 
models. An approach to do this for frequency-domain models was recently presented in Ref.~\cite{Ghosh:2023mhc}, and this or other methods
need to be developed for any waveform model intended for use on signals beyond moderate SNRs, where in-plane spin information may be
measurable.

\section{Acknowledgements}
We thank Charlie Hoy for discussions and assistance with parameter estimation analyses, and Lionel London for sharing his numerical-relativity 
data processing tools. We also thank Steve Fairhurst, Vivien Raymond, Frank Ohme, Patricia Schmidt and Geraint Pratten for discussions. 

The authors were supported in part by Science and Technology Facilities Council (STFC) grant ST/V00154X/1 and European Research Council (ERC)
Consolidator Grant 647839. P.K was also supported by the GW consolidated grant: STFC grant ST/V005677/1.
J.T.\ acknowledges support from the NASA LISA Preparatory Science grant 20-LPS20-0005.

This research was undertaken using the supercomputing facilities at Cardiff University operated by Advanced Research Computing at Cardiff (ARCCA) on behalf of the Cardiff Supercomputing Facility and the HPC Wales and Supercomputing Wales (SCW) projects. We acknowledge the support of the latter, which is part-funded by the European Regional Development Fund (ERDF) via the Welsh Government.

Plots were prepared with \texttt{Matplotlib}~\cite{Matplotlib} and {\texttt{PESummary}}~\cite{Hoy:PESummary}.
Parameter estimation was performed with the 
\texttt{LALInference} software library~\cite{veitch2015parameter}.
{\texttt{NumPy}}~\cite{numpy} and {\texttt{Scipy}}~\cite{Scipy} were also used during our analysis.

\bibliography{references.bib}

\begin{thebibliography}{66}%
\makeatletter
\providecommand \@ifxundefined [1]{%
 \@ifx{#1\undefined}
}%
\providecommand \@ifnum [1]{%
 \ifnum #1\expandafter \@firstoftwo
 \else \expandafter \@secondoftwo
 \fi
}%
\providecommand \@ifx [1]{%
 \ifx #1\expandafter \@firstoftwo
 \else \expandafter \@secondoftwo
 \fi
}%
\providecommand \natexlab [1]{#1}%
\providecommand \enquote  [1]{``#1''}%
\providecommand \bibnamefont  [1]{#1}%
\providecommand \bibfnamefont [1]{#1}%
\providecommand \citenamefont [1]{#1}%
\providecommand \href@noop [0]{\@secondoftwo}%
\providecommand \href [0]{\begingroup \@sanitize@url \@href}%
\providecommand \@href[1]{\@@startlink{#1}\@@href}%
\providecommand \@@href[1]{\endgroup#1\@@endlink}%
\providecommand \@sanitize@url [0]{\catcode `\\12\catcode `\$12\catcode
  `\&12\catcode `\#12\catcode `\^12\catcode `\_12\catcode `\%12\relax}%
\providecommand \@@startlink[1]{}%
\providecommand \@@endlink[0]{}%
\providecommand \url  [0]{\begingroup\@sanitize@url \@url }%
\providecommand \@url [1]{\endgroup\@href {#1}{\urlprefix }}%
\providecommand \urlprefix  [0]{URL }%
\providecommand \Eprint [0]{\href }%
\providecommand \doibase [0]{http://dx.doi.org/}%
\providecommand \selectlanguage [0]{\@gobble}%
\providecommand \bibinfo  [0]{\@secondoftwo}%
\providecommand \bibfield  [0]{\@secondoftwo}%
\providecommand \translation [1]{[#1]}%
\providecommand \BibitemOpen [0]{}%
\providecommand \bibitemStop [0]{}%
\providecommand \bibitemNoStop [0]{.\EOS\space}%
\providecommand \EOS [0]{\spacefactor3000\relax}%
\providecommand \BibitemShut  [1]{\csname bibitem#1\endcsname}%
\let\auto@bib@innerbib\@empty
\bibitem [{\citenamefont {Abbott}\ \emph
  {et~al.}(2021{\natexlab{a}})\citenamefont {Abbott} \emph
  {et~al.}}]{LIGOScientific:2020kqk}%
  \BibitemOpen
  \bibfield  {author} {\bibinfo {author} {\bibfnamefont {R.}~\bibnamefont
  {Abbott}} \emph {et~al.} (\bibinfo {collaboration} {LIGO Scientific,
  Virgo}),\ }\href {\doibase 10.3847/2041-8213/abe949} {\bibfield  {journal}
  {\bibinfo  {journal} {Astrophys. J. Lett.}\ }\textbf {\bibinfo {volume}
  {913}},\ \bibinfo {pages} {L7} (\bibinfo {year} {2021}{\natexlab{a}})},\
  \Eprint {http://arxiv.org/abs/2010.14533} {arXiv:2010.14533 [astro-ph.HE]}
  \BibitemShut {NoStop}%
\bibitem [{\citenamefont {Abbott}\ \emph {et~al.}(2023)\citenamefont {Abbott}
  \emph {et~al.}}]{KAGRA:2021duu}%
  \BibitemOpen
  \bibfield  {author} {\bibinfo {author} {\bibfnamefont {R.}~\bibnamefont
  {Abbott}} \emph {et~al.} (\bibinfo {collaboration} {KAGRA, VIRGO, LIGO
  Scientific}),\ }\href {\doibase 10.1103/PhysRevX.13.011048} {\bibfield
  {journal} {\bibinfo  {journal} {Phys. Rev. X}\ }\textbf {\bibinfo {volume}
  {13}},\ \bibinfo {pages} {011048} (\bibinfo {year} {2023})},\ \Eprint
  {http://arxiv.org/abs/2111.03634} {arXiv:2111.03634 [astro-ph.HE]}
  \BibitemShut {NoStop}%
\bibitem [{\citenamefont {Abbott}\ \emph
  {et~al.}(2016{\natexlab{a}})\citenamefont {Abbott} \emph
  {et~al.}}]{LIGOScientific:2016vpg}%
  \BibitemOpen
  \bibfield  {author} {\bibinfo {author} {\bibfnamefont {B.~P.}\ \bibnamefont
  {Abbott}} \emph {et~al.} (\bibinfo {collaboration} {LIGO Scientific,
  Virgo}),\ }\href {\doibase 10.3847/2041-8205/818/2/L22} {\bibfield  {journal}
  {\bibinfo  {journal} {Astrophys. J. Lett.}\ }\textbf {\bibinfo {volume}
  {818}},\ \bibinfo {pages} {L22} (\bibinfo {year} {2016}{\natexlab{a}})},\
  \Eprint {http://arxiv.org/abs/1602.03846} {arXiv:1602.03846 [astro-ph.HE]}
  \BibitemShut {NoStop}%
\bibitem [{\citenamefont {Abbott}\ \emph
  {et~al.}(2019{\natexlab{a}})\citenamefont {Abbott} \emph
  {et~al.}}]{LIGOScientific:2018jsj}%
  \BibitemOpen
  \bibfield  {author} {\bibinfo {author} {\bibfnamefont {B.~P.}\ \bibnamefont
  {Abbott}} \emph {et~al.} (\bibinfo {collaboration} {LIGO Scientific,
  Virgo}),\ }\href {\doibase 10.3847/2041-8213/ab3800} {\bibfield  {journal}
  {\bibinfo  {journal} {Astrophys. J. Lett.}\ }\textbf {\bibinfo {volume}
  {882}},\ \bibinfo {pages} {L24} (\bibinfo {year} {2019}{\natexlab{a}})},\
  \Eprint {http://arxiv.org/abs/1811.12940} {arXiv:1811.12940 [astro-ph.HE]}
  \BibitemShut {NoStop}%
\bibitem [{\citenamefont {{LIGO Scientific Collaboration}}\ \emph
  {et~al.}(2015)\citenamefont {{LIGO Scientific Collaboration}}, \citenamefont
  {{Aasi}}, \citenamefont {{Abbott}}, \citenamefont {{Abbott}}, \citenamefont
  {{Abbott}}, \citenamefont {{Abernathy}}, \citenamefont {{Ackley}},
  \citenamefont {{Adams}}, \citenamefont {{Adams}}, \citenamefont {{Addesso}},\
  and\ \citenamefont {et~al.}}]{2015DEFrange}%
  \BibitemOpen
  \bibfield  {author} {\bibinfo {author} {\bibnamefont {{LIGO Scientific
  Collaboration}}}, \bibinfo {author} {\bibfnamefont {J.}~\bibnamefont
  {{Aasi}}}, \bibinfo {author} {\bibfnamefont {B.~P.}\ \bibnamefont
  {{Abbott}}}, \bibinfo {author} {\bibfnamefont {R.}~\bibnamefont {{Abbott}}},
  \bibinfo {author} {\bibfnamefont {T.}~\bibnamefont {{Abbott}}}, \bibinfo
  {author} {\bibfnamefont {M.~R.}\ \bibnamefont {{Abernathy}}}, \bibinfo
  {author} {\bibfnamefont {K.}~\bibnamefont {{Ackley}}}, \bibinfo {author}
  {\bibfnamefont {C.}~\bibnamefont {{Adams}}}, \bibinfo {author} {\bibfnamefont
  {T.}~\bibnamefont {{Adams}}}, \bibinfo {author} {\bibfnamefont
  {P.}~\bibnamefont {{Addesso}}}, \ and\ \bibinfo {author} {\bibnamefont
  {et~al.}},\ }\href {\doibase 10.1088/0264-9381/32/7/074001} {\bibfield
  {journal} {\bibinfo  {journal} {Classical and Quantum Gravity}\ }\textbf
  {\bibinfo {volume} {32}},\ \bibinfo {eid} {074001} (\bibinfo {year}
  {2015})},\ \Eprint {http://arxiv.org/abs/1411.4547} {arXiv:1411.4547 [gr-qc]}
  \BibitemShut {NoStop}%
\bibitem [{\citenamefont {Acernese}\ \emph {et~al.}(2014)\citenamefont
  {Acernese}, \citenamefont {Agathos}, \citenamefont {Agatsuma}, \citenamefont
  {Aisa}, \citenamefont {Allemandou}, \citenamefont {Allocca}, \citenamefont
  {Amarni}, \citenamefont {Astone}, \citenamefont {Balestri}, \citenamefont
  {Ballardin} \emph {et~al.}}]{acernese:2014advanced}%
  \BibitemOpen
  \bibfield  {author} {\bibinfo {author} {\bibfnamefont {F.~a.}\ \bibnamefont
  {Acernese}}, \bibinfo {author} {\bibfnamefont {M.}~\bibnamefont {Agathos}},
  \bibinfo {author} {\bibfnamefont {K.}~\bibnamefont {Agatsuma}}, \bibinfo
  {author} {\bibfnamefont {D.}~\bibnamefont {Aisa}}, \bibinfo {author}
  {\bibfnamefont {N.}~\bibnamefont {Allemandou}}, \bibinfo {author}
  {\bibfnamefont {A.}~\bibnamefont {Allocca}}, \bibinfo {author} {\bibfnamefont
  {J.}~\bibnamefont {Amarni}}, \bibinfo {author} {\bibfnamefont
  {P.}~\bibnamefont {Astone}}, \bibinfo {author} {\bibfnamefont
  {G.}~\bibnamefont {Balestri}}, \bibinfo {author} {\bibfnamefont
  {G.}~\bibnamefont {Ballardin}},  \emph {et~al.},\ }\href@noop {} {\bibfield
  {journal} {\bibinfo  {journal} {Classical and Quantum Gravity}\ }\textbf
  {\bibinfo {volume} {32}},\ \bibinfo {pages} {024001} (\bibinfo {year}
  {2014})}\BibitemShut {NoStop}%
\bibitem [{\citenamefont {{Kagra Collaboration}}\ \emph
  {et~al.}(2019)\citenamefont {{Kagra Collaboration}}, \citenamefont
  {{Akutsu}}, \citenamefont {{Ando}},\ and\ \citenamefont
  {et~al.}}]{2019kagra}%
  \BibitemOpen
  \bibfield  {author} {\bibinfo {author} {\bibnamefont {{Kagra
  Collaboration}}}, \bibinfo {author} {\bibfnamefont {T.}~\bibnamefont
  {{Akutsu}}}, \bibinfo {author} {\bibfnamefont {M.}~\bibnamefont {{Ando}}}, \
  and\ \bibinfo {author} {\bibnamefont {et~al.}},\ }\href {\doibase
  10.1038/s41550-018-0658-y} {\bibfield  {journal} {\bibinfo  {journal} {Nature
  Astronomy}\ }\textbf {\bibinfo {volume} {3}},\ \bibinfo {pages} {35}
  (\bibinfo {year} {2019})}\BibitemShut {NoStop}%
\bibitem [{\citenamefont {Ajith}\ \emph {et~al.}(2011)\citenamefont {Ajith}
  \emph {et~al.}}]{Ajith:2009bn}%
  \BibitemOpen
  \bibfield  {author} {\bibinfo {author} {\bibfnamefont {P.}~\bibnamefont
  {Ajith}} \emph {et~al.},\ }\href {\doibase 10.1103/PhysRevLett.106.241101}
  {\bibfield  {journal} {\bibinfo  {journal} {Phys. Rev. Lett.}\ }\textbf
  {\bibinfo {volume} {106}},\ \bibinfo {pages} {241101} (\bibinfo {year}
  {2011})},\ \Eprint {http://arxiv.org/abs/0909.2867} {arXiv:0909.2867 [gr-qc]}
  \BibitemShut {NoStop}%
\bibitem [{\citenamefont {{Apostolatos}}\ \emph {et~al.}(1994)\citenamefont
  {{Apostolatos}}, \citenamefont {{Cutler}}, \citenamefont {{Sussman}},\ and\
  \citenamefont {{Thorne}}}]{Apostolatos:1994sio}%
  \BibitemOpen
  \bibfield  {author} {\bibinfo {author} {\bibfnamefont {T.~A.}\ \bibnamefont
  {{Apostolatos}}}, \bibinfo {author} {\bibfnamefont {C.}~\bibnamefont
  {{Cutler}}}, \bibinfo {author} {\bibfnamefont {G.~J.}\ \bibnamefont
  {{Sussman}}}, \ and\ \bibinfo {author} {\bibfnamefont {K.~S.}\ \bibnamefont
  {{Thorne}}},\ }\href {\doibase 10.1103/PhysRevD.49.6274} {\bibfield
  {journal} {\bibinfo  {journal} {\prd}\ }\textbf {\bibinfo {volume} {49}},\
  \bibinfo {pages} {6274} (\bibinfo {year} {1994})}\BibitemShut {NoStop}%
\bibitem [{\citenamefont {{Kidder}}(1995)}]{Kidder:1995cbsv}%
  \BibitemOpen
  \bibfield  {author} {\bibinfo {author} {\bibfnamefont {L.~E.}\ \bibnamefont
  {{Kidder}}},\ }\href {\doibase 10.1103/PhysRevD.52.821} {\bibfield  {journal}
  {\bibinfo  {journal} {\prd}\ }\textbf {\bibinfo {volume} {52}},\ \bibinfo
  {pages} {821} (\bibinfo {year} {1995})},\ \Eprint
  {http://arxiv.org/abs/gr-qc/9506022} {arXiv:gr-qc/9506022 [gr-qc]}
  \BibitemShut {NoStop}%
\bibitem [{\citenamefont {{Schmidt}}\ \emph {et~al.}(2012)\citenamefont
  {{Schmidt}}, \citenamefont {{Hannam}},\ and\ \citenamefont
  {{Husa}}}]{Schmidt:2012tmg}%
  \BibitemOpen
  \bibfield  {author} {\bibinfo {author} {\bibfnamefont {P.}~\bibnamefont
  {{Schmidt}}}, \bibinfo {author} {\bibfnamefont {M.}~\bibnamefont {{Hannam}}},
  \ and\ \bibinfo {author} {\bibfnamefont {S.}~\bibnamefont {{Husa}}},\ }\href
  {\doibase 10.1103/PhysRevD.86.104063} {\bibfield  {journal} {\bibinfo
  {journal} {\prd}\ }\textbf {\bibinfo {volume} {86}},\ \bibinfo {eid} {104063}
  (\bibinfo {year} {2012})},\ \Eprint {http://arxiv.org/abs/1207.3088}
  {arXiv:1207.3088 [gr-qc]} \BibitemShut {NoStop}%
\bibitem [{\citenamefont {Hannam}\ \emph {et~al.}(2014)\citenamefont {Hannam},
  \citenamefont {Schmidt}, \citenamefont {Boh{\'e}}, \citenamefont {Haegel},
  \citenamefont {Husa}, \citenamefont {Ohme}, \citenamefont {Pratten},\ and\
  \citenamefont {P{\"u}rrer}}]{Hannam:2013oca}%
  \BibitemOpen
  \bibfield  {author} {\bibinfo {author} {\bibfnamefont {M.}~\bibnamefont
  {Hannam}}, \bibinfo {author} {\bibfnamefont {P.}~\bibnamefont {Schmidt}},
  \bibinfo {author} {\bibfnamefont {A.}~\bibnamefont {Boh{\'e}}}, \bibinfo
  {author} {\bibfnamefont {L.}~\bibnamefont {Haegel}}, \bibinfo {author}
  {\bibfnamefont {S.}~\bibnamefont {Husa}}, \bibinfo {author} {\bibfnamefont
  {F.}~\bibnamefont {Ohme}}, \bibinfo {author} {\bibfnamefont {G.}~\bibnamefont
  {Pratten}}, \ and\ \bibinfo {author} {\bibfnamefont {M.}~\bibnamefont
  {P{\"u}rrer}},\ }\href {\doibase 10.1103/PhysRevLett.113.151101} {\bibfield
  {journal} {\bibinfo  {journal} {Phys. Rev. Lett.}\ }\textbf {\bibinfo
  {volume} {113}},\ \bibinfo {pages} {151101} (\bibinfo {year} {2014})},\
  \Eprint {http://arxiv.org/abs/1308.3271} {arXiv:1308.3271 [gr-qc]}
  \BibitemShut {NoStop}%
\bibitem [{\citenamefont {Khan}\ \emph {et~al.}(2019)\citenamefont {Khan},
  \citenamefont {Chatziioannou}, \citenamefont {Hannam},\ and\ \citenamefont
  {Ohme}}]{Khan:2018fmp}%
  \BibitemOpen
  \bibfield  {author} {\bibinfo {author} {\bibfnamefont {S.}~\bibnamefont
  {Khan}}, \bibinfo {author} {\bibfnamefont {K.}~\bibnamefont {Chatziioannou}},
  \bibinfo {author} {\bibfnamefont {M.}~\bibnamefont {Hannam}}, \ and\ \bibinfo
  {author} {\bibfnamefont {F.}~\bibnamefont {Ohme}},\ }\href {\doibase
  10.1103/PhysRevD.100.024059} {\bibfield  {journal} {\bibinfo  {journal}
  {Phys. Rev.}\ }\textbf {\bibinfo {volume} {D100}},\ \bibinfo {pages} {024059}
  (\bibinfo {year} {2019})},\ \Eprint {http://arxiv.org/abs/1809.10113}
  {arXiv:1809.10113 [gr-qc]} \BibitemShut {NoStop}%
\bibitem [{\citenamefont {Khan}\ \emph {et~al.}(2020)\citenamefont {Khan},
  \citenamefont {Ohme}, \citenamefont {Chatziioannou},\ and\ \citenamefont
  {Hannam}}]{Khan:2019kot}%
  \BibitemOpen
  \bibfield  {author} {\bibinfo {author} {\bibfnamefont {S.}~\bibnamefont
  {Khan}}, \bibinfo {author} {\bibfnamefont {F.}~\bibnamefont {Ohme}}, \bibinfo
  {author} {\bibfnamefont {K.}~\bibnamefont {Chatziioannou}}, \ and\ \bibinfo
  {author} {\bibfnamefont {M.}~\bibnamefont {Hannam}},\ }\href {\doibase
  10.1103/PhysRevD.101.024056} {\bibfield  {journal} {\bibinfo  {journal}
  {Phys. Rev. D}\ }\textbf {\bibinfo {volume} {101}},\ \bibinfo {pages}
  {024056} (\bibinfo {year} {2020})},\ \Eprint
  {http://arxiv.org/abs/1911.06050} {arXiv:1911.06050 [gr-qc]} \BibitemShut
  {NoStop}%
\bibitem [{\citenamefont {Pratten}\ \emph {et~al.}(2021)\citenamefont {Pratten}
  \emph {et~al.}}]{Pratten:2020ceb}%
  \BibitemOpen
  \bibfield  {author} {\bibinfo {author} {\bibfnamefont {G.}~\bibnamefont
  {Pratten}} \emph {et~al.},\ }\href {\doibase 10.1103/PhysRevD.103.104056}
  {\bibfield  {journal} {\bibinfo  {journal} {Phys. Rev. D}\ }\textbf {\bibinfo
  {volume} {103}},\ \bibinfo {pages} {104056} (\bibinfo {year} {2021})},\
  \Eprint {http://arxiv.org/abs/2004.06503} {arXiv:2004.06503 [gr-qc]}
  \BibitemShut {NoStop}%
\bibitem [{\citenamefont {Estell\'es}\ \emph {et~al.}(2022)\citenamefont
  {Estell\'es}, \citenamefont {Colleoni}, \citenamefont {Garc\'\i{}a-Quir\'os},
  \citenamefont {Husa}, \citenamefont {Keitel}, \citenamefont {Mateu-Lucena},
  \citenamefont {Planas},\ and\ \citenamefont
  {Ramos-Buades}}]{Estelles:2021gvs}%
  \BibitemOpen
  \bibfield  {author} {\bibinfo {author} {\bibfnamefont {H.}~\bibnamefont
  {Estell\'es}}, \bibinfo {author} {\bibfnamefont {M.}~\bibnamefont
  {Colleoni}}, \bibinfo {author} {\bibfnamefont {C.}~\bibnamefont
  {Garc\'\i{}a-Quir\'os}}, \bibinfo {author} {\bibfnamefont {S.}~\bibnamefont
  {Husa}}, \bibinfo {author} {\bibfnamefont {D.}~\bibnamefont {Keitel}},
  \bibinfo {author} {\bibfnamefont {M.}~\bibnamefont {Mateu-Lucena}}, \bibinfo
  {author} {\bibfnamefont {M.~d.~L.}\ \bibnamefont {Planas}}, \ and\ \bibinfo
  {author} {\bibfnamefont {A.}~\bibnamefont {Ramos-Buades}},\ }\href {\doibase
  10.1103/PhysRevD.105.084040} {\bibfield  {journal} {\bibinfo  {journal}
  {Phys. Rev. D}\ }\textbf {\bibinfo {volume} {105}},\ \bibinfo {pages}
  {084040} (\bibinfo {year} {2022})},\ \Eprint
  {http://arxiv.org/abs/2105.05872} {arXiv:2105.05872 [gr-qc]} \BibitemShut
  {NoStop}%
\bibitem [{\citenamefont {Hamilton}\ \emph {et~al.}(2021)\citenamefont
  {Hamilton}, \citenamefont {London}, \citenamefont {Thompson}, \citenamefont
  {Fauchon-Jones}, \citenamefont {Hannam}, \citenamefont {Kalaghatgi},
  \citenamefont {Khan}, \citenamefont {Pannarale},\ and\ \citenamefont
  {Vano-Vinuales}}]{Hamilton:2021pkf}%
  \BibitemOpen
  \bibfield  {author} {\bibinfo {author} {\bibfnamefont {E.}~\bibnamefont
  {Hamilton}}, \bibinfo {author} {\bibfnamefont {L.}~\bibnamefont {London}},
  \bibinfo {author} {\bibfnamefont {J.~E.}\ \bibnamefont {Thompson}}, \bibinfo
  {author} {\bibfnamefont {E.}~\bibnamefont {Fauchon-Jones}}, \bibinfo {author}
  {\bibfnamefont {M.}~\bibnamefont {Hannam}}, \bibinfo {author} {\bibfnamefont
  {C.}~\bibnamefont {Kalaghatgi}}, \bibinfo {author} {\bibfnamefont
  {S.}~\bibnamefont {Khan}}, \bibinfo {author} {\bibfnamefont {F.}~\bibnamefont
  {Pannarale}}, \ and\ \bibinfo {author} {\bibfnamefont {A.}~\bibnamefont
  {Vano-Vinuales}},\ }\href {\doibase 10.1103/PhysRevD.104.124027} {\bibfield
  {journal} {\bibinfo  {journal} {Phys. Rev. D}\ }\textbf {\bibinfo {volume}
  {104}},\ \bibinfo {pages} {124027} (\bibinfo {year} {2021})},\ \Eprint
  {http://arxiv.org/abs/2107.08876} {arXiv:2107.08876 [gr-qc]} \BibitemShut
  {NoStop}%
\bibitem [{\citenamefont {{Pan}}\ \emph {et~al.}(2014)\citenamefont {{Pan}},
  \citenamefont {{Buonanno}}, \citenamefont {{Taracchini}}, \citenamefont
  {{Kidder}}, \citenamefont {{Mrou{\'e}}}, \citenamefont {{Pfeiffer}},
  \citenamefont {{Scheel}},\ and\ \citenamefont
  {{Szil{\'a}gyi}}}]{Pan:2014yaq}%
  \BibitemOpen
  \bibfield  {author} {\bibinfo {author} {\bibfnamefont {Y.}~\bibnamefont
  {{Pan}}}, \bibinfo {author} {\bibfnamefont {A.}~\bibnamefont {{Buonanno}}},
  \bibinfo {author} {\bibfnamefont {A.}~\bibnamefont {{Taracchini}}}, \bibinfo
  {author} {\bibfnamefont {L.~E.}\ \bibnamefont {{Kidder}}}, \bibinfo {author}
  {\bibfnamefont {A.~H.}\ \bibnamefont {{Mrou{\'e}}}}, \bibinfo {author}
  {\bibfnamefont {H.~P.}\ \bibnamefont {{Pfeiffer}}}, \bibinfo {author}
  {\bibfnamefont {M.~A.}\ \bibnamefont {{Scheel}}}, \ and\ \bibinfo {author}
  {\bibfnamefont {B.}~\bibnamefont {{Szil{\'a}gyi}}},\ }\href {\doibase
  10.1103/PhysRevD.89.084006} {\bibfield  {journal} {\bibinfo  {journal}
  {\prd}\ }\textbf {\bibinfo {volume} {89}},\ \bibinfo {eid} {084006} (\bibinfo
  {year} {2014})},\ \Eprint {http://arxiv.org/abs/1307.6232} {arXiv:1307.6232
  [gr-qc]} \BibitemShut {NoStop}%
\bibitem [{\citenamefont {{Taracchini}}\ \emph {et~al.}(2014)\citenamefont
  {{Taracchini}}, \citenamefont {{Buonanno}}, \citenamefont {{Pan}},
  \citenamefont {{Hinderer}}, \citenamefont {{Boyle}}, \citenamefont
  {{Hemberger}}, \citenamefont {{Kidder}}, \citenamefont {{Lovelace}},
  \citenamefont {{Mrou{\'e}}}, \citenamefont {{Pfeiffer}}, \citenamefont
  {{Scheel}}, \citenamefont {{Szil{\'a}gyi}}, \citenamefont {{Taylor}},\ and\
  \citenamefont {{Zenginoglu}}}]{Taracchini:2014hgd}%
  \BibitemOpen
  \bibfield  {author} {\bibinfo {author} {\bibfnamefont {A.}~\bibnamefont
  {{Taracchini}}}, \bibinfo {author} {\bibfnamefont {A.}~\bibnamefont
  {{Buonanno}}}, \bibinfo {author} {\bibfnamefont {Y.}~\bibnamefont {{Pan}}},
  \bibinfo {author} {\bibfnamefont {T.}~\bibnamefont {{Hinderer}}}, \bibinfo
  {author} {\bibfnamefont {M.}~\bibnamefont {{Boyle}}}, \bibinfo {author}
  {\bibfnamefont {D.~A.}\ \bibnamefont {{Hemberger}}}, \bibinfo {author}
  {\bibfnamefont {L.~E.}\ \bibnamefont {{Kidder}}}, \bibinfo {author}
  {\bibfnamefont {G.}~\bibnamefont {{Lovelace}}}, \bibinfo {author}
  {\bibfnamefont {A.~H.}\ \bibnamefont {{Mrou{\'e}}}}, \bibinfo {author}
  {\bibfnamefont {H.~P.}\ \bibnamefont {{Pfeiffer}}}, \bibinfo {author}
  {\bibfnamefont {M.~A.}\ \bibnamefont {{Scheel}}}, \bibinfo {author}
  {\bibfnamefont {B.}~\bibnamefont {{Szil{\'a}gyi}}}, \bibinfo {author}
  {\bibfnamefont {N.~W.}\ \bibnamefont {{Taylor}}}, \ and\ \bibinfo {author}
  {\bibfnamefont {A.}~\bibnamefont {{Zenginoglu}}},\ }\href {\doibase
  10.1103/PhysRevD.89.061502} {\bibfield  {journal} {\bibinfo  {journal}
  {\prd}\ }\textbf {\bibinfo {volume} {89}},\ \bibinfo {eid} {061502} (\bibinfo
  {year} {2014})},\ \Eprint {http://arxiv.org/abs/1311.2544} {arXiv:1311.2544
  [gr-qc]} \BibitemShut {NoStop}%
\bibitem [{\citenamefont {{Ossokine}}\ \emph {et~al.}(2020)\citenamefont
  {{Ossokine}}, \citenamefont {{Buonanno}}, \citenamefont {{Marsat}},
  \citenamefont {{Cotesta}}, \citenamefont {{Babak}}, \citenamefont
  {{Dietrich}}, \citenamefont {{Haas}}, \citenamefont {{Hinder}}, \citenamefont
  {{Pfeiffer}}, \citenamefont {{P{\"u}rrer}}, \citenamefont {{Woodford}},
  \citenamefont {{Boyle}}, \citenamefont {{Kidder}}, \citenamefont {{Scheel}},\
  and\ \citenamefont {{Szil{\'a}gyi}}}]{Ossokine:2020thd}%
  \BibitemOpen
  \bibfield  {author} {\bibinfo {author} {\bibfnamefont {S.}~\bibnamefont
  {{Ossokine}}}, \bibinfo {author} {\bibfnamefont {A.}~\bibnamefont
  {{Buonanno}}}, \bibinfo {author} {\bibfnamefont {S.}~\bibnamefont
  {{Marsat}}}, \bibinfo {author} {\bibfnamefont {R.}~\bibnamefont {{Cotesta}}},
  \bibinfo {author} {\bibfnamefont {S.}~\bibnamefont {{Babak}}}, \bibinfo
  {author} {\bibfnamefont {T.}~\bibnamefont {{Dietrich}}}, \bibinfo {author}
  {\bibfnamefont {R.}~\bibnamefont {{Haas}}}, \bibinfo {author} {\bibfnamefont
  {I.}~\bibnamefont {{Hinder}}}, \bibinfo {author} {\bibfnamefont {H.~P.}\
  \bibnamefont {{Pfeiffer}}}, \bibinfo {author} {\bibfnamefont
  {M.}~\bibnamefont {{P{\"u}rrer}}}, \bibinfo {author} {\bibfnamefont {C.~J.}\
  \bibnamefont {{Woodford}}}, \bibinfo {author} {\bibfnamefont
  {M.}~\bibnamefont {{Boyle}}}, \bibinfo {author} {\bibfnamefont {L.~E.}\
  \bibnamefont {{Kidder}}}, \bibinfo {author} {\bibfnamefont {M.~A.}\
  \bibnamefont {{Scheel}}}, \ and\ \bibinfo {author} {\bibfnamefont
  {B.}~\bibnamefont {{Szil{\'a}gyi}}},\ }\href {\doibase
  10.1103/PhysRevD.102.044055} {\bibfield  {journal} {\bibinfo  {journal}
  {\prd}\ }\textbf {\bibinfo {volume} {102}},\ \bibinfo {eid} {044055}
  (\bibinfo {year} {2020})},\ \Eprint {http://arxiv.org/abs/2004.09442}
  {arXiv:2004.09442 [gr-qc]} \BibitemShut {NoStop}%
\bibitem [{\citenamefont {Ramos-Buades}\ \emph
  {et~al.}(2023{\natexlab{a}})\citenamefont {Ramos-Buades}, \citenamefont
  {Buonanno}, \citenamefont {Estell\'es}, \citenamefont {Khalil}, \citenamefont
  {Mihaylov}, \citenamefont {Ossokine}, \citenamefont {Pompili},\ and\
  \citenamefont {Shiferaw}}]{Ramos-Buades:2023ehm}%
  \BibitemOpen
  \bibfield  {author} {\bibinfo {author} {\bibfnamefont {A.}~\bibnamefont
  {Ramos-Buades}}, \bibinfo {author} {\bibfnamefont {A.}~\bibnamefont
  {Buonanno}}, \bibinfo {author} {\bibfnamefont {H.}~\bibnamefont
  {Estell\'es}}, \bibinfo {author} {\bibfnamefont {M.}~\bibnamefont {Khalil}},
  \bibinfo {author} {\bibfnamefont {D.~P.}\ \bibnamefont {Mihaylov}}, \bibinfo
  {author} {\bibfnamefont {S.}~\bibnamefont {Ossokine}}, \bibinfo {author}
  {\bibfnamefont {L.}~\bibnamefont {Pompili}}, \ and\ \bibinfo {author}
  {\bibfnamefont {M.}~\bibnamefont {Shiferaw}},\ }\href@noop {} {\  (\bibinfo
  {year} {2023}{\natexlab{a}})},\ \Eprint {http://arxiv.org/abs/2303.18046}
  {arXiv:2303.18046 [gr-qc]} \BibitemShut {NoStop}%
\bibitem [{\citenamefont {Boyle}\ \emph {et~al.}(2014)\citenamefont {Boyle},
  \citenamefont {Kidder}, \citenamefont {Ossokine},\ and\ \citenamefont
  {Pfeiffer}}]{Boyle:2014ioa}%
  \BibitemOpen
  \bibfield  {author} {\bibinfo {author} {\bibfnamefont {M.}~\bibnamefont
  {Boyle}}, \bibinfo {author} {\bibfnamefont {L.~E.}\ \bibnamefont {Kidder}},
  \bibinfo {author} {\bibfnamefont {S.}~\bibnamefont {Ossokine}}, \ and\
  \bibinfo {author} {\bibfnamefont {H.~P.}\ \bibnamefont {Pfeiffer}},\
  }\href@noop {} {\  (\bibinfo {year} {2014})},\ \Eprint
  {http://arxiv.org/abs/1409.4431} {arXiv:1409.4431 [gr-qc]} \BibitemShut
  {NoStop}%
\bibitem [{\citenamefont {{Islam}}\ \emph {et~al.}(2021)\citenamefont
  {{Islam}}, \citenamefont {{Field}}, \citenamefont {{Haster}},\ and\
  \citenamefont {{Smith}}}]{Islam:2021PhI}%
  \BibitemOpen
  \bibfield  {author} {\bibinfo {author} {\bibfnamefont {T.}~\bibnamefont
  {{Islam}}}, \bibinfo {author} {\bibfnamefont {S.~E.}\ \bibnamefont
  {{Field}}}, \bibinfo {author} {\bibfnamefont {C.-J.}\ \bibnamefont
  {{Haster}}}, \ and\ \bibinfo {author} {\bibfnamefont {R.}~\bibnamefont
  {{Smith}}},\ }\href {\doibase 10.1103/PhysRevD.103.104027} {\bibfield
  {journal} {\bibinfo  {journal} {\prd}\ }\textbf {\bibinfo {volume} {103}},\
  \bibinfo {eid} {104027} (\bibinfo {year} {2021})},\ \Eprint
  {http://arxiv.org/abs/2010.04848} {arXiv:2010.04848 [gr-qc]} \BibitemShut
  {NoStop}%
\bibitem [{\citenamefont {Green}\ \emph {et~al.}(2021)\citenamefont {Green},
  \citenamefont {Hoy}, \citenamefont {Fairhurst}, \citenamefont {Hannam},
  \citenamefont {Pannarale},\ and\ \citenamefont {Thomas}}]{Green:2020ptm}%
  \BibitemOpen
  \bibfield  {author} {\bibinfo {author} {\bibfnamefont {R.}~\bibnamefont
  {Green}}, \bibinfo {author} {\bibfnamefont {C.}~\bibnamefont {Hoy}}, \bibinfo
  {author} {\bibfnamefont {S.}~\bibnamefont {Fairhurst}}, \bibinfo {author}
  {\bibfnamefont {M.}~\bibnamefont {Hannam}}, \bibinfo {author} {\bibfnamefont
  {F.}~\bibnamefont {Pannarale}}, \ and\ \bibinfo {author} {\bibfnamefont
  {C.}~\bibnamefont {Thomas}},\ }\href {\doibase 10.1103/PhysRevD.103.124023}
  {\bibfield  {journal} {\bibinfo  {journal} {Phys. Rev. D}\ }\textbf {\bibinfo
  {volume} {103}},\ \bibinfo {pages} {124023} (\bibinfo {year} {2021})},\
  \Eprint {http://arxiv.org/abs/2010.04131} {arXiv:2010.04131 [gr-qc]}
  \BibitemShut {NoStop}%
\bibitem [{\citenamefont {Kalaghatgi}\ and\ \citenamefont
  {Hannam}(2021)}]{Kalaghatgi:2020gsq}%
  \BibitemOpen
  \bibfield  {author} {\bibinfo {author} {\bibfnamefont {C.}~\bibnamefont
  {Kalaghatgi}}\ and\ \bibinfo {author} {\bibfnamefont {M.}~\bibnamefont
  {Hannam}},\ }\href {\doibase 10.1103/PhysRevD.103.024024} {\bibfield
  {journal} {\bibinfo  {journal} {Phys. Rev. D}\ }\textbf {\bibinfo {volume}
  {103}},\ \bibinfo {pages} {024024} (\bibinfo {year} {2021})},\ \Eprint
  {http://arxiv.org/abs/2008.09957} {arXiv:2008.09957 [gr-qc]} \BibitemShut
  {NoStop}%
\bibitem [{\citenamefont {{Gonz{\'a}lez}}\ \emph {et~al.}(2007)\citenamefont
  {{Gonz{\'a}lez}}, \citenamefont {{Hannam}}, \citenamefont {{Sperhake}},
  \citenamefont {{Br{\"u}gmann}},\ and\ \citenamefont
  {{Husa}}}]{Gonzalez:2007srv}%
  \BibitemOpen
  \bibfield  {author} {\bibinfo {author} {\bibfnamefont {J.~A.}\ \bibnamefont
  {{Gonz{\'a}lez}}}, \bibinfo {author} {\bibfnamefont {M.}~\bibnamefont
  {{Hannam}}}, \bibinfo {author} {\bibfnamefont {U.}~\bibnamefont
  {{Sperhake}}}, \bibinfo {author} {\bibfnamefont {B.}~\bibnamefont
  {{Br{\"u}gmann}}}, \ and\ \bibinfo {author} {\bibfnamefont {S.}~\bibnamefont
  {{Husa}}},\ }\href {\doibase 10.1103/PhysRevLett.98.231101} {\bibfield
  {journal} {\bibinfo  {journal} {\prl}\ }\textbf {\bibinfo {volume} {98}},\
  \bibinfo {eid} {231101} (\bibinfo {year} {2007})},\ \Eprint
  {http://arxiv.org/abs/gr-qc/0702052} {arXiv:gr-qc/0702052 [gr-qc]}
  \BibitemShut {NoStop}%
\bibitem [{\citenamefont {Campanelli}\ \emph {et~al.}(2007)\citenamefont
  {Campanelli}, \citenamefont {Lousto}, \citenamefont {Zlochower},\ and\
  \citenamefont {Merritt}}]{Campanelli:2007ew}%
  \BibitemOpen
  \bibfield  {author} {\bibinfo {author} {\bibfnamefont {M.}~\bibnamefont
  {Campanelli}}, \bibinfo {author} {\bibfnamefont {C.~O.}\ \bibnamefont
  {Lousto}}, \bibinfo {author} {\bibfnamefont {Y.}~\bibnamefont {Zlochower}}, \
  and\ \bibinfo {author} {\bibfnamefont {D.}~\bibnamefont {Merritt}},\ }\href
  {\doibase 10.1086/516712} {\bibfield  {journal} {\bibinfo  {journal}
  {Astrophys. J. Lett.}\ }\textbf {\bibinfo {volume} {659}},\ \bibinfo {pages}
  {L5} (\bibinfo {year} {2007})},\ \Eprint {http://arxiv.org/abs/gr-qc/0701164}
  {arXiv:gr-qc/0701164} \BibitemShut {NoStop}%
\bibitem [{\citenamefont {Bruegmann}\ \emph {et~al.}(2008)\citenamefont
  {Bruegmann}, \citenamefont {Gonzalez}, \citenamefont {Hannam}, \citenamefont
  {Husa},\ and\ \citenamefont {Sperhake}}]{Bruegmann:2007bri}%
  \BibitemOpen
  \bibfield  {author} {\bibinfo {author} {\bibfnamefont {B.}~\bibnamefont
  {Bruegmann}}, \bibinfo {author} {\bibfnamefont {J.~A.}\ \bibnamefont
  {Gonzalez}}, \bibinfo {author} {\bibfnamefont {M.}~\bibnamefont {Hannam}},
  \bibinfo {author} {\bibfnamefont {S.}~\bibnamefont {Husa}}, \ and\ \bibinfo
  {author} {\bibfnamefont {U.}~\bibnamefont {Sperhake}},\ }\href {\doibase
  10.1103/PhysRevD.77.124047} {\bibfield  {journal} {\bibinfo  {journal} {Phys.
  Rev. D}\ }\textbf {\bibinfo {volume} {77}},\ \bibinfo {pages} {124047}
  (\bibinfo {year} {2008})},\ \Eprint {http://arxiv.org/abs/0707.0135}
  {arXiv:0707.0135 [gr-qc]} \BibitemShut {NoStop}%
\bibitem [{\citenamefont {Ghosh}\ \emph {et~al.}(2023)\citenamefont {Ghosh},
  \citenamefont {Kolitsidou},\ and\ \citenamefont {Hannam}}]{Ghosh:2023mhc}%
  \BibitemOpen
  \bibfield  {author} {\bibinfo {author} {\bibfnamefont {S.}~\bibnamefont
  {Ghosh}}, \bibinfo {author} {\bibfnamefont {P.}~\bibnamefont {Kolitsidou}}, \
  and\ \bibinfo {author} {\bibfnamefont {M.}~\bibnamefont {Hannam}},\
  }\href@noop {} {\  (\bibinfo {year} {2023})},\ \Eprint
  {http://arxiv.org/abs/2310.16980} {arXiv:2310.16980 [gr-qc]} \BibitemShut
  {NoStop}%
\bibitem [{\citenamefont {Varma}\ \emph {et~al.}(2019)\citenamefont {Varma},
  \citenamefont {Field}, \citenamefont {Scheel}, \citenamefont {Blackman},
  \citenamefont {Gerosa}, \citenamefont {Stein}, \citenamefont {Kidder},\ and\
  \citenamefont {Pfeiffer}}]{Varma:2019csw}%
  \BibitemOpen
  \bibfield  {author} {\bibinfo {author} {\bibfnamefont {V.}~\bibnamefont
  {Varma}}, \bibinfo {author} {\bibfnamefont {S.~E.}\ \bibnamefont {Field}},
  \bibinfo {author} {\bibfnamefont {M.~A.}\ \bibnamefont {Scheel}}, \bibinfo
  {author} {\bibfnamefont {J.}~\bibnamefont {Blackman}}, \bibinfo {author}
  {\bibfnamefont {D.}~\bibnamefont {Gerosa}}, \bibinfo {author} {\bibfnamefont
  {L.~C.}\ \bibnamefont {Stein}}, \bibinfo {author} {\bibfnamefont {L.~E.}\
  \bibnamefont {Kidder}}, \ and\ \bibinfo {author} {\bibfnamefont {H.~P.}\
  \bibnamefont {Pfeiffer}},\ }\href {\doibase 10.1103/PhysRevResearch.1.033015}
  {\bibfield  {journal} {\bibinfo  {journal} {Phys. Rev. Research.}\ }\textbf
  {\bibinfo {volume} {1}},\ \bibinfo {pages} {033015} (\bibinfo {year}
  {2019})},\ \Eprint {http://arxiv.org/abs/1905.09300} {arXiv:1905.09300
  [gr-qc]} \BibitemShut {NoStop}%
\bibitem [{\citenamefont {Abbott}\ \emph
  {et~al.}(2021{\natexlab{b}})\citenamefont {Abbott} \emph
  {et~al.}}]{LIGOScientific:2021djp}%
  \BibitemOpen
  \bibfield  {author} {\bibinfo {author} {\bibfnamefont {R.}~\bibnamefont
  {Abbott}} \emph {et~al.} (\bibinfo {collaboration} {LIGO Scientific, VIRGO,
  KAGRA}),\ }\href@noop {} {\  (\bibinfo {year} {2021}{\natexlab{b}})},\
  \Eprint {http://arxiv.org/abs/2111.03606} {arXiv:2111.03606 [gr-qc]}
  \BibitemShut {NoStop}%
\bibitem [{\citenamefont {Hannam}\ \emph {et~al.}(2022)\citenamefont {Hannam}
  \emph {et~al.}}]{Hannam:2021pit}%
  \BibitemOpen
  \bibfield  {author} {\bibinfo {author} {\bibfnamefont {M.}~\bibnamefont
  {Hannam}} \emph {et~al.},\ }\href {\doibase 10.1038/s41586-022-05212-z}
  {\bibfield  {journal} {\bibinfo  {journal} {Nature}\ }\textbf {\bibinfo
  {volume} {610}},\ \bibinfo {pages} {652} (\bibinfo {year} {2022})},\ \Eprint
  {http://arxiv.org/abs/2112.11300} {arXiv:2112.11300 [gr-qc]} \BibitemShut
  {NoStop}%
\bibitem [{\citenamefont {Varma}\ \emph {et~al.}(2022)\citenamefont {Varma},
  \citenamefont {Biscoveanu}, \citenamefont {Islam}, \citenamefont {Shaik},
  \citenamefont {Haster}, \citenamefont {Isi}, \citenamefont {Farr},
  \citenamefont {Field},\ and\ \citenamefont {Vitale}}]{Varma:2022pld}%
  \BibitemOpen
  \bibfield  {author} {\bibinfo {author} {\bibfnamefont {V.}~\bibnamefont
  {Varma}}, \bibinfo {author} {\bibfnamefont {S.}~\bibnamefont {Biscoveanu}},
  \bibinfo {author} {\bibfnamefont {T.}~\bibnamefont {Islam}}, \bibinfo
  {author} {\bibfnamefont {F.~H.}\ \bibnamefont {Shaik}}, \bibinfo {author}
  {\bibfnamefont {C.-J.}\ \bibnamefont {Haster}}, \bibinfo {author}
  {\bibfnamefont {M.}~\bibnamefont {Isi}}, \bibinfo {author} {\bibfnamefont
  {W.~M.}\ \bibnamefont {Farr}}, \bibinfo {author} {\bibfnamefont {S.~E.}\
  \bibnamefont {Field}}, \ and\ \bibinfo {author} {\bibfnamefont
  {S.}~\bibnamefont {Vitale}},\ }\href {\doibase
  10.1103/PhysRevLett.128.191102} {\bibfield  {journal} {\bibinfo  {journal}
  {Phys. Rev. Lett.}\ }\textbf {\bibinfo {volume} {128}},\ \bibinfo {pages}
  {191102} (\bibinfo {year} {2022})},\ \Eprint
  {http://arxiv.org/abs/2201.01302} {arXiv:2201.01302 [astro-ph.HE]}
  \BibitemShut {NoStop}%
\bibitem [{\citenamefont {Cutler}\ and\ \citenamefont
  {Flanagan}(1994)}]{Cutler:1994ys}%
  \BibitemOpen
  \bibfield  {author} {\bibinfo {author} {\bibfnamefont {C.}~\bibnamefont
  {Cutler}}\ and\ \bibinfo {author} {\bibfnamefont {E.~E.}\ \bibnamefont
  {Flanagan}},\ }\href {\doibase 10.1103/PhysRevD.49.2658} {\bibfield
  {journal} {\bibinfo  {journal} {Phys. Rev.}\ }\textbf {\bibinfo {volume}
  {D49}},\ \bibinfo {pages} {2658} (\bibinfo {year} {1994})},\ \Eprint
  {http://arxiv.org/abs/gr-qc/9402014} {arXiv:gr-qc/9402014 [gr-qc]}
  \BibitemShut {NoStop}%
\bibitem [{\citenamefont {Poisson}\ and\ \citenamefont
  {Will}(1995)}]{Poisson:1995ef}%
  \BibitemOpen
  \bibfield  {author} {\bibinfo {author} {\bibfnamefont {E.}~\bibnamefont
  {Poisson}}\ and\ \bibinfo {author} {\bibfnamefont {C.~M.}\ \bibnamefont
  {Will}},\ }\href {\doibase 10.1103/PhysRevD.52.848} {\bibfield  {journal}
  {\bibinfo  {journal} {Phys. Rev.}\ }\textbf {\bibinfo {volume} {D52}},\
  \bibinfo {pages} {848} (\bibinfo {year} {1995})},\ \Eprint
  {http://arxiv.org/abs/gr-qc/9502040} {arXiv:gr-qc/9502040 [gr-qc]}
  \BibitemShut {NoStop}%
\bibitem [{\citenamefont {Ajith}(2011)}]{Ajith:2011ec}%
  \BibitemOpen
  \bibfield  {author} {\bibinfo {author} {\bibfnamefont {P.}~\bibnamefont
  {Ajith}},\ }\href {\doibase 10.1103/PhysRevD.84.084037} {\bibfield  {journal}
  {\bibinfo  {journal} {Phys. Rev. D}\ }\textbf {\bibinfo {volume} {84}},\
  \bibinfo {pages} {084037} (\bibinfo {year} {2011})},\ \Eprint
  {http://arxiv.org/abs/1107.1267} {arXiv:1107.1267 [gr-qc]} \BibitemShut
  {NoStop}%
\bibitem [{\citenamefont {{Ruiz}}\ \emph {et~al.}(2008)\citenamefont {{Ruiz}},
  \citenamefont {{Alcubierre}}, \citenamefont {{N{\'u}{\~n}ez}},\ and\
  \citenamefont {{Takahashi}}}]{Ruiz:2008GRe}%
  \BibitemOpen
  \bibfield  {author} {\bibinfo {author} {\bibfnamefont {M.}~\bibnamefont
  {{Ruiz}}}, \bibinfo {author} {\bibfnamefont {M.}~\bibnamefont
  {{Alcubierre}}}, \bibinfo {author} {\bibfnamefont {D.}~\bibnamefont
  {{N{\'u}{\~n}ez}}}, \ and\ \bibinfo {author} {\bibfnamefont {R.}~\bibnamefont
  {{Takahashi}}},\ }\href {\doibase 10.1007/s10714-007-0570-8} {\bibfield
  {journal} {\bibinfo  {journal} {General Relativity and Gravitation}\ }\textbf
  {\bibinfo {volume} {40}},\ \bibinfo {pages} {1705} (\bibinfo {year}
  {2008})},\ \Eprint {http://arxiv.org/abs/0707.4654} {arXiv:0707.4654 [gr-qc]}
  \BibitemShut {NoStop}%
\bibitem [{\citenamefont {{Hannam}}\ \emph {et~al.}(2014)\citenamefont
  {{Hannam}}, \citenamefont {{Schmidt}}, \citenamefont {{Boh{\'e}}},
  \citenamefont {{Haegel}}, \citenamefont {{Husa}}, \citenamefont {{Ohme}},
  \citenamefont {{Pratten}},\ and\ \citenamefont
  {{P{\"u}rrer}}}]{Hannam:2014ghd}%
  \BibitemOpen
  \bibfield  {author} {\bibinfo {author} {\bibfnamefont {M.}~\bibnamefont
  {{Hannam}}}, \bibinfo {author} {\bibfnamefont {P.}~\bibnamefont {{Schmidt}}},
  \bibinfo {author} {\bibfnamefont {A.}~\bibnamefont {{Boh{\'e}}}}, \bibinfo
  {author} {\bibfnamefont {L.}~\bibnamefont {{Haegel}}}, \bibinfo {author}
  {\bibfnamefont {S.}~\bibnamefont {{Husa}}}, \bibinfo {author} {\bibfnamefont
  {F.}~\bibnamefont {{Ohme}}}, \bibinfo {author} {\bibfnamefont
  {G.}~\bibnamefont {{Pratten}}}, \ and\ \bibinfo {author} {\bibfnamefont
  {M.}~\bibnamefont {{P{\"u}rrer}}},\ }\href {\doibase
  10.1103/PhysRevLett.113.151101} {\bibfield  {journal} {\bibinfo  {journal}
  {\prl}\ }\textbf {\bibinfo {volume} {113}},\ \bibinfo {eid} {151101}
  (\bibinfo {year} {2014})},\ \Eprint {http://arxiv.org/abs/1308.3271}
  {arXiv:1308.3271 [gr-qc]} \BibitemShut {NoStop}%
\bibitem [{\citenamefont {{Khan}}\ \emph {et~al.}(2020)\citenamefont {{Khan}},
  \citenamefont {{Ohme}}, \citenamefont {{Chatziioannou}},\ and\ \citenamefont
  {{Hannam}}}]{Khan:2020hdg}%
  \BibitemOpen
  \bibfield  {author} {\bibinfo {author} {\bibfnamefont {S.}~\bibnamefont
  {{Khan}}}, \bibinfo {author} {\bibfnamefont {F.}~\bibnamefont {{Ohme}}},
  \bibinfo {author} {\bibfnamefont {K.}~\bibnamefont {{Chatziioannou}}}, \ and\
  \bibinfo {author} {\bibfnamefont {M.}~\bibnamefont {{Hannam}}},\ }\href
  {\doibase 10.1103/PhysRevD.101.024056} {\bibfield  {journal} {\bibinfo
  {journal} {\prd}\ }\textbf {\bibinfo {volume} {101}},\ \bibinfo {eid}
  {024056} (\bibinfo {year} {2020})},\ \Eprint
  {http://arxiv.org/abs/1911.06050} {arXiv:1911.06050 [gr-qc]} \BibitemShut
  {NoStop}%
\bibitem [{\citenamefont {P\"urrer}(2016)}]{Purrer:2015tud}%
  \BibitemOpen
  \bibfield  {author} {\bibinfo {author} {\bibfnamefont {M.}~\bibnamefont
  {P\"urrer}},\ }\href {\doibase 10.1103/PhysRevD.93.064041} {\bibfield
  {journal} {\bibinfo  {journal} {Phys. Rev. D}\ }\textbf {\bibinfo {volume}
  {93}},\ \bibinfo {pages} {064041} (\bibinfo {year} {2016})},\ \Eprint
  {http://arxiv.org/abs/1512.02248} {arXiv:1512.02248 [gr-qc]} \BibitemShut
  {NoStop}%
\bibitem [{\citenamefont {{Boh{\'e}}}\ \emph {et~al.}(2017)\citenamefont
  {{Boh{\'e}}}, \citenamefont {{Shao}}, \citenamefont {{Taracchini}},
  \citenamefont {{Buonanno}}, \citenamefont {{Babak}}, \citenamefont {{Harry}},
  \citenamefont {{Hinder}}, \citenamefont {{Ossokine}}, \citenamefont
  {{P{\"u}rrer}}, \citenamefont {{Raymond}}, \citenamefont {{Chu}},
  \citenamefont {{Fong}}, \citenamefont {{Kumar}}, \citenamefont {{Pfeiffer}},
  \citenamefont {{Boyle}}, \citenamefont {{Hemberger}}, \citenamefont
  {{Kidder}}, \citenamefont {{Lovelace}}, \citenamefont {{Scheel}},\ and\
  \citenamefont {{Szil{\'a}gyi}}}]{Bohe:2017hgs}%
  \BibitemOpen
  \bibfield  {author} {\bibinfo {author} {\bibfnamefont {A.}~\bibnamefont
  {{Boh{\'e}}}}, \bibinfo {author} {\bibfnamefont {L.}~\bibnamefont {{Shao}}},
  \bibinfo {author} {\bibfnamefont {A.}~\bibnamefont {{Taracchini}}}, \bibinfo
  {author} {\bibfnamefont {A.}~\bibnamefont {{Buonanno}}}, \bibinfo {author}
  {\bibfnamefont {S.}~\bibnamefont {{Babak}}}, \bibinfo {author} {\bibfnamefont
  {I.~W.}\ \bibnamefont {{Harry}}}, \bibinfo {author} {\bibfnamefont
  {I.}~\bibnamefont {{Hinder}}}, \bibinfo {author} {\bibfnamefont
  {S.}~\bibnamefont {{Ossokine}}}, \bibinfo {author} {\bibfnamefont
  {M.}~\bibnamefont {{P{\"u}rrer}}}, \bibinfo {author} {\bibfnamefont
  {V.}~\bibnamefont {{Raymond}}}, \bibinfo {author} {\bibfnamefont
  {T.}~\bibnamefont {{Chu}}}, \bibinfo {author} {\bibfnamefont
  {H.}~\bibnamefont {{Fong}}}, \bibinfo {author} {\bibfnamefont
  {P.}~\bibnamefont {{Kumar}}}, \bibinfo {author} {\bibfnamefont {H.~P.}\
  \bibnamefont {{Pfeiffer}}}, \bibinfo {author} {\bibfnamefont
  {M.}~\bibnamefont {{Boyle}}}, \bibinfo {author} {\bibfnamefont {D.~A.}\
  \bibnamefont {{Hemberger}}}, \bibinfo {author} {\bibfnamefont {L.~E.}\
  \bibnamefont {{Kidder}}}, \bibinfo {author} {\bibfnamefont {G.}~\bibnamefont
  {{Lovelace}}}, \bibinfo {author} {\bibfnamefont {M.~A.}\ \bibnamefont
  {{Scheel}}}, \ and\ \bibinfo {author} {\bibfnamefont {B.}~\bibnamefont
  {{Szil{\'a}gyi}}},\ }\href {\doibase 10.1103/PhysRevD.95.044028} {\bibfield
  {journal} {\bibinfo  {journal} {\prd}\ }\textbf {\bibinfo {volume} {95}},\
  \bibinfo {eid} {044028} (\bibinfo {year} {2017})},\ \Eprint
  {http://arxiv.org/abs/1611.03703} {arXiv:1611.03703 [gr-qc]} \BibitemShut
  {NoStop}%
\bibitem [{\citenamefont {{Blackman}}\ \emph
  {et~al.}(2017{\natexlab{a}})\citenamefont {{Blackman}}, \citenamefont
  {{Field}}, \citenamefont {{Scheel}}, \citenamefont {{Galley}}, \citenamefont
  {{Hemberger}}, \citenamefont {{Schmidt}},\ and\ \citenamefont
  {{Smith}}}]{Blackman:2017sdr}%
  \BibitemOpen
  \bibfield  {author} {\bibinfo {author} {\bibfnamefont {J.}~\bibnamefont
  {{Blackman}}}, \bibinfo {author} {\bibfnamefont {S.~E.}\ \bibnamefont
  {{Field}}}, \bibinfo {author} {\bibfnamefont {M.~A.}\ \bibnamefont
  {{Scheel}}}, \bibinfo {author} {\bibfnamefont {C.~R.}\ \bibnamefont
  {{Galley}}}, \bibinfo {author} {\bibfnamefont {D.~A.}\ \bibnamefont
  {{Hemberger}}}, \bibinfo {author} {\bibfnamefont {P.}~\bibnamefont
  {{Schmidt}}}, \ and\ \bibinfo {author} {\bibfnamefont {R.}~\bibnamefont
  {{Smith}}},\ }\href {\doibase 10.1103/PhysRevD.95.104023} {\bibfield
  {journal} {\bibinfo  {journal} {\prd}\ }\textbf {\bibinfo {volume} {95}},\
  \bibinfo {eid} {104023} (\bibinfo {year} {2017}{\natexlab{a}})},\ \Eprint
  {http://arxiv.org/abs/1701.00550} {arXiv:1701.00550 [gr-qc]} \BibitemShut
  {NoStop}%
\bibitem [{\citenamefont {{Blackman}}\ \emph
  {et~al.}(2017{\natexlab{b}})\citenamefont {{Blackman}}, \citenamefont
  {{Field}}, \citenamefont {{Scheel}}, \citenamefont {{Galley}}, \citenamefont
  {{Ott}}, \citenamefont {{Boyle}}, \citenamefont {{Kidder}}, \citenamefont
  {{Pfeiffer}},\ and\ \citenamefont {{Szil{\'a}gyi}}}]{Blackman:2017nrw}%
  \BibitemOpen
  \bibfield  {author} {\bibinfo {author} {\bibfnamefont {J.}~\bibnamefont
  {{Blackman}}}, \bibinfo {author} {\bibfnamefont {S.~E.}\ \bibnamefont
  {{Field}}}, \bibinfo {author} {\bibfnamefont {M.~A.}\ \bibnamefont
  {{Scheel}}}, \bibinfo {author} {\bibfnamefont {C.~R.}\ \bibnamefont
  {{Galley}}}, \bibinfo {author} {\bibfnamefont {C.~D.}\ \bibnamefont {{Ott}}},
  \bibinfo {author} {\bibfnamefont {M.}~\bibnamefont {{Boyle}}}, \bibinfo
  {author} {\bibfnamefont {L.~E.}\ \bibnamefont {{Kidder}}}, \bibinfo {author}
  {\bibfnamefont {H.~P.}\ \bibnamefont {{Pfeiffer}}}, \ and\ \bibinfo {author}
  {\bibfnamefont {B.}~\bibnamefont {{Szil{\'a}gyi}}},\ }\href {\doibase
  10.1103/PhysRevD.96.024058} {\bibfield  {journal} {\bibinfo  {journal}
  {\prd}\ }\textbf {\bibinfo {volume} {96}},\ \bibinfo {eid} {024058} (\bibinfo
  {year} {2017}{\natexlab{b}})},\ \Eprint {http://arxiv.org/abs/1705.07089}
  {arXiv:1705.07089 [gr-qc]} \BibitemShut {NoStop}%
\bibitem [{\citenamefont {{LIGO Scientific Collaboration}}(2018)}]{lalsuite}%
  \BibitemOpen
  \bibfield  {author} {\bibinfo {author} {\bibnamefont {{LIGO Scientific
  Collaboration}}},\ }\href {https://doi.org/10.7935/GT1W-FZ16} {\emph
  {\bibinfo {title} {LIGO Algorithm Library}}},\ \bibinfo {organization}
  {doi.org/10.7935/GT1W-FZ16} (\bibinfo {year} {2018})\BibitemShut {NoStop}%
\bibitem [{\citenamefont {Payne}\ \emph {et~al.}(2022)\citenamefont {Payne},
  \citenamefont {Hourihane}, \citenamefont {Golomb}, \citenamefont {Udall},
  \citenamefont {Udall}, \citenamefont {Davis},\ and\ \citenamefont
  {Chatziioannou}}]{Payne:2022spz}%
  \BibitemOpen
  \bibfield  {author} {\bibinfo {author} {\bibfnamefont {E.}~\bibnamefont
  {Payne}}, \bibinfo {author} {\bibfnamefont {S.}~\bibnamefont {Hourihane}},
  \bibinfo {author} {\bibfnamefont {J.}~\bibnamefont {Golomb}}, \bibinfo
  {author} {\bibfnamefont {R.}~\bibnamefont {Udall}}, \bibinfo {author}
  {\bibfnamefont {R.}~\bibnamefont {Udall}}, \bibinfo {author} {\bibfnamefont
  {D.}~\bibnamefont {Davis}}, \ and\ \bibinfo {author} {\bibfnamefont
  {K.}~\bibnamefont {Chatziioannou}},\ }\href {\doibase
  10.1103/PhysRevD.106.104017} {\bibfield  {journal} {\bibinfo  {journal}
  {Phys. Rev. D}\ }\textbf {\bibinfo {volume} {106}},\ \bibinfo {pages}
  {104017} (\bibinfo {year} {2022})},\ \Eprint
  {http://arxiv.org/abs/2206.11932} {arXiv:2206.11932 [gr-qc]} \BibitemShut
  {NoStop}%
\bibitem [{\citenamefont {Hourihane}\ \emph {et~al.}(2022)\citenamefont
  {Hourihane}, \citenamefont {Chatziioannou}, \citenamefont {Wijngaarden},
  \citenamefont {Davis}, \citenamefont {Littenberg},\ and\ \citenamefont
  {Cornish}}]{Hourihane:2022doe}%
  \BibitemOpen
  \bibfield  {author} {\bibinfo {author} {\bibfnamefont {S.}~\bibnamefont
  {Hourihane}}, \bibinfo {author} {\bibfnamefont {K.}~\bibnamefont
  {Chatziioannou}}, \bibinfo {author} {\bibfnamefont {M.}~\bibnamefont
  {Wijngaarden}}, \bibinfo {author} {\bibfnamefont {D.}~\bibnamefont {Davis}},
  \bibinfo {author} {\bibfnamefont {T.}~\bibnamefont {Littenberg}}, \ and\
  \bibinfo {author} {\bibfnamefont {N.}~\bibnamefont {Cornish}},\ }\href
  {\doibase 10.1103/PhysRevD.106.042006} {\bibfield  {journal} {\bibinfo
  {journal} {Phys. Rev. D}\ }\textbf {\bibinfo {volume} {106}},\ \bibinfo
  {pages} {042006} (\bibinfo {year} {2022})},\ \Eprint
  {http://arxiv.org/abs/2205.13580} {arXiv:2205.13580 [gr-qc]} \BibitemShut
  {NoStop}%
\bibitem [{\citenamefont {Macas}\ \emph {et~al.}(2023)\citenamefont {Macas},
  \citenamefont {Lundgren},\ and\ \citenamefont {Ashton}}]{Macas:2023wiw}%
  \BibitemOpen
  \bibfield  {author} {\bibinfo {author} {\bibfnamefont {R.}~\bibnamefont
  {Macas}}, \bibinfo {author} {\bibfnamefont {A.}~\bibnamefont {Lundgren}}, \
  and\ \bibinfo {author} {\bibfnamefont {G.}~\bibnamefont {Ashton}},\
  }\href@noop {} {\  (\bibinfo {year} {2023})},\ \Eprint
  {http://arxiv.org/abs/2311.09921} {arXiv:2311.09921 [gr-qc]} \BibitemShut
  {NoStop}%
\bibitem [{\citenamefont {Veitch}\ \emph {et~al.}(2015)\citenamefont {Veitch},
  \citenamefont {Raymond}, \citenamefont {Farr}, \citenamefont {Farr},
  \citenamefont {Graff}, \citenamefont {Vitale}, \citenamefont {Aylott},
  \citenamefont {Blackburn}, \citenamefont {Christensen}, \citenamefont
  {Coughlin} \emph {et~al.}}]{veitch2015parameter}%
  \BibitemOpen
  \bibfield  {author} {\bibinfo {author} {\bibfnamefont {J.}~\bibnamefont
  {Veitch}}, \bibinfo {author} {\bibfnamefont {V.}~\bibnamefont {Raymond}},
  \bibinfo {author} {\bibfnamefont {B.}~\bibnamefont {Farr}}, \bibinfo {author}
  {\bibfnamefont {W.}~\bibnamefont {Farr}}, \bibinfo {author} {\bibfnamefont
  {P.}~\bibnamefont {Graff}}, \bibinfo {author} {\bibfnamefont
  {S.}~\bibnamefont {Vitale}}, \bibinfo {author} {\bibfnamefont
  {B.}~\bibnamefont {Aylott}}, \bibinfo {author} {\bibfnamefont
  {K.}~\bibnamefont {Blackburn}}, \bibinfo {author} {\bibfnamefont
  {N.}~\bibnamefont {Christensen}}, \bibinfo {author} {\bibfnamefont
  {M.}~\bibnamefont {Coughlin}},  \emph {et~al.},\ }\href@noop {} {\bibfield
  {journal} {\bibinfo  {journal} {Physical Review D}\ }\textbf {\bibinfo
  {volume} {91}},\ \bibinfo {pages} {042003} (\bibinfo {year}
  {2015})}\BibitemShut {NoStop}%
\bibitem [{\citenamefont {Abbott}\ \emph
  {et~al.}(2016{\natexlab{b}})\citenamefont {Abbott} \emph
  {et~al.}}]{LIGOScientific:2016aoc}%
  \BibitemOpen
  \bibfield  {author} {\bibinfo {author} {\bibfnamefont {B.~P.}\ \bibnamefont
  {Abbott}} \emph {et~al.} (\bibinfo {collaboration} {LIGO Scientific,
  Virgo}),\ }\href {\doibase 10.1103/PhysRevLett.116.061102} {\bibfield
  {journal} {\bibinfo  {journal} {Phys. Rev. Lett.}\ }\textbf {\bibinfo
  {volume} {116}},\ \bibinfo {pages} {061102} (\bibinfo {year}
  {2016}{\natexlab{b}})},\ \Eprint {http://arxiv.org/abs/1602.03837}
  {arXiv:1602.03837 [gr-qc]} \BibitemShut {NoStop}%
\bibitem [{\citenamefont {Abbott}\ \emph
  {et~al.}(2019{\natexlab{b}})\citenamefont {Abbott} \emph
  {et~al.}}]{LIGOScientific:2018mvr}%
  \BibitemOpen
  \bibfield  {author} {\bibinfo {author} {\bibfnamefont {B.~P.}\ \bibnamefont
  {Abbott}} \emph {et~al.} (\bibinfo {collaboration} {LIGO Scientific,
  Virgo}),\ }\href {\doibase 10.1103/PhysRevX.9.031040} {\bibfield  {journal}
  {\bibinfo  {journal} {Phys. Rev. X}\ }\textbf {\bibinfo {volume} {9}},\
  \bibinfo {pages} {031040} (\bibinfo {year} {2019}{\natexlab{b}})},\ \Eprint
  {http://arxiv.org/abs/1811.12907} {arXiv:1811.12907 [astro-ph.HE]}
  \BibitemShut {NoStop}%
\bibitem [{\citenamefont {Abbott}\ \emph
  {et~al.}(2021{\natexlab{c}})\citenamefont {Abbott} \emph
  {et~al.}}]{LIGOScientific:2020ibl}%
  \BibitemOpen
  \bibfield  {author} {\bibinfo {author} {\bibfnamefont {R.}~\bibnamefont
  {Abbott}} \emph {et~al.} (\bibinfo {collaboration} {LIGO Scientific,
  Virgo}),\ }\href {\doibase 10.1103/PhysRevX.11.021053} {\bibfield  {journal}
  {\bibinfo  {journal} {Phys. Rev. X}\ }\textbf {\bibinfo {volume} {11}},\
  \bibinfo {pages} {021053} (\bibinfo {year} {2021}{\natexlab{c}})},\ \Eprint
  {http://arxiv.org/abs/2010.14527} {arXiv:2010.14527 [gr-qc]} \BibitemShut
  {NoStop}%
\bibitem [{\citenamefont {Abbott}\ \emph
  {et~al.}(2021{\natexlab{d}})\citenamefont {Abbott} \emph
  {et~al.}}]{Abbott:2021GWTC21}%
  \BibitemOpen
  \bibfield  {author} {\bibinfo {author} {\bibfnamefont {B.~P.}\ \bibnamefont
  {Abbott}} \emph {et~al.} (\bibinfo {collaboration} {LIGO Scientific,
  Virgo}),\ }\href {\doibase 10.48550/arXiv.2108.01045} {\bibfield  {journal}
  {\bibinfo  {journal} {arXiv e-prints}\ ,\ \bibinfo {eid} {arXiv:2108.01045}}
  (\bibinfo {year} {2021}{\natexlab{d}})},\ \Eprint
  {http://arxiv.org/abs/2108.01045} {arXiv:2108.01045 [gr-qc]} \BibitemShut
  {NoStop}%
\bibitem [{\citenamefont {Gelman}\ and\ \citenamefont
  {Rubin}(1992)}]{gelman1992inference}%
  \BibitemOpen
  \bibfield  {author} {\bibinfo {author} {\bibfnamefont {A.}~\bibnamefont
  {Gelman}}\ and\ \bibinfo {author} {\bibfnamefont {D.~B.}\ \bibnamefont
  {Rubin}},\ }\href@noop {} {\bibfield  {journal} {\bibinfo  {journal}
  {Statistical science}\ }\textbf {\bibinfo {volume} {7}},\ \bibinfo {pages}
  {457} (\bibinfo {year} {1992})}\BibitemShut {NoStop}%
\bibitem [{\citenamefont {{Cutler}}\ and\ \citenamefont
  {{Flanagan}}(1994)}]{Cutler:1994gmc}%
  \BibitemOpen
  \bibfield  {author} {\bibinfo {author} {\bibfnamefont {C.}~\bibnamefont
  {{Cutler}}}\ and\ \bibinfo {author} {\bibfnamefont {{\'E}.~E.}\ \bibnamefont
  {{Flanagan}}},\ }\href {\doibase 10.1103/PhysRevD.49.2658} {\bibfield
  {journal} {\bibinfo  {journal} {\prd}\ }\textbf {\bibinfo {volume} {49}},\
  \bibinfo {pages} {2658} (\bibinfo {year} {1994})},\ \Eprint
  {http://arxiv.org/abs/gr-qc/9402014} {arXiv:gr-qc/9402014 [gr-qc]}
  \BibitemShut {NoStop}%
\bibitem [{\citenamefont {{Finn}}(1992)}]{Finn:1992dmg}%
  \BibitemOpen
  \bibfield  {author} {\bibinfo {author} {\bibfnamefont {L.~S.}\ \bibnamefont
  {{Finn}}},\ }\href {\doibase 10.1103/PhysRevD.46.5236} {\bibfield  {journal}
  {\bibinfo  {journal} {\prd}\ }\textbf {\bibinfo {volume} {46}},\ \bibinfo
  {pages} {5236} (\bibinfo {year} {1992})},\ \Eprint
  {http://arxiv.org/abs/gr-qc/9209010} {arXiv:gr-qc/9209010 [gr-qc]}
  \BibitemShut {NoStop}%
\bibitem [{\citenamefont {Schmidt}\ \emph {et~al.}(2017)\citenamefont
  {Schmidt}, \citenamefont {Harry},\ and\ \citenamefont
  {Pfeiffer}}]{schmidt2017numerical}%
  \BibitemOpen
  \bibfield  {author} {\bibinfo {author} {\bibfnamefont {P.}~\bibnamefont
  {Schmidt}}, \bibinfo {author} {\bibfnamefont {I.~W.}\ \bibnamefont {Harry}},
  \ and\ \bibinfo {author} {\bibfnamefont {H.~P.}\ \bibnamefont {Pfeiffer}},\
  }\href@noop {} {\bibfield  {journal} {\bibinfo  {journal} {arXiv preprint
  arXiv:1703.01076}\ } (\bibinfo {year} {2017})}\BibitemShut {NoStop}%
\bibitem [{\citenamefont {Ruiz}\ \emph {et~al.}(2008)\citenamefont {Ruiz},
  \citenamefont {Takahashi}, \citenamefont {Alcubierre},\ and\ \citenamefont
  {Nunez}}]{Ruiz:2007yx}%
  \BibitemOpen
  \bibfield  {author} {\bibinfo {author} {\bibfnamefont {M.}~\bibnamefont
  {Ruiz}}, \bibinfo {author} {\bibfnamefont {R.}~\bibnamefont {Takahashi}},
  \bibinfo {author} {\bibfnamefont {M.}~\bibnamefont {Alcubierre}}, \ and\
  \bibinfo {author} {\bibfnamefont {D.}~\bibnamefont {Nunez}},\ }\href
  {\doibase 10.1007/s10714-007-0570-8} {\bibfield  {journal} {\bibinfo
  {journal} {Gen. Rel. Grav.}\ }\textbf {\bibinfo {volume} {40}},\ \bibinfo
  {pages} {2467} (\bibinfo {year} {2008})},\ \Eprint
  {http://arxiv.org/abs/0707.4654} {arXiv:0707.4654 [gr-qc]} \BibitemShut
  {NoStop}%
\bibitem [{\citenamefont {{Schmidt}}\ \emph {et~al.}(2011)\citenamefont
  {{Schmidt}}, \citenamefont {{Hannam}}, \citenamefont {{Husa}},\ and\
  \citenamefont {{Ajith}}}]{Schmidt:2011tpc}%
  \BibitemOpen
  \bibfield  {author} {\bibinfo {author} {\bibfnamefont {P.}~\bibnamefont
  {{Schmidt}}}, \bibinfo {author} {\bibfnamefont {M.}~\bibnamefont {{Hannam}}},
  \bibinfo {author} {\bibfnamefont {S.}~\bibnamefont {{Husa}}}, \ and\ \bibinfo
  {author} {\bibfnamefont {P.}~\bibnamefont {{Ajith}}},\ }\href {\doibase
  10.1103/PhysRevD.84.024046} {\bibfield  {journal} {\bibinfo  {journal}
  {\prd}\ }\textbf {\bibinfo {volume} {84}},\ \bibinfo {eid} {024046} (\bibinfo
  {year} {2011})},\ \Eprint {http://arxiv.org/abs/1012.2879} {arXiv:1012.2879
  [gr-qc]} \BibitemShut {NoStop}%
\bibitem [{\citenamefont {{Baird}}\ \emph {et~al.}(2013)\citenamefont
  {{Baird}}, \citenamefont {{Fairhurst}}, \citenamefont {{Hannam}},\ and\
  \citenamefont {{Murphy}}}]{Baird:2013dbm}%
  \BibitemOpen
  \bibfield  {author} {\bibinfo {author} {\bibfnamefont {E.}~\bibnamefont
  {{Baird}}}, \bibinfo {author} {\bibfnamefont {S.}~\bibnamefont
  {{Fairhurst}}}, \bibinfo {author} {\bibfnamefont {M.}~\bibnamefont
  {{Hannam}}}, \ and\ \bibinfo {author} {\bibfnamefont {P.}~\bibnamefont
  {{Murphy}}},\ }\href {\doibase 10.1103/PhysRevD.87.024035} {\bibfield
  {journal} {\bibinfo  {journal} {\prd}\ }\textbf {\bibinfo {volume} {87}},\
  \bibinfo {eid} {024035} (\bibinfo {year} {2013})},\ \Eprint
  {http://arxiv.org/abs/1211.0546} {arXiv:1211.0546 [gr-qc]} \BibitemShut
  {NoStop}%
\bibitem [{\citenamefont {{Hoy}}(2022)}]{Hoy:202203H}%
  \BibitemOpen
  \bibfield  {author} {\bibinfo {author} {\bibfnamefont {C.}~\bibnamefont
  {{Hoy}}},\ }\href {\doibase 10.1103/PhysRevD.106.083003} {\bibfield
  {journal} {\bibinfo  {journal} {\prd}\ }\textbf {\bibinfo {volume} {106}},\
  \bibinfo {eid} {083003} (\bibinfo {year} {2022})},\ \Eprint
  {http://arxiv.org/abs/2208.00106} {arXiv:2208.00106 [gr-qc]} \BibitemShut
  {NoStop}%
\bibitem [{\citenamefont {Ramos-Buades}\ \emph
  {et~al.}(2023{\natexlab{b}})\citenamefont {Ramos-Buades}, \citenamefont
  {Buonanno}, \citenamefont {Estell{\'e}s}, \citenamefont {Khalil},
  \citenamefont {Mihaylov}, \citenamefont {Ossokine}, \citenamefont {Pompili},\
  and\ \citenamefont {Shiferaw}}]{ramos2023seobnrv5phm}%
  \BibitemOpen
  \bibfield  {author} {\bibinfo {author} {\bibfnamefont {A.}~\bibnamefont
  {Ramos-Buades}}, \bibinfo {author} {\bibfnamefont {A.}~\bibnamefont
  {Buonanno}}, \bibinfo {author} {\bibfnamefont {H.}~\bibnamefont
  {Estell{\'e}s}}, \bibinfo {author} {\bibfnamefont {M.}~\bibnamefont
  {Khalil}}, \bibinfo {author} {\bibfnamefont {D.~P.}\ \bibnamefont
  {Mihaylov}}, \bibinfo {author} {\bibfnamefont {S.}~\bibnamefont {Ossokine}},
  \bibinfo {author} {\bibfnamefont {L.}~\bibnamefont {Pompili}}, \ and\
  \bibinfo {author} {\bibfnamefont {M.}~\bibnamefont {Shiferaw}},\ }\href@noop
  {} {\bibfield  {journal} {\bibinfo  {journal} {arXiv preprint
  arXiv:2303.18046}\ } (\bibinfo {year} {2023}{\natexlab{b}})}\BibitemShut
  {NoStop}%
\bibitem [{\citenamefont {Thompson}\ \emph {et~al.}(2023)\citenamefont
  {Thompson}, \citenamefont {Hamilton}, \citenamefont {London}, \citenamefont
  {Ghosh}, \citenamefont {Kolitsidou}, \citenamefont {Hoy},\ and\ \citenamefont
  {Hannam}}]{Thompson:2023}%
  \BibitemOpen
  \bibfield  {author} {\bibinfo {author} {\bibfnamefont {J.~E.}\ \bibnamefont
  {Thompson}}, \bibinfo {author} {\bibfnamefont {E.}~\bibnamefont {Hamilton}},
  \bibinfo {author} {\bibfnamefont {L.}~\bibnamefont {London}}, \bibinfo
  {author} {\bibfnamefont {S.}~\bibnamefont {Ghosh}}, \bibinfo {author}
  {\bibfnamefont {P.}~\bibnamefont {Kolitsidou}}, \bibinfo {author}
  {\bibfnamefont {C.}~\bibnamefont {Hoy}}, \ and\ \bibinfo {author}
  {\bibfnamefont {M.}~\bibnamefont {Hannam}},\ }\href@noop {} {\  (\bibinfo
  {year} {2023})},\ \bibinfo {note} {in preparation}\BibitemShut {NoStop}%
\bibitem [{\citenamefont {Hunter}(2007)}]{Matplotlib}%
  \BibitemOpen
  \bibfield  {author} {\bibinfo {author} {\bibfnamefont {J.~D.}\ \bibnamefont
  {Hunter}},\ }\href {\doibase 10.1109/MCSE.2007.55} {\bibfield  {journal}
  {\bibinfo  {journal} {Computing in Science \& Engineering}\ }\textbf
  {\bibinfo {volume} {9}},\ \bibinfo {pages} {90} (\bibinfo {year}
  {2007})}\BibitemShut {NoStop}%
\bibitem [{\citenamefont {Hoy}\ and\ \citenamefont
  {Raymond}(2021)}]{Hoy:PESummary}%
  \BibitemOpen
  \bibfield  {author} {\bibinfo {author} {\bibfnamefont {C.}~\bibnamefont
  {Hoy}}\ and\ \bibinfo {author} {\bibfnamefont {V.}~\bibnamefont {Raymond}},\
  }\href {\doibase 10.1016/j.softx.2021.100765} {\bibfield  {journal} {\bibinfo
   {journal} {SoftwareX}\ }\textbf {\bibinfo {volume} {15}},\ \bibinfo {pages}
  {100765} (\bibinfo {year} {2021})},\ \Eprint
  {http://arxiv.org/abs/2006.06639} {arXiv:2006.06639 [astro-ph.IM]}
  \BibitemShut {NoStop}%
\bibitem [{\citenamefont {Harris}\ \emph {et~al.}(2020)\citenamefont {Harris},
  \citenamefont {Millman}, \citenamefont {van~der Walt}, \citenamefont
  {Gommers}, \citenamefont {Virtanen}, \citenamefont {Cournapeau},
  \citenamefont {Wieser}, \citenamefont {Taylor}, \citenamefont {Berg},
  \citenamefont {Smith}, \citenamefont {Kern}, \citenamefont {Picus},
  \citenamefont {Hoyer}, \citenamefont {van Kerkwijk}, \citenamefont {Brett},
  \citenamefont {Haldane}, \citenamefont {del R{\'{i}}o}, \citenamefont
  {Wiebe}, \citenamefont {Peterson}, \citenamefont {G{\'{e}}rard-Marchant},
  \citenamefont {Sheppard}, \citenamefont {Reddy}, \citenamefont {Weckesser},
  \citenamefont {Abbasi}, \citenamefont {Gohlke},\ and\ \citenamefont
  {Oliphant}}]{numpy}%
  \BibitemOpen
  \bibfield  {author} {\bibinfo {author} {\bibfnamefont {C.~R.}\ \bibnamefont
  {Harris}}, \bibinfo {author} {\bibfnamefont {K.~J.}\ \bibnamefont {Millman}},
  \bibinfo {author} {\bibfnamefont {S.~J.}\ \bibnamefont {van~der Walt}},
  \bibinfo {author} {\bibfnamefont {R.}~\bibnamefont {Gommers}}, \bibinfo
  {author} {\bibfnamefont {P.}~\bibnamefont {Virtanen}}, \bibinfo {author}
  {\bibfnamefont {D.}~\bibnamefont {Cournapeau}}, \bibinfo {author}
  {\bibfnamefont {E.}~\bibnamefont {Wieser}}, \bibinfo {author} {\bibfnamefont
  {J.}~\bibnamefont {Taylor}}, \bibinfo {author} {\bibfnamefont
  {S.}~\bibnamefont {Berg}}, \bibinfo {author} {\bibfnamefont {N.~J.}\
  \bibnamefont {Smith}}, \bibinfo {author} {\bibfnamefont {R.}~\bibnamefont
  {Kern}}, \bibinfo {author} {\bibfnamefont {M.}~\bibnamefont {Picus}},
  \bibinfo {author} {\bibfnamefont {S.}~\bibnamefont {Hoyer}}, \bibinfo
  {author} {\bibfnamefont {M.~H.}\ \bibnamefont {van Kerkwijk}}, \bibinfo
  {author} {\bibfnamefont {M.}~\bibnamefont {Brett}}, \bibinfo {author}
  {\bibfnamefont {A.}~\bibnamefont {Haldane}}, \bibinfo {author} {\bibfnamefont
  {J.~F.}\ \bibnamefont {del R{\'{i}}o}}, \bibinfo {author} {\bibfnamefont
  {M.}~\bibnamefont {Wiebe}}, \bibinfo {author} {\bibfnamefont
  {P.}~\bibnamefont {Peterson}}, \bibinfo {author} {\bibfnamefont
  {P.}~\bibnamefont {G{\'{e}}rard-Marchant}}, \bibinfo {author} {\bibfnamefont
  {K.}~\bibnamefont {Sheppard}}, \bibinfo {author} {\bibfnamefont
  {T.}~\bibnamefont {Reddy}}, \bibinfo {author} {\bibfnamefont
  {W.}~\bibnamefont {Weckesser}}, \bibinfo {author} {\bibfnamefont
  {H.}~\bibnamefont {Abbasi}}, \bibinfo {author} {\bibfnamefont
  {C.}~\bibnamefont {Gohlke}}, \ and\ \bibinfo {author} {\bibfnamefont {T.~E.}\
  \bibnamefont {Oliphant}},\ }\href {\doibase 10.1038/s41586-020-2649-2}
  {\bibfield  {journal} {\bibinfo  {journal} {Nature}\ }\textbf {\bibinfo
  {volume} {585}},\ \bibinfo {pages} {357} (\bibinfo {year}
  {2020})}\BibitemShut {NoStop}%
\bibitem [{\citenamefont {Virtanen}\ \emph {et~al.}(2020)\citenamefont
  {Virtanen}, \citenamefont {Gommers}, \citenamefont {Oliphant}, \citenamefont
  {Haberland}, \citenamefont {Reddy}, \citenamefont {Cournapeau}, \citenamefont
  {Burovski}, \citenamefont {Peterson}, \citenamefont {Weckesser},
  \citenamefont {Bright}, \citenamefont {{van der Walt}}, \citenamefont
  {Brett}, \citenamefont {Wilson}, \citenamefont {Millman}, \citenamefont
  {Mayorov}, \citenamefont {Nelson}, \citenamefont {Jones}, \citenamefont
  {Kern}, \citenamefont {Larson}, \citenamefont {Carey}, \citenamefont {Polat},
  \citenamefont {Feng}, \citenamefont {Moore}, \citenamefont {{VanderPlas}},
  \citenamefont {Laxalde}, \citenamefont {Perktold}, \citenamefont {Cimrman},
  \citenamefont {Henriksen}, \citenamefont {Quintero}, \citenamefont {Harris},
  \citenamefont {Archibald}, \citenamefont {Ribeiro}, \citenamefont
  {Pedregosa}, \citenamefont {{van Mulbregt}},\ and\ \citenamefont {{SciPy 1.0
  Contributors}}}]{Scipy}%
  \BibitemOpen
  \bibfield  {author} {\bibinfo {author} {\bibfnamefont {P.}~\bibnamefont
  {Virtanen}}, \bibinfo {author} {\bibfnamefont {R.}~\bibnamefont {Gommers}},
  \bibinfo {author} {\bibfnamefont {T.~E.}\ \bibnamefont {Oliphant}}, \bibinfo
  {author} {\bibfnamefont {M.}~\bibnamefont {Haberland}}, \bibinfo {author}
  {\bibfnamefont {T.}~\bibnamefont {Reddy}}, \bibinfo {author} {\bibfnamefont
  {D.}~\bibnamefont {Cournapeau}}, \bibinfo {author} {\bibfnamefont
  {E.}~\bibnamefont {Burovski}}, \bibinfo {author} {\bibfnamefont
  {P.}~\bibnamefont {Peterson}}, \bibinfo {author} {\bibfnamefont
  {W.}~\bibnamefont {Weckesser}}, \bibinfo {author} {\bibfnamefont
  {J.}~\bibnamefont {Bright}}, \bibinfo {author} {\bibfnamefont {S.~J.}\
  \bibnamefont {{van der Walt}}}, \bibinfo {author} {\bibfnamefont
  {M.}~\bibnamefont {Brett}}, \bibinfo {author} {\bibfnamefont
  {J.}~\bibnamefont {Wilson}}, \bibinfo {author} {\bibfnamefont {K.~J.}\
  \bibnamefont {Millman}}, \bibinfo {author} {\bibfnamefont {N.}~\bibnamefont
  {Mayorov}}, \bibinfo {author} {\bibfnamefont {A.~R.~J.}\ \bibnamefont
  {Nelson}}, \bibinfo {author} {\bibfnamefont {E.}~\bibnamefont {Jones}},
  \bibinfo {author} {\bibfnamefont {R.}~\bibnamefont {Kern}}, \bibinfo {author}
  {\bibfnamefont {E.}~\bibnamefont {Larson}}, \bibinfo {author} {\bibfnamefont
  {C.~J.}\ \bibnamefont {Carey}}, \bibinfo {author} {\bibfnamefont
  {{\.I}.}~\bibnamefont {Polat}}, \bibinfo {author} {\bibfnamefont
  {Y.}~\bibnamefont {Feng}}, \bibinfo {author} {\bibfnamefont {E.~W.}\
  \bibnamefont {Moore}}, \bibinfo {author} {\bibfnamefont {J.}~\bibnamefont
  {{VanderPlas}}}, \bibinfo {author} {\bibfnamefont {D.}~\bibnamefont
  {Laxalde}}, \bibinfo {author} {\bibfnamefont {J.}~\bibnamefont {Perktold}},
  \bibinfo {author} {\bibfnamefont {R.}~\bibnamefont {Cimrman}}, \bibinfo
  {author} {\bibfnamefont {I.}~\bibnamefont {Henriksen}}, \bibinfo {author}
  {\bibfnamefont {E.~A.}\ \bibnamefont {Quintero}}, \bibinfo {author}
  {\bibfnamefont {C.~R.}\ \bibnamefont {Harris}}, \bibinfo {author}
  {\bibfnamefont {A.~M.}\ \bibnamefont {Archibald}}, \bibinfo {author}
  {\bibfnamefont {A.~H.}\ \bibnamefont {Ribeiro}}, \bibinfo {author}
  {\bibfnamefont {F.}~\bibnamefont {Pedregosa}}, \bibinfo {author}
  {\bibfnamefont {P.}~\bibnamefont {{van Mulbregt}}}, \ and\ \bibinfo {author}
  {\bibnamefont {{SciPy 1.0 Contributors}}},\ }\href {\doibase
  10.1038/s41592-019-0686-2} {\bibfield  {journal} {\bibinfo  {journal} {Nature
  Methods}\ }\textbf {\bibinfo {volume} {17}},\ \bibinfo {pages} {261}
  (\bibinfo {year} {2020})}\BibitemShut {NoStop}%
\end{thebibliography}%

\end{document}